\documentclass[graybox]{svmult}

\usepackage{type1cm}        
\usepackage{makeidx}         
\usepackage{graphicx}        
\usepackage{multicol}        
\usepackage[bottom]{footmisc}

\usepackage{newtxtext}       %
\usepackage{newtxmath}       
\UseRawInputEncoding
\def\bra#1{\langle#1 |}
\def\ket#1{| #1\rangle}

\makeindex             

\let\ACMmaketitle=\maketitle
\renewcommand{\maketitle}{\begingroup\let\footnote=\thanks \ACMmaketitle\endgroup}

\begin{document}

\title*{Subgap states in semiconductor-superconductor devices for quantum technologies: Andreev qubits and minimal Majorana chains \footnote{This is a preprint of
the following chapter: Rub\'en Seoane and Ram\'on Aguado, "\emph{Subgap States in Semiconductor-Superconductor Devices for Quantum Technologies: Andreev Qubits and Minimal Majorana Chains}", 
to be published in \emph{New Trends and Platforms for Quantum Technologies}, edited by Ram\'on Aguado, Roberta Citro, Maciej Lewenstein and Michael Stern, 2024, Springer reproduced with permission of Springer Nature Switzerland AG. The final authenticated version is available online at: http://dx.doi.org/[insert DOI]}}
\titlerunning{Subgap states in semiconductor-superconductor devices for quantum technologies}
\author{Rub\'en Seoane Souto and Ram\'on Aguado}
\institute{Rub\'en Seoane Souto \at Instituto de Ciencia de Materiales de Madrid (ICMM), Consejo Superior de Investigaciones Cient\'ificas (CSIC), Sor Juana In\'es de la Cruz 3, 28049 Madrid, Spain, \email{ruben.seoane@csic.es}
\and Ram\'on Aguado \at Instituto de Ciencia de Materiales de Madrid (ICMM), Consejo Superior de Investigaciones Cient\'ificas (CSIC), Sor Juana In\'es de la Cruz 3, 28049 Madrid, Spain, \email{ramon.aguado@csic.es}}
\maketitle

\abstract*{Each chapter should be preceded by an abstract (no more than 200 words) that summarizes the content. The abstract will appear \textit{online} at \url{www.SpringerLink.com} and be available with unrestricted access. This allows unregistered users to read the abstract as a teaser for the complete chapter.
Please use the 'starred' version of the \texttt{abstract} command for typesetting the text of the online abstracts (cf. source file of this chapter template \texttt{abstract}) and include them with the source files of your manuscript. Use the plain \texttt{abstract} command if the abstract is also to appear in the printed version of the book.}

\abstract{In recent years, experimental advances have made it possible to achieve an unprecedented degree of control over the properties of subgap bound states in hybrid nanoscale superconducting structures. This research has been driven by the promise of engineering subgap states  for quantum applications, which includes Majorana zero modes predicted to appear at the interface of superconductor and other materials, like topological insulators or semiconductors. In this chapter, we revise the status of the field towards the engineering of quantum devices in controllable semiconductor-superconductor heterostructures. We begin the chapter with a brief introduction about subgap states, focusing on their mathematical formulation. After introducing topological superconductivity using the Kitaev model, we discuss the advances in the search for Majorana states over the last few years, highlighting the difficulties of unambiguously distinguish these states from nontopological subgap states. In recent years, the precise engineering of bound states by a bottom-up approach using quantum dots has led to unprecedented experimental advances, including experimental demonstrations of an Andreev qubits based on a quantum dot Josephson junction and a minimal Kitaev chain based on two quantum dots coherently coupled by the bound states of an intermediate superconducting segment. These experimental advances have revitalized the field and helped to understand that, far from being a disadvantage, the presence of subgap bound states can be exploited for new qubit designs and quantum coherence experiments, including Majorana-based qubits.}

\begin{overview}{Why hybrid semiconductor-superconductor qubits based on subgap states?}
\label{sec:1}
Despite the amazing progress of the last twenty years in qubit research \cite{CommPhys2023}, going beyond the actual sizes, with a few tens of noisy qubits \cite{Preskill2018}, to scalable quantum processors with thousands of qubits is far from trivial and represents an engineering challenge \cite{Kjaergaard:2020}. If we focus on solid state implementations, transmon qubits in superconducting circuits \cite{PhysRevA.76.042319,RevModPhys.93.025005} and spin qubits \cite{Chatterjee:2021,Stano:2022,RevModPhys.95.025003} in semiconductors are currently two of the most promising platforms for quantum computing, but both have their pros and cons. Spin qubits, on one hand, are promising from a scalability standpoint and compatibility with industrial semiconductor processing. However, realizing spin-spin interactions over extended distances still represents a challenge.  Transmon-based circuits, on the other hand, can be readily controlled, read and coupled over long distances using circuit quantum electrodynamics (QED) techniques. However, they have a limited speed of qubit operations and they are relatively large, making them prone to unwanted cross-coupling with other circuit elements. 

Hybrid combinations of materials provide an interesting change of paradigm, trying to benefit from combining superconductors with semiconductors. This platform provides can be controlled by voltages that are less susceptible to heating and crosstalk. The first demonstration of this concept was the so-called gatemon qubit: a superconducting transmon qubit based on Josephson junctions with gate-tunable semiconducting elements \cite{PhysRevLett.115.127001,PhysRevLett.115.127002,Casparis-Nature-Nano2018,Wang-NatureNano2019}. Following this demonstration, new routes have been opened for the study of various forms of hybrid semiconductor–superconductor qubits. Among them, one interesting option if to exploit fermionic-like Bogoliubov de Gennes quasiparticle degrees of freedom in the Josephson junction, which manifest as subgap states generically known as Andreev bound states. The combination of such fermionic degrees of freedom with the bosonic ones associated to the superconducting circuit results in a plethora of novel physics and functionalities.

Using this general concept, various designs have been proposed.  One interesting option is to encode a qubit in the spin degree of freedom of a quasiparticle occupying a so-called Andreev bound state in the junction, leading to an effective spin 1/2 degree of freedom where information can be encoded. This qubit is the superconducting counterpart of conventional spin qubits, realized in quantum dots~\cite{RevModPhys.95.025003}, and is hence dubbed superconducting spin qubit \cite{PhysRevB.81.144519} or Andreev spin qubit. An important point that must be stressed is that the qubit uses true spin states in the odd fermionic parity sector of the junction as opposed to the Andreev pair qubit \cite{PhysRevLett.90.087003,doi:10.1126/science.aab2179} which uses the ground and excited states of a standard Andreev level (even parity) as a qubit. 
Importantly, the intrinsic spin-supercurrent coupling of the Andreev spin qubit facilitates fast, high-fidelity readout and monitoring of the state of the spin qubit using circuit quantum electrodynamics techniques \cite{HaysNaturePhysics2020,PRXQuantum.3.030311,PhysRevLett.131.097001,PhysRevLett.129.227701}. Recent developments that consolidate this architecture as an interesting alternative to other more mature platforms include the demonstration of strong coherent qubit-qubit coupling between a transmon and an Andreev spin qubit~\cite{Pita-Vidal_NatPhys2023}, as well as strong tunable coupling between two distant qubits \cite{Pita-Vidal_arXiv2023}.

The generalization of this idea to other types of bound states in superconductors and even to "pseudo-spin" degrees of freedom based on topological states with Majorana zero modes, undoubtedly makes the field of semiconductor-superconductor hybrid qubits one of the most vibrant in recent years. In particular, a large part of the interest in hybrids is motivated by the recent experimental demonstration ~\cite{Dvir2023} of bottom-up engineering of a minimal Kitaev chain \cite{Kitaev_2001}. The degree of control on these devices is exquisite. This has allowed the technical challenge of inducing triplet-like equal spin superconducting electron pairing, both in nanowires \cite{Wang_Nat2022,Bordoloi_Nat2022} and in two-dimensional electron gases~\cite{Wang_Natcom2023}. Remarkably, already a minimal Kitaev chain with only two quantum dots separated by a superconductor can
host Majorana bound states (MBS)~\cite{Leijnse_PRB2012}.
These experiments have given new impetus to a field that seemed somewhat stagnant after more than ten years trying to detect Majorana bound states in hybrid semiconducting-superconducting nanowires \cite{Prada_review}. 

Using such minimal Kitaev chains, the field seems ripe for the demonstration of the basic steps toward a minimal Majorana qubit. In the long run, however, two key steps must be completed before Majorana qubits can be used in topological quantum computation \cite{10.1063/PT.3.4499}. First, researchers must unambiguously establish that fully nonlocal Majoranas can be fabricated in the laboratory in longer quantum dot chains, with the necessary topological protection. More importantly, unambiguous demonstration of non-abelian braiding of Majoranas would represent an unprecedented breakthrough for fundamental physics.

\begin{table}
\caption{For the shake of clarity and readibility, we present here a list of acronyms used along the chapter.}
\begin{center}
\begin{tabular}{ |c|c|c| } 
 \hline
 Acronym & Meaning\\
 \hline
 ABS & Andreev bound state\\
 CAR & Crossed Andreeev reflection\\
 ECT & elastic cotunneling\\
 MBS & Majorana bound state\\
 MP & Majorana polarization\\
 MZM & Majorana zero mode\\
 QD & quantum dot\\
 PMM & Poor man's Majorana\\ 
 YSR & Yu-Shiba-Rusinov\\ 
 \hline
\end{tabular}
\end{center}
\end{table}
\end{overview}

\section{Subgap states in superconductors}
\label{sec2}
Discrete states may appear below the superconducting gap for various reasons, such as the existence of normal (non-superconducting) regions within the SC, the presence of magnetic fields or magnetic impurities \cite{RevModPhys.78.373}, which develop so-called Yu-Shiba-Rusinov (YSR) states \cite{Yu:APS65,Shiba:POTP68,Rusinov:JL69}, disorder, etc. All these subgap states are generically known as Andreev bound states (ABS) \cite{Sauls:PTRSA18,Meden_2019,Alvaro-Alfredo-Review}. In the last decades, the discovery of topological materials has allowed for a new twist in the possibilities to engineer subgap states in superconductors. Inspired by notions of topology such as topological invariants and bound states at topological defects, several authors have predicted the existence of a new phase of matter known as the topological superconducting state. This topological phase is characterized by the emergence of a rather special type of bound states occurring at precisely zero energy, where the electron and the hole contribute with exactly half the amplitude to form a quasiparticle. Such a zero energy quasiparticle is equal to its own antiparticle and has therefore Majorana character. As opposed to standard ABSs that can be pushed out of the gap by continuous gap-preserving deformations of the Hamiltonian, well-separated MBSs cannot be removed from zero energy by any local perturbation or noise. This is possible owing to the bulk-boundary correspondence principle of band topology, which predicts that at the boundaries of topological materials with non-topological ones, edge states must appear that are protected against perturbations by the topology of the bulk. Quite remarkably, MBSs do not follow fermion statistics, unlike the original particles predicted by Ettore Majorana, but rather possess non-Abelian exchange statistics which, together with their topological protection against local noise, holds promise for applications in fault-tolerant quantum computing.
Majorana quasiparticles were first predicted to occur in so-called p-wave superconductors, which are characterized by rare triplet-like pairing instead of the singlet- like pairing of standard s-wave superconductors. These ideas date back to almost two decades, with various seminal papers including the prediction of topological superconductivity in one-dimensional p-wave superconductors by Alexei Kitaev~\cite{Kitaev_2001}. 

After a short review about the basic properties of Bogolioubov de Gennes (BdG) quasiparticle excitations in superconductors (subsection \ref{BdG}), ABSs (subsection \ref{subsectionABS}) and other types of subgap states, includings YSR in quantum dot junctions (subsection \ref{subsectionQDs}), Majorana zero modes will be explained in some detail in the context of the Kitaev model (section \ref{sec:3}). 
After a short discussion about the state-of-the-art in Majorana detection in hybrid semiconducting-superconducting nanostructures (section \ref{sec:4}), we will describe how the Kitaev model, apart from its pedagogical value, has recent direct applicability in current experiments demonstrating a minimal Kitaev chain in hybrid semiconductor-superconductor nanostructures based on quantum dots (section \ref{sec:5}). In section 5, we discuss recent advances in hybrid semiconductor-based superconducting qubits based on subgap states, including the Andreev spin qubit. We finish the chapter with a short review in section 6 on various theoretical ideas for coherent operating and manipulation of Majorana states. 
\begin{figure}[t]
\begin{center}
\includegraphics[width=0.70\linewidth]{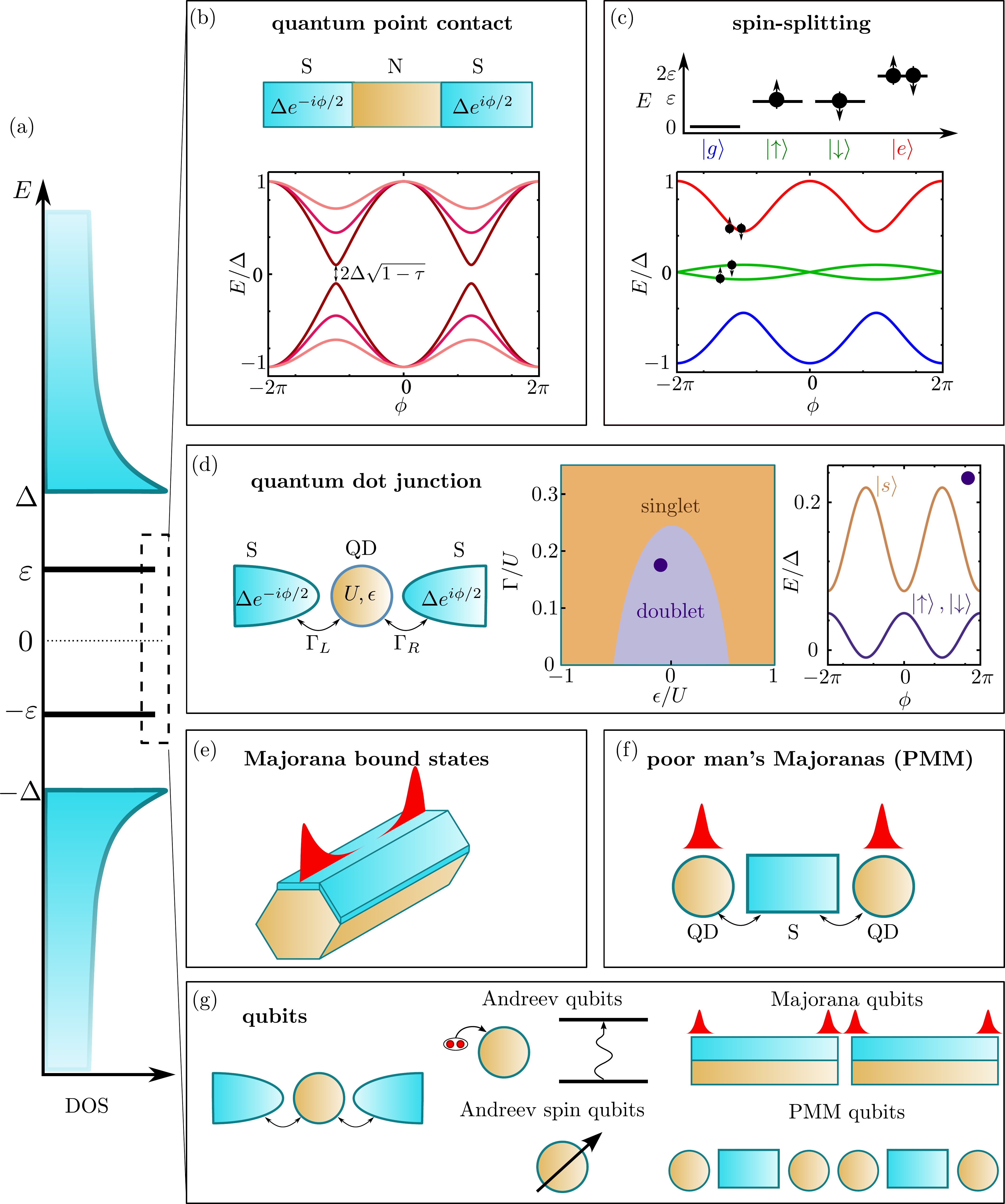}
\caption{(a) Sketch of the density of states of a superconducting system with an energy spectrum  containing subgap states (black lines), separated from the continuum of states (blue). (b) sketch (top) and subgap spectrum as a function of the superconducting phase difference, $\phi$ (bottom) of a short SNS juntion of transparency $\tau$ (different colours denote decreasing transparency as the energies get closer to the gap edges at $\pm\Delta$). When $\tau\approx 1$ and near $\phi\approx \pi$ the Andreev subgap levels are well detached from the continuum of excitations near $\Delta$ and can be used as a qubit system. (c) Many-body eigenergies (top), that the include the four possible occupations of subgap states. The intermediate states $|\uparrow\rangle$ and $|\downarrow\rangle$ are correspond to odd fermionic parity (doublet) when only a single quasiparticle populates the subgap state. Bottom panel shows the two possible qubits that can be realized depending on fermionic parity. Blue and red lines correspond to standard Andreev excitations in the even manifold and constitute a so-called Andreev pair qubit. In the odd fermionic sector, a spin splitting leads to spin-resolved subgap states (green lines) that can be used as another type of qubit (Andreev spin qubit). (d) Quantum dot junction, showing the sketch of the junction. Middle panel shows the phase diagram as a function of the coupling to the superconducting leads ($\Gamma$) and the dot level position ($\epsilon)$, showing two possible ground states: singlet and doublet. Right panel: energy spectrum at a a place where the ground state is a doublet (black dot in the middle panel). (e) Sketch of a Majorana wire. (f) Sketch of a Poor man's Majorana system. (g) Different semiconductor-superconductor qubits summarized in this chapter.}
\label{fig:1.1}
\end{center}
\end{figure}
\subsection{Bogoliubov de Gennes quasiparticles in superconductors}
\label{BdG}
In the so-called Bogoliubov de Gennes formulation of the mean-field BCS Hamiltonian for superconductivity, the excitation spectrum is symmetric with respect to the superconductor's Fermi level (taken as zero energy as a reference), with $\pm E$ energy levels corresponding to opposite occupancies of the same fermion electronic state. This makes the Hamiltonian intrinsically particle-hole symmetric, at the price of a doubling of the spectrum. This can be understood 
by considering a generic Hamiltonian with BCS mean-field pairing:
\begin{equation}
\label{Hinicial}
{\cal H}=\sum_{\sigma,\sigma'}H_0^{\sigma,\sigma'} c^\dagger_{\sigma}c_{\sigma '}+(\Delta c^\dagger_{\uparrow}c^\dagger_{\downarrow}+H.c.).
\end{equation}
Here $H_0^{\sigma,\sigma'}$ is a generic spin-dependent quadratic Hamiltonian (kinetic + single electron potentials), $\Delta$ is the superconducting pair potential and $c^\dagger_{\sigma}$ creates an electron with spin $\sigma=\uparrow,\downarrow$ (an implicit spatial dependence is assumed in this notation). By using the four-component Nambu spinor
\begin{eqnarray}
\label{Nambu-spinor}
\hat\Psi=\left( \begin{array}{ccc}
 c_{\uparrow} \\
c_{\downarrow}\\  c^\dagger_{\downarrow} \\
-c^\dagger_{\uparrow}\end{array} \right),
\end{eqnarray}
the BCS Hamiltonian in Eq. \eqref{Hinicial} can be cast in the so-called BdG form

\begin{equation}
\label{BdG Hamiltonian1}
{\cal H}=\frac{1}{2}\hat\Psi^\dagger H_{BdG}\hat\Psi,
\end{equation}
with 
\begin{eqnarray}
\label{BdG Hamiltonian2}
H_{BdG}=\left( \begin{array}{ccc}
H_0&\Delta \\
\Delta^*&-\sigma^y H_0^{*}\sigma^y \end{array} \right).
\end{eqnarray}
Crucially, the term $-\sigma^y H_0^{*}\sigma^y$ is the time-reversal of $H_0$ and appears since holes are the time-reversed version of electrons. The problem defined by Eqs. (\ref{BdG Hamiltonian1}) and (\ref{BdG Hamiltonian2}) can be solved by seeking for eigenvalues fulfilling the equation 
\begin{equation}
\label{BdG-eigen}
H_{BdG}\Phi_n=E_n\Phi_n,
\end{equation}
with $\Phi_n=[u_{n\uparrow},u_{n\downarrow},v_{n\uparrow},v_{n\downarrow}]^T$, such that the diagonalised Hamiltonian becomes
\begin{equation}
{\cal H}=\frac{1}{2}\sum_nE_n\gamma^\dagger_n \gamma_n,
\end{equation}
with BdG quasiparticle operators defined as
\begin{equation}
\label{BdG-quasiparticle}
\gamma_n=\int d{\bf r}\,\Phi^\dagger_n({\bf r})\hat\Psi({\bf r})=\int d{\bf r} [u^*_{n,\uparrow}({\bf r})c_{{\bf r},\uparrow}+u^*_{n,\downarrow}({\bf r})c_{{\bf r},\downarrow}-v^*_{n,\uparrow}({\bf r})c^\dagger_{{\bf r},\downarrow}+v^*_{n,\downarrow}({\bf r})c^\dagger_{{\bf r},\uparrow}],
\end{equation}
where we have restored the spatial dependence for clarity.
\newpage
\begin{svgraybox}
Note that, since we have explicitly included the hole states, thus doubling the dimension of the Hamiltonian, the BdG description is redundant. Therefore, there must necessarily be some symmetry constraint between eigenstates that fixes the number of independent solutions. This constraint reads:
\begin{eqnarray}
PH_{BdG}({\bf r})P^\dagger=-H_{BdG}({\bf r}),
\label{Eq:PH_def}
\end{eqnarray}
where the operator $P\equiv \mathcal{C}K$, defined in terms of the charge conjugation operator $\mathcal{C}=\tau^y\otimes\sigma^y$ and the complex conjugation operator $K$, reflects electron-hole symmetry. This means that if there is a solution $\Phi_n({\bf r})$ at positive energy $E_n$, then there is also a solution $ \Phi_{m}({\bf r})$
 at $E_{m}\equiv-E_n$ fulfilling 
 \begin{equation}
 \label{eigenstate-ehsymmetry}
 \Phi_{m}({\bf r})=P\Phi_n({\bf r}),
 \end{equation}
 with 
 \begin{equation}
 \label{positive-negative energy}
  E_{m}\equiv-E_n.
 \end{equation}
 Equivalently, $\gamma^\dagger_n=\gamma_{m}$ which, in other words, means that creating a Bogoliubov quasiparticle with energy $E$ or removing one with energy $-E$ are identical operations.
\end{svgraybox}
\subsection{Andreev bound states}
\label{subsectionABS}
At high transparent normal-superconducting (NS) interfaces, a process known as Andreev reflection may take place, whereby electrons are coherently retro-reflected back as holes with inverted
spin and momentum, while transferring a Cooper pair into the superconductor~\cite{tinkham_book}. In superconductor-normal-superconductor (SNS) junctions, constructive interference between Andreev processes at both NS interfaces leads to a coherent electron-hole superposition, known as Andreev bound states (ABSs). Such superpositions result in standing waves with quantized energy leading to subgap states at energies below the superconducting gap. Intuitively, they can be viewed as the electron-hole counterparts of particle-in-a-box states of a quantum well where the boundaries are replaced by superconducting walls. The specific energies of this particular case of BdG excitations depend not only on the superconducting phase difference between both superconductors but also on specific properties of the normal region, such as the transmission probability of electron transport across the normal link, and on the ratio between the length of the normal segment $L_N$ and the superconducting coherence length $\xi$. For a ballistic system, the latter reads $\xi=\hbar v_F/\Delta$⁠, written in terms of the Fermi velocity $v_F$ of the quasiparticles within the normal region and the superconducting gap $\Delta$. 
\begin{example}{Short junction limit}
The simplest limit is the so-called short junction limit ($L_N\ll \xi$) at zero magnetic field and in the Andreev approximation (valid for high density systems, where the Fermi energy is much greater that $\Delta$). In this case, two ABSs appear as subgap BdG excitations in the junction at energies below the gap:
\begin{equation}
\label{ABS}
E_A^\pm(\phi)=\pm\Delta\sqrt{1-\tau \sin^2(\phi/2)},
\end{equation}
where $\tau$ is the normal transmission probability of the conducting channel in the junction. The electrodynamics of such Josephson weak link in a circuit depends not only on the ABSs energy but also on their occupation. 
\end{example}
Importantly, in the absence of spin-splitting, ABSs are spin degenerate. Specifically, there is a manifold of even fermionic parity spanned by the ground state and an excited state with two BdG quasiparticles of opposite spin and total energy $2E_A^+(\phi)$. This Hilbert space can be used to define a so-called Andreev pair qubit. There is another possibility corresponding to odd fermionic parity in which only one spinful BdG quasiparticle occupies the junction. Using the spin degrees of freedom of such single quasiparticle one can define a spin qubit inside the superconducting junction. A full discussion of how to use the different Hilbert spaces defined by the fermionic parity of an SNS junction in order to create different Andreev qubit versions will be presented in section \ref{Sec:qubits}. 

\subsection{Subgap levels in quantum dot junctions}
\label{subsectionQDs}
In the previous discussion about ABSs, we have just assumed that the channel connecting the superconducting electrodes allows for coherent transport through ballistic segments. In contrast, its is quite common that in SNS junctions defined in semiconductors near depletion quantum dots (QDs) are formed by e.g. inducing barriers with electrostatic gates. In this situation, charges localize in the channel and the effects of confinement and electrostatic charging become important. 
\begin{example}{The superconducting Anderson model}
The standard theoretical model that describes a Josephson junction containing a QD is the single impurity Anderson model coupled to two superconducting reservoirs, the so-called superconducting Anderson model. Despite its apparent simplicity, this model correctly captures the interplay between different physical mechanisms that appear owing to the competition between electron repulsion, originated from Coulomb interactions, and electron pairing, coming from the coupling to the superconductor.
The Hamiltonian of the superconducting Anderson model reads
\begin{equation}
    H = H_{QD} +H^L_S + +H^R_S + H_T^S\,.
    \label{Eq:H_SAnderson}
\end{equation}
Here, $H_{QD}$ refers to the decoupled dot Hamiltonian given by
\begin{equation} 
H_{QD} =  \sum_\sigma \epsilon_\sigma d^\dagger_\sigma d_\sigma + U n_\uparrow n_\downarrow,
\end{equation} 
where $d_\sigma$ ($d_\sigma^\dagger$) annihilates (creates) an electron with spin $\sigma=\{ \uparrow, \downarrow \}$ and energy $\epsilon_\sigma = \epsilon_0 \mp V_Z$, where $V_Z=g\mu_BB/2$ refers to the Zeeman energy owing to the external magnetic field $B$ (with $g$ and $\mu_B$ being, respectively, the gyromagnetic factor and the Bohr's magneton). Electron-electron correlations in the dot occupations $n_\sigma = d^\dagger_\sigma d_\sigma $ are included by means of the charging energy $U$. 

The BCS pairing Hamiltonian at each lead is given by
\begin{equation} 
\label{Eq:H_leads_SQDS}
H^{L/R}_{S} =  \sum_{ k_S \sigma}  \varepsilon^{L/R}_{k_S} c^\dagger_{k_S \sigma} c_{k_S \sigma} + \sum_{k_S} \Delta^{L/R}(c^\dagger_{k_S \uparrow} c^\dagger_{-k_S \downarrow} + h.c.),
\end{equation} 
where $c^\dagger_{k_S}$ ($c_{k_S}$) denotes the creation (destruction) operator in the lead with momentum $k_S$, spin $\sigma$ and energy $\varepsilon_{k_S}$, which is measured with respect to the chemical potential of the superconductor. The complex  superconducting order parameter on each lead is $\Delta^{L/R}= |\Delta^{L/R}|e^{i\phi_{L/R}}$, where $\phi_{L/R}$ is the superconducting phase of the left/right lead. The remaining two terms in the Hamiltonian are the couplings between the quantum dot and the left/right superconducting lead:
\begin{equation} 
H_{T}^S =  \sum_{i\in L,R}\sum_{k_S \sigma} \left(V^i_{k_S} c^\dagger_{k_S \sigma} d_\sigma + h.c. \right),
\label{Eq:H_tunnel_SQDS}
\end{equation}  
These two tunneling terms define the two tunneling rates: $\Gamma_{L/R} = \pi \sum_{k_{S} }|V^{L/R}_{k_{S}}|^2 \delta(\omega - \epsilon_{k_{S}})=\pi|V^{L/R}_{{S}}|^2\rho^{L/R}_{{S}}$, with $\rho^{L/R}_{{S}}$ being the normal density of states of the S leads evaluated at the Fermi energy. The total coupling between the QD and the superconducting electrodes is $\Gamma=\Gamma_{L}+\Gamma_{R}$ and the superconducting phase bias is $\phi=\phi_R-\phi_L$.
\end{example}
\subsubsection{Non-interacting limit: resonant level model}
For negligible charging effects $U=0$, the resonant state inside the QD dominates the physical properties of the system. In the single resonant level limit, Eq. \eqref{ABS} becomes
\begin{equation}
\label{ABS_QD}
\varepsilon_{A}(\phi)=\pm\tilde\Delta\sqrt{1-\tau \sin^2(\phi/2)},
\end{equation}
where $\tau=\frac{\Gamma^2}{\epsilon^2_0+\Gamma^2}$ is now the transmission trough a Breit-Wigner resonance of energy $\epsilon_0$ and width $\Gamma$ corresponding to the total tunneling rate between the two superconducting electrodes and the resonant
level. The energy scale $\tilde\Delta$ corresponds to the position of the subgap states at $\phi=0,2\pi$ and can be obtained from the transcendental equation:
\begin{eqnarray}
 \label{ABSphi0}
\tilde\Delta^2 \left[ 1+\tau\frac{\Delta^2-\tilde\Delta^2}{\Gamma^2}+2\tau\sqrt{\frac{\Delta^2-\tilde\Delta^2}{\Gamma^2}}\right]=\Delta^2,
\end{eqnarray}
which defines various interesting limits. Specifically, it goes from $\tilde\Delta\approx\Delta$  for $\Gamma\gg\Delta$ and $\tilde\Delta\approx\Gamma$ in the opposite $\Gamma\ll\Delta$ limit. For $|\epsilon_0|\ll\Delta$, Eq. \eqref{ABS_QD} can be conveniently rewritten as \cite{PhysRevB.104.174517,PhysRevB.107.195405} 
\begin{equation}
\label{ABS_QD2}
\varepsilon_{A}(\phi)=\pm\frac{\Delta}{\Delta+\Gamma}\sqrt{\epsilon_0^2+\Gamma^2\cos^2(\phi/2)+(\Gamma_L-\Gamma_R)^2\sin^2(\phi/2)},
\end{equation}
which is valid for $\Gamma\ll\Delta$ at any $\phi$ or $\Gamma\gtrsim\Delta$ provided that $|\pi-\phi|\ll 1$. The prefactor $\frac{\Delta}{\Delta+\Gamma}$  characterizes the extent to which
the wave function of the ABS is localized at the quantum
dot. If the tunneling between the superconducting leads and the dot is weak, $\Gamma\ll\Delta$, the wave function is predominantly localized at the
dot and $\frac{\Delta}{\Delta+\Gamma}\approx 1$. In the opposite limit, $\Gamma\gg\Delta$, the support of the wave function is mostly in the leads.
Interestingly, Eq. \eqref{ABS_QD2} can be written as an ABS in a short junction (see Eq. \eqref{ABS}) by noticing that it can be rewritten as  \cite{Tanaka_2007,PhysRevB.104.174517}
\begin{equation}
\label{ABS_QD3}
\varepsilon_{A}(\phi)=\pm\Gamma_A\frac{\Delta}{\Delta+\Gamma}\sqrt{1-\tau\sin^2(\phi/2)},
\end{equation}
with $\Gamma_A^2=\epsilon_0^2+\Gamma^2$ and $\tau=1-|r|^2$ the transparency of the junction controlled by the reflection coefficient $r=(\epsilon_0+i\delta\Gamma)/\Gamma_A$, which depends on both the level position $\epsilon_0$ and the rate asymmetry $\delta\Gamma=(\Gamma_L-\Gamma_R)$. At perfect transparency $\tau=1$, which is achieved for $\epsilon_0=\delta\Gamma=0$, the subgap levels appear at
\begin{equation}
\label{ABS_QD3}
\varepsilon_{A}(\phi)=\pm \Gamma\frac{\Delta}{\Delta+\Gamma}\cos(\phi/2),
\end{equation}
defining , $4\pi$-periodic branches with a zero-energy level crossing at $\phi=\pi$. Note that for $\Gamma\ll\Delta$ the subgap levels lie very deep inside the gap of the superconductor even at phases $\phi=0,2\pi$.

\subsubsection{Interacting quantum dots: some limits \label{InteractingQDS}}
The physics of QDs coupled to superconductors is richer if interactions are included. Electron-electron interactions tend to fix the number of electrons in the QD to a value that can be tuned using electrostatic gates. On the other hand, superconducting correlations coming from the superconductor tend to pair electrons into singlet Cooper pair. This leads to an interesting competition between these two effects.

Therefore, in the interacting case, $U\neq 0$, two ground states are possible: a spin doublet, $ |D\rangle$, with spin $\frac{1}{2}$, and a spin singlet, $  |S\rangle$, with spin 0. Transitions between the ground state and the first excited state of the system, i.e., between a doublet and a singlet state or vice-versa, are manifested as a subgap resonance at energies $\pm\varepsilon_A$. Changes in the parity of the ground state of the system appear as points in parameter space where $\varepsilon_A$ changes sign (signaled by zero energy crossing of the subgap states).
Whether the system is a doublet or a singlet state is determined by the nontrivial interplay between interactions and quantum fluctuations induced by the coupling $\Gamma_S$: Coulomb repulsion enforces a one by one electron filling, favoring odd electron occupations with a doublet ground state. The coupling to the superconductor, $\Gamma_S$, on the other hand, privileges  a singlet ground state. Its physical nature crucially depends on the ratio between the induced pair correlations and interaction strength, $\Delta/U$, as we discuss now. 

\begin{example}{Large gap limit $\Delta\gg U$: BCS-like charge singlets}
In the large  $\Delta/U$ limit, the coupling to the superconductor mainly induces local superconducting correlations in the QD, which acquires local superconducting pairing terms owing to the proximity effect. Such local superconducting pairing leads to Bogoliubov-type singlets, which are BCS-like superpositions of the empty $|0\rangle$ and doubly-occupied 
$|\uparrow\downarrow\rangle$ states in the QD. 
In this so-called atomic limit \cite{Tanaka_2007,Bauer_2007,PhysRevB.79.224521}, the quasiparticles of the superconducting electrodes are removed from the problem and the coupling to the superconductor $\Gamma_S$ mainly induces local superconducting correlations in the quantum dot, resulting in the Hamiltonian:
\begin{equation} 
\label{SC-QD}
H^S_{QD} =  \sum_\sigma \epsilon_\sigma d^\dagger_\sigma d_\sigma + U n_\uparrow n_\downarrow+\Delta_\phi(d^\dagger_\uparrow d^\dagger_\downarrow+h.c),
\end{equation} 
with $\Delta_\phi=\Gamma_Le^{i\phi_L}+\Gamma_R e^{i\phi_R}$, which for $\Gamma_L=\Gamma_R$ can be written in a compact form as $\Delta_\phi=2\Gamma \cos{\phi/2}\equiv\Gamma_S \cos{\phi/2}$.

Without quasiparticles, the Hilbert space is just four dimensional spanned by the local states in the QD: the empty and doubly occupied states $|0\rangle$, $|2\rangle$ (even fermionic parity) and the doublet $|\uparrow\rangle$, $|\downarrow\rangle$ (odd fermionic parity). Owing to the Pauli exclusion principle, the doubly occupied level in the QD must be a singlet $|2\rangle=1/\sqrt 2(|\uparrow\downarrow\rangle-|\downarrow\uparrow\rangle)$. Using this four dimensional Hilbert space, the Hamiltonian in Eq. \eqref{SC-QD} can be written in a compact form as:
\begin{equation}
H=\begin{pmatrix}
0 & \Delta_\phi & 0 & 0\\
\Delta_\phi & \epsilon_\uparrow+\epsilon_\downarrow+U & 0 & 0\\
0 & 0 & \epsilon_\uparrow & 0\\
0 & 0 & 0 & \epsilon_\downarrow
\end{pmatrix}\,,
\label{matrix-SC-QD}
\end{equation}
which is block-diagonalized into even and odd fermion parity subspaces. The eigenvalues of the odd subspace are obviously the same as the spin doublet states, where we allow to have spin-degeneracy breaking by e.g. a Zeeman term $\epsilon_{\uparrow,\downarrow}=\epsilon_0\pm E_Z$. The even parity eigenvalues correspond to bonding and anti-bonding combinations of empty and doubly occupied states in the QD, namely $|S_\pm\rangle=u|0\rangle\pm v|2\rangle$, with energies
\begin{equation}
E_{e\pm}=\frac{(\epsilon_\uparrow+\epsilon_\downarrow+U)}{2}\pm\sqrt{(\epsilon_\uparrow+\epsilon_\downarrow+U)^2/4+\Delta^2_\phi}.
\end{equation}
The $u$ and $v$ are BCS coherence factors are
\begin{eqnarray}
u^2=\frac{1}{2}[1+\frac{(\epsilon_\uparrow+\epsilon_\downarrow+U)}{2\sqrt{(\epsilon_\uparrow+\epsilon_\downarrow+U)^2/4+\Delta^2_\phi}}],\nonumber\\
v^2=\frac{1}{2}[1-\frac{(\epsilon_\uparrow+\epsilon_\downarrow+U)}{2\sqrt{(\epsilon_\uparrow+\epsilon_\downarrow+U)^2/4+\Delta^2_\phi}}],
\end{eqnarray}

In this atomic limit, the singlet-doublet (SD) boundary can be easily obtained as the degeneracy condition between even and odd states. This defining a parity crossing at $\xi_\downarrow=\epsilon_\downarrow-E_{e_-}=0$, resulting in the boundary
\begin{equation}
\label{single-doublet-large Delta}
\frac{(\epsilon_\downarrow-\epsilon_\uparrow)}{2}+\sqrt{(\epsilon_\uparrow+\epsilon_\downarrow+U)^2/4+\Delta^2_\phi}-\frac{U}{2} = 0.
\end{equation}
This parity crossing occurs as a quantum phase transition and can be experimentally tuned and detected by means of the Andreev conductance in QDs coupled to superconductors \cite{Lee-NatureNano14,PhysRevB.95.180502,doi:10.1126/science.abf1513}.

For $\phi=0$, this parity crossing condition reads \cite{PhysRevB.91.045441}
\begin{equation}
\label{parity-crossingQD}
E_Z+\frac{U}{2}=\sqrt{(\epsilon_0+\frac{U}{2})^2+\Gamma_S^2}
\end{equation}
Interestingly, this condition for a Zeeman-induced zero-energy crossing in a QD is essentially  the same as the condition for obtaining MZMs in a semiconducting nanowire with proximity-induced superconductivity \cite{Lutchyn_PRL2010,Oreg_PRL2010,AguadoReview,LutchynReview}. This expression illustrates the profound connection between fermionic parity crossings and MZMs, even in this zero-dimensional example.

Finally, we end this part by just mentioning that finite $\Delta$ corrections to the above atomic limit boundary are captured by the so-called generalized atomic limit where the original parameters of the atomic limit are rescaled by the quantity 
$1/(1+\Gamma_S/\Delta)$
\cite{Zonda1,Zonda2,Zonda3}.
\end{example}
\begin{example}{Large charging limit $U\gg\Delta$: Yu-Shiba-Rusinov singlets}
In the opposite limit, $U\gg\Delta$, the QD is in the so-called
Coulomb blockade regime with a well defined number of
electrons. The emergence of subgap states can be understood by considering just one spinful quantum dot level coupled to a superconducting electrode. Conceptually, this scenario is identical to having an isolated magnetic impurity in a superconducting host, a problem well known since the 60s for classical spin impuritites resulting in so-called YSR subgap states \cite{Yu:APS65,Shiba:POTP68,Rusinov:JL69}. For QDs the problem is richer owing to quantum fluctuations induced by $\Gamma_S$. Quantum fluctuations lead to a very complex scenario since exchange is mediated by Kondo processes, as the the unpaired spin in the QD couples to the BdG quasiparticles outside the gap of the SC, with an exchange interaction $J\sim 2\Gamma_S/U$. This exchange interaction creates YSR singlets well beyond the original classical limit since they originate from Kondo-like correlations \cite{PhysRevB.91.045441}. The main difference with the standard Kondo effect is that in the superconducting electrode, no quasiparticles are available below the gap, hence Kondo screening is incomplete. There are two main physical consequences of being in the YSR regime with $U\gg\Delta$ as compared to the BCS-like regime with $U\ll\Delta$: i) instead of the fully \emph{local} BCS-like superpositions of zero and doubly occupied states in the QD, the singlets in this limit are \emph{nonlocal}, corresponding to Kondo-like superpositions between the spin doublet and Bogoliubov quasiparticles in the superconductor. In the presence of phase bias $\phi\neq 0$, the degree of deslocalization of the screening quasiparticles depends on the phase; ii) the boundary of the singlet-doublet transition in Eq. \eqref{single-doublet-large Delta} gets replaced by the condition $T_K\sim 0.3\Delta$, where $T_K$ is the Kondo temperature of the problem. Early work on hybrid QDs indicated the importance of Kondo-like correlations \cite{PhysRevLett.89.256801,PhysRevLett.99.126603}, while more recent experimental work
has provided precise boundaries for the transition \cite{PhysRevB.95.180502} and many other aspects \cite{PhysRevLett.118.117001,Estrada-2020,PhysRevResearch.2.012065,PhysRevLett.130.136004}.

While a full solution of this intricate problem needs sophisticated numerics, such as e.g. the numerical renormalization group method \cite{PhysRevB.91.045441,RevModPhys.80.395}, deep in the Kondo limit, with $U\rightarrow\infty$, one can use resormalised parameters using e.g. slave boson theory and write an effective resonant level model as
\begin{eqnarray}
 \label{slave}
\varepsilon_A \approx\pm\tilde\Delta\sqrt{ 1-\tilde\tau sin^2(\phi/2)},
\end{eqnarray}
with $\tilde\tau=\frac{\tilde\Gamma^2_S}{\tilde\epsilon^2_0+\tilde\Gamma^2_S}$ and renormalised parameters defined through the constraint $\sqrt{\tilde\epsilon^2_0+\tilde\Gamma^2_S}=T_K\approx De^{-\frac{\pi |\epsilon_0|}{\Gamma_S}}$, with $D$ a high-energy bandwidth cutoff \cite{aguadoslave,alfredoslave}. In the unitary limit $\tilde\epsilon_0\rightarrow 0$ and $\tilde\Gamma_S\rightarrow T_K$, we can write the ABSs as \cite{alfredoslave,Josephsonslave1,Josephsonslave2}:
\begin{eqnarray}
 \label{Kondo}
\varepsilon_A \approx\pm\Delta\left[1-2\left(\frac{\Delta}{T_K}\right)^2\right]\cos(\phi/2),
\end{eqnarray}
which is essentially Eq. \eqref{ABS_QD3} for a resonant level model written in terms of Kondo parameters.
\end{example}
\subsubsection{Competition between pairing and charging energy \label{InteractingQDS_numerics}}
The interacting problem becomes analytically unsolvable in the limit where there is not a clear separation between the different energy scales. In this limit, only a numerical treatment can provide an exact solution to the Hamiltonian in Eq.~\eqref{Eq:H_SAnderson}. In equilibrium conditions, the numerical renormalization group method lead to an exact description of the low-energy properties of the system. The me is based on a logarithmic discretization of the superconductor's band that has an exponential resolution at low energies, close to the superconductor's Fermi level. Precise results on the energy of the ABSs, the phase diagram, and the supercurrent have been reported, see for example Refs.~\cite{Yoshioka_JPSJ2000,Choi_PRB2004,Oguri_JPSJ2004,Tanaka_NJP2007,Lim_JPCM2008,Bauer_JPCM2007,Karrasch_PRB2008,Zitko_PRB2011,Rodero_JPCM2012,Pillet_PRB2013,Zitko_PRB2015}. Quantum Monte Carlo~\cite{Siano_PRL2004,Luitz_PRB2010,Luitz_PRL2012,Pokorny_PRR2021} and functional renormalization~\cite{Kadlecova_PRB2017,Karrasch_PRB2008} group are alternative numerical methods to describe the system. The system under non-equilibrium conditions has also been studied by means of exact numerical techniques, like time-dependent density matrix renormalization group~\cite{Souto_PRB2021}.

Apart from the previously mentioned numerical methods, there exists a series of approximations that allow for a simpler treatment of the problem. These methods are based on reducing the number of relevant degrees of freedom, so the system can be diagonalized exactly, or perturbation expansion son some of the parameters. One of these approximations is the infinite gap approximation, given in Eq.~\eqref{matrix-SC-QD}, that only considers the induced gap in the superconductor. Below we describe other approximations that have been used to understand the properties of the system. 

\begin{example}{The zero bandwidth approximation}
The zero-bandwidth approximation is based on reducing the infinite number of degrees of freedom of the superconductor to one spin-degenerate state with pairing amplitude. Under this approximation, the Hamiltonian of the superconducting leads, Eq.~\eqref{Eq:H_leads_SQDS}, transforms into 
\begin{equation}
H^{L/R}_{S} =  \varepsilon^{L/R} c^\dagger_{\sigma} c_{\sigma} + \Delta^{L/R}(c^\dagger_{\uparrow} c^\dagger_{\downarrow} + h.c.),
\end{equation} 
that describes two quasiparticle states at energies $\pm\Delta$. Consistently, the hoping between the QD and the superconductors has the same form as in Eq.~\eqref{Eq:H_tunnel_SQDS}.

The model provides a good qualitative description of the system properties in a broad range of parameters, including the $0-\pi$ transition induced by the many-body interactions taking place at the QD~\cite{Vecino_PRB2003,Bergeret_PRB2007,Zazunov_PRB2010}. Extensions of this model including more states in the leads have been demonstrated to improve results from the zero-bandwidth model, see Ref.~\cite{Baran_PRB2023} for a discussion. However, the absence of a real continuum in the superconductor imposes constrains on the validity range of the approximation. For instance, the approximations fails at describing the regime where the QD is strongly coupled to the superconductor, predicting an nonphysical anti-crossing between the QD and the states at $\pm \Delta$. It also fails at describing non-equilibrium effects, as they rely on the properties of the continuum.
\end{example}
\begin{example}{Perturbation theory}Perturbation theory provides another way to treat interactions in the system. 
At the lowest-order expansion in $U$, the electron-electron interaction renormalizes self-consistently the level and the induced pairing amplitude at the dot as~\cite{Pillet_NatPhys2010}
\begin{eqnarray}
    \tilde{\varepsilon}_\sigma&=&\varepsilon_\sigma+U\left\langle n_{\bar{\sigma}}\right\rangle\,,\nonumber\\
    \tilde{\Delta}&=&-U \left\langle d_{\uparrow}d_{\downarrow}\right\rangle\,.
\end{eqnarray}
This approximation reproduces the position and supercurrent through the system in the small- $U$ limit~\cite{Rodero_JPCM2012}.  
However, it also predicts an spontaneous breaking of the spin-degeneracy that is unphysical. Higher orders in the expansion with $U/\Gamma_S$ describe quantum fluctuations, essential to accurately describe the properties of the system. In fact, second-order expansions have been shown to correctly describe properties of the system up to a moderate values of $U/\Gamma_S$~\cite{Zonda1,Zonda2,Souto_PRB2021}.
\end{example}

\section{Kitaev chain: basics}
\label{sec:3}
 \subsection{Majorana zero modes}
There is an interesting  connection between the BdG description of superconductivity and Majoranas. Most importantly, the four-component Nambu spinor in Eq. (\ref{Nambu-spinor}) is nothing but an operator version of the Majorana wave functions. Indeed, by particle-hole symmetry, the Nambu spinor fulfils the so-called Majorana pseudo-reality condition \cite{PhysRevB.81.224515}, 
$\hat\Psi({\bf r})=P\hat\Psi({\bf r})=C\hat\Psi^*({\bf r})$,
where $P$ and $C$ operations are defined around Eq.~\eqref{Eq:PH_def}. Therefore, the BdG theory for quasiparticle excitations in a superconductor already possesses all the key properties of Majorana fermions. Note that this Majorana character is satisfied by the entire (time-dependent) quantum field but not by the eigenmodes of well defined energy. This can be, in principle, a serious problem towards the observability of the Majorana character of BdG quasiparticles. To avoid the full time dependence of the Nambu spinor, one can focus instead on zero-energy BdG quasiparticles.

\begin{svgraybox}
 The profound meaning of zero-energy BdG quasiparticles can be understood by analyzing the properties of the 
 eigenstates of Eq. (\ref{BdG-eigen}) that are self-conjugate, namely $\gamma_n=\gamma^\dagger_n$. According to Eqs. (\ref{eigenstate-ehsymmetry}) and (\ref{positive-negative energy}), this can only happen at $E=0$, so that a stationary solution with Majorana character fulfils necessarily,
\begin{equation}
\label{BdG-Majorana}
H_{BdG}({\bf r})\Phi_0({\bf r})=0.
\end{equation}
The corresponding real space spinor satisfies $\Phi_{0}({\bf r})=P\Phi_0({\bf r})$, 
which implies 
\begin{equation}
\gamma_0=i\int d{\bf r} [u^*_{0,\uparrow}({\bf r})c_{{\bf r},\uparrow}+u^*_{0,\downarrow}({\bf r})c_{{\bf r},\downarrow}-u_{0,\downarrow}({\bf r})c^\dagger_{{\bf r},\downarrow}-u_{0,\uparrow}({\bf r})c^\dagger_{{\bf r},\uparrow}],
\end{equation}
which is clearly self-conjugate. Such exotic self-conjugate BdG excitations receive the name of Majorana zero modes or Majorana bound states (MBSs).
\end{svgraybox}
\subsection{The Kitaev model}
In the previous subsection, we have just assumed that the MBS  exists but that this is a rather peculiar situation. Indeed, the general symmetry obeyed by BdG eigenstates is 
$\Phi_{m}({\bf r})=P\Phi_n({\bf r})$ with $E_m=-E_n$, see Eqs. (\ref{eigenstate-ehsymmetry}),(\ref{positive-negative energy}), and not $\Phi_{n}({\bf r})=P\Phi_n({\bf r})$ with $E_n=-E_n=0$. More importantly, if one of these zero modes exists, it \emph{cannot} acquire a nonzero energy $E$ by any smooth deformation of the Hamiltonian since \emph{finite energy BdG excitations always occur in pairs}. Namely, symmetry would require \emph{another mode} to appear at energy -$E$, in violation of unitarity. There is, however, a way out to the previous conundrum: if the gap separating the zero mode from other quasiparticle excitations \emph{closes}, nothing prevents the system to host pairs of standard BdG excitations as the gap reopens. Such closing and re-opening of the superconducting gap is an instance of a \emph{topological transition}: broadly speaking, a transition that separates two phases characterised by the value of a topological invariant (instead of a broken symmetry). In this particular case, the topological invariant counts the number of MBSs. The superconductors hosting such exotic zero modes with Majorana character are called topological superconductors and the fact that the zero mode cannot acquire a nonzero energy without closing the gap is called topological protection. In this section, we explain how this MBSs appear in, arguably, the simplest model exhibiting topological superconductivity: the Kitaev model \cite{Kitaev_2001}. 

\begin{figure}[t]
\includegraphics[width=1\linewidth]{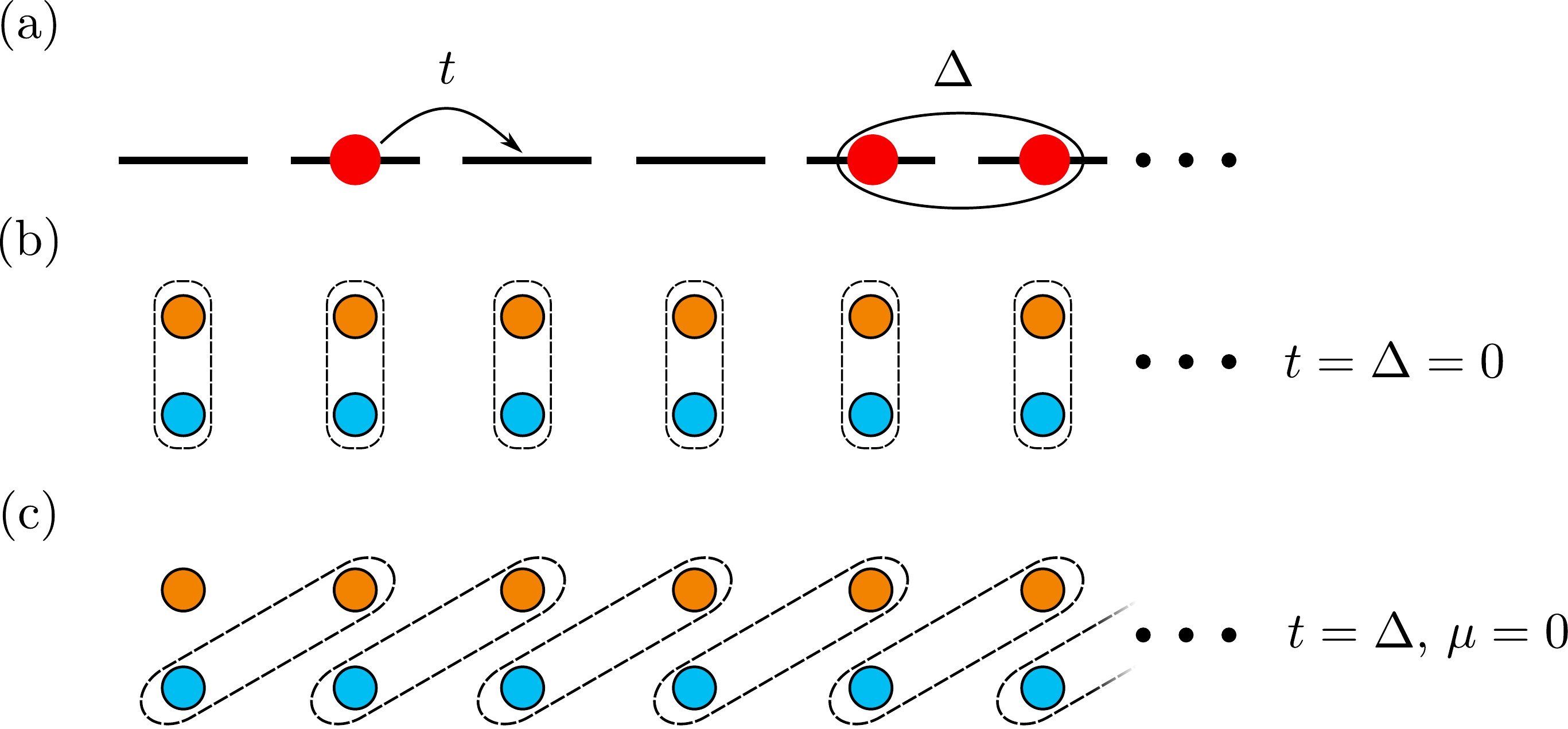}
\caption{(a) Sketch of the Kitaev chain, composed by spinless fermions that are allowed to tunnel ($t$) and pair ($\Delta$) between neighboring sites. (b,c) Represents pictorially the decomposition of the local fermionic operators into Majorana operators (orange and blue dots). Panel (b) shows the paradigmatic trivial situation with $t=\Delta$, where Majorana operators couple onsite. (c) Topological scenario where Majoranas couple between nearest neighbor sites, leaving two unpaired MBSs at the ends.}
\label{fig:KitaevChain}
\end{figure}
\begin{example}{The Kitaev model as a tight-binding lattice}
The Kitaev model is essentially a lattice model that describes a one-dimensional $p$-wave superconductor, see Fig~\ref{fig:KitaevChain}(a). Owing to its simplicity, the model grasps the main property we are seeking, a topological phase with emergent MBSs,  in a rather intuitive fashion. Specifically, the model describes a chain with N sites of spinless fermions with long-range $p$-wave superconductivity
\begin{equation}
\label{kiatev0}
H=-\mu\sum_{j=1}^{N}\Big(c^{\dagger}_{j}c_{j}-\frac{1}{2}\Big)\,+\,\sum_{j=1}^{N-1}\Big[- t\,\big(c^{\dagger}_{j}c_{j+1}+c^{\dagger}_{j+1}c_{j}\big)\,+\,\Delta\,(c_{j}c_{j+1}\,+\,c^{\dagger}_{j+1}c^{\dagger}_{j})\Big]\,,
\end{equation}
where $\mu$ represents the onsite energy, $t$ is the nearest-neighbor hopping amplitude and $\Delta$ is the is the p-wave pairing amplitude (assumed real for the moment). This model is quite simple but already contains all the relevant ingredients for topological superconductivity. First note that time-reversal symmetry is broken (since electrons do not have spin degeneracy) Furthermore, the superconducting pairing is rather non-standard (it couples electrons with the same spin in contrast to standard s-wave pairing). Note also that electrons on adjacent sites are paired.

In order to reveal the nontrivial properties of the model, let us first consider a chain with open boundary conditions, and then write the fermionic operators in terms of two new operators $\gamma_{j}^{A}$ and $\gamma_{j}^{B}$ as
\begin{equation}
\label{fermionMajoTrans}
c_{j}=\frac{1}{2}\Big(\gamma_{j}^{A}+i\gamma_{j}^{B}\Big)\,,\quad c_{j}^{\dagger}=\frac{1}{2}\Big(\gamma_{j}^{A}-i\gamma_{j}^{B}\Big)\,.
\end{equation}
Using standard fermionic anticommutation algebra for $c_{j}$, it is very easy to verify that the new operators satisfy the following algebra
\begin{equation}
\label{MajAlgebra}
\{\gamma_{i}^{A},\gamma_{j}^{B}\}=2\delta_{ij}\delta_{AB}\,,\quad \gamma_{j}=\gamma^{\dagger}_{j}\,,\quad \gamma_{j}^{2}=\gamma^{\dagger2}_{j}=1\,.
\end{equation}
Thus $\gamma_{j}^{A}$ and $\gamma_{j}^{B}$ are Majorana operators.

This decomposition can be understood as the decomposition of a complex Dirac fermion into real and imaginary parts that correspond to Majorana fermions. The inverse transformation gives us the Majorana operators,
\begin{equation}
\label{MajTrans}
\gamma_{j}^{A}=c_{j}+c_{j}^{\dagger}\,,\quad \gamma_{j}^{B}=i(c^{\dagger}-c_{j})\,.
\end{equation}
In terms of these new operators, the Hamiltonian in Eq.\,(\ref{kiatev0}) reads
\begin{equation}
\label{kitaev0a}
H=-\frac{i\mu}{2}\sum_{j=1}^{N}\gamma^{A}_{j}\gamma^{B}_{j}+\frac{i}{2}\sum_{j=1}^{N-1}\Big[\omega_{+}\gamma^{B}_{j}\gamma^{A}_{j+1}+ \omega_{-}\gamma^{A}_{j}\gamma^{B}_{j+1}\Big]\,,
\end{equation}
where $\omega_{-}=\Delta-t$ and $\omega_{+}=\Delta+t$ represent hopping amplitudes between MBSs in neighbouring sites, while $\mu$ pairs MBSs in the same site. Let us consider now different possibilities depending on the value of $\omega_{-}$ and $\omega_{+}$. For $t=\Delta=0$,  the Hamiltonian is trivial, coupling Majorana operators in the same site, Fig.~\ref{fig:KitaevChain}(b), and given by 
\begin{equation}
\label{kitaev0b}
H=-\frac{i\mu}{2}\sum_{j=1}^{N}\gamma^{A}_{j}\gamma^{B}_{j}\,.
\end{equation}
This Hamiltonian just expresses the fact that Majorana operators from the same physical site are paired together to form a standard fermion. A less obvious case occurs for $t=\Delta$, namely $\omega_{-}=0$ and $\mu=0$. In this case, Eq.\,(\ref{kitaev0a}) becomes,
\begin{equation}
\label{kitaev0c}
H=it\sum_{j=1}^{N-1}\gamma^{B}_{j}\gamma^{A}_{j+1}\,.
\end{equation}
Despite its innocent-looking form, Eq. (\ref{kitaev0c}) is rather nontrivial. First, notice that Majorana operators on the same site are now decoupled  (remember that they originally represented a single fermionic degree of freedom!). Furthermore, long range coupling is established since Majorana operators on neighbouring sites are now coupled, Fig.~\ref{fig:KitaevChain}(c). Finally,  the Majorana operators at the end of chain $\gamma_{1}^{A}$ and $\gamma_{N}^{B}$ seem to have disappeared from the problem. In order to reveal the deep meaning of all these features in full, let us rewrite the Hamiltonian by defining a new set of fermionic operators 
\begin{equation}
\begin{split}
d_{j}&=\frac{1}{2}\Big(\gamma_{j}^{B}+i\gamma_{j+1}^{A} \Big)\,,\quad d_{j}^{\dagger}=\frac{1}{2}\Big(\gamma_{j}^{B}-i\gamma_{j+1}^{A} \Big)\,.
\end{split}
\end{equation}

In terms of these new operators, Eq.\,(\ref{kitaev0a}) reads
\begin{equation}
\label{fermioMan}
H=2t\sum_{j=1}^{N-1}\Big(d_{j}^{\dagger}d_{j}-\frac{1}{2} \Big)\,.
\end{equation}
The fermionic operators $d_j$ diagonalise the superconducting problem and therefore describe Bogoliubov quasiparticles with energy $t=\Delta$. Importantly, the diagonalised Hamiltonian contains $N-1$ quasiparticle  operators while the original problem has $N$ sites. The missing fermionic degree of freedom is hiding in the highly delocalised combination
\begin{equation}
\label{newopMaj}
f=\frac{1}{2}\Big(\gamma_{1}^{A}+i\gamma_{N}^{B}\Big)\,,\quad f^{\dagger}=\frac{1}{2}\Big(\gamma_{1}^{A}-i\gamma_{N}^{B}\Big)\,.
\end{equation} 
This fermionic operator does not appear in the Hamiltonian and thus has zero energy. This is an obvious consequence of the fact that the Majorana operators at the ends of the chain commute with the Hamiltonian
$[H,\gamma_{1}^{A}]=[H,\gamma_{N}^{B}]=0$. Furthermore, this is a standard fermion operator which, as usual, can be empty or occupied. However, this fermionic state is special since both the empty and occupied configurations are degenerate, owing to their zero energy. Note that this ground state degeneracy is rather peculiar since both states differ in fermion parity. This is very different from standard superconductors where, despite the breakdown of particle number conservation, fermion parity is conserved (with even being the parity of the ground state). In contrast, the ground state degeneracy found here, corresponding to different fermionic parities, is unique to topological superconductors and has profound consequences as we shall discuss in the next subsection. 

In general, for a small but non-zero $\mu$, the Majorana bound states are not really localised at the ends of the wire, but their wave-functions exhibit an exponential decay into the bulk of the wire. The non-zero spatial overlap of the two Majorana wave-functions results in a non-zero energy splitting between the two Majorana states. Of course, for long wire's lengths, the splitting can be so small that the two Majorana states can be considered to be degenerate. 
Moreover, the Majoranas can also split when the higher-energy states in the bulk
come very close to zero energy, hence the Majorana modes are protected as long as the bulk energy gap is finite. 
This follows from the particle-hole symmetry involved in the problem, where the spectrum has to be symmetric around zero energy. Therefore, trying to move the Majorana zero modes 
from zero energy individually is impossible, as it would violate particle-hole symmetry.

\end{example}
\begin{example}{The Kitaev model in momentum space}
To make connection with the theory of topological bands, nontrivial topological numbers and gap closings in terms of bulk properties, we now consider periodic boundary conditions which, using translational invariance $c_j=\frac{1}{\sqrt{N}}\sum_p e^{ipj}c_p$, allows to write Eq.\,(\ref{kiatev0}) in momentum space $p$ as,
\begin{equation}
H=\sum_p\xi_p(c^\dagger_pc_p-\frac{1}{2})-\sum_p t\cos p\,a+\Delta\sum_p(c^\dagger_pc^\dagger_{-p}e^{ip\,a}+c_kc_{-p}e^{-ip\,a}),
\end{equation}
with $\xi_p=-(\mu+2t\cos pa)$ and $a$ the lattice spacing. Dropping the unimportant constant, the above Hamiltonian can be written in BdG form as
\begin{equation}
H=\frac{1}{2}\sum_{p}\psi_{p}^{\dagger}H_{BdG}\psi_{p}\,\quad 
\psi_{p}\,=\,\begin{pmatrix}
c_{p}\\
c_{-p}^{\dagger}
 \end{pmatrix}\,,
\end{equation}
with $H_{BdG}=\xi_{p}\tau_{z}+\Delta_{p}\tau_{y}=\mathbf{h}\cdot\mathbf{\tau}\,,$
where $\mathbf{h}=(0,\Delta_{p},\xi_{p})$, $\Delta_{p}=-2\Delta\sin pa$ and $\mathbf{\tau}=(\tau_{x},\tau_{y},\tau_{z})$ the Pauli matrices in electron-hole space. The excitation spectrum of $H_{BdG}$ is then given by
\begin{equation}
\label{quasipKiatev}
E_{p,\pm}=\pm\sqrt{(\mu+2t\cos pa)^{2}+4\Delta^{2}\sin^{2}pa}\,.
\end{equation}
This spectrum is mostly gapped except in special cases: for $\Delta\neq0$, the energy gap closes when both elements inside the square root  vanish simultaneously. The normal-state dispersion $\xi_{p}$ vanishes at $\pm p_F$, where the Fermi wavevector is determined by the condition $\mu+2t\cos p_Fa=0$. The pairing term, on the other hand, vanishes at $p=0$ and $p=\pm\pi/a$, which is a direct consequence of its $p$-wave nature. Thus the system is gapless only when the Fermi wavevector equals $0$ or $p=\pm\pi/a$. This happens when the chemical potential is at the edges of the normal state dispersion, namely $\mu=-2t$ (for $p_F=0$) or $\mu=2t$ (for $p_F=\pm\pi/a$. The lines $\mu=\pm 2t$ define phase boundaries corresponding to two distinct topological phases. These phases are distinguished by the presence or absence of unpaired
MBSs at the ends in the geometry with open boundary conditions (bulk-boundary correspondence). These phases are characterised by a $\mathbb{Z}_2$ topological invariant, the Majorana number $M=(-1)^{\nu}$, where $\nu$ represents the number of pairs of Fermi points. The topological superconducting phase occurs for an odd number of pairs of Fermi points, namely $M=-1$, while an even number corresponds to the trivial one. Owing to the bulk-boundary correspondence, one thus expects unpaired MBSs for the open chain when $M=-1$. Indeed, the special point $\mu=0$ and $\Delta=t$ discussed above for the open chain is well within the topological phase. 
\end{example}

\subsection{Majorana non-Abelian properties}
One of the most fascinating aspect of Majorana states appearing in 1-dimensional topological superconductors are their non-Abelian properties. In 3 dimensions, particles obey either the Fermi-Dirac or the Bose-Einstein distribution. The distribution determines the sign of the wavefunction after the exchange of two identical particles, $\ket{\ldots,\Psi_N,\ldots\Psi_{N'},\ldots}=\pm\ket{\ldots,\Psi_{N'},\ldots\Psi_{N},\ldots}$, where the $\pm$ sign denotes the result for bosons/fermions. Anyons in low-dimensional systems allow to generalize these commutation relations to more complex ones. The simplest extension to the commutation relation is the one where the wavefunction acquires an arbitrary phase after the exchange of two particles, {\it i.e.} $\ket{\ldots,\Psi_N,\ldots\Psi_{N'},\ldots}=e^{i\theta}\ket{\ldots,\Psi_{N'},\ldots\Psi_{N},\ldots}$. Even more interesting is the case where the ground state of the system is degenerate. In this case, the anyon exchange can transform one ground state into another via an unitary transformation. This transformation can depend on the way the anyons are exchanged, but not on the exact path they follow and timescales, as long as they are sufficiently slow for the system to remain in the ground state, $\ket{\ldots,\Psi_N,\ldots\Psi_{N'},\ldots}=U(A\to B)\ket{\ldots,\Psi_{N'},\ldots\Psi_{N},\ldots}$. For these reasons, these anyons are often referred to as non-Abelian, as exchange operations do not commute.

Majorana bound states are quasiparticles with non-Abelian exchange properties. To understand this property, we introduce the concept of fermion parity, related to the number of electrons being even or odd in a given system or part of the system. In a superconducting system, the total fermion parity is a conserved quantity. In the ground states, the parity of the topological superconductor is determined by the Majorana parity,  defined as
\begin{equation}
    P_{12}=1-2n=1-2f^\dagger f=i\gamma^A_1\gamma_2^B\,,
    \label{Eq:MMBS_parity}
\end{equation}
where we have used Eq.~\eqref{newopMaj} for the definition of the non-local fermion operators as a function of MBS operators. The eigenvalues of the parity operator, $P$, are $\pm1$, corresponding to even/odd fermion parity. In case of having additional MBS pairs, the total fermion parity is the sum of the one of each pair. The ground state defined by a pair of MBSs is twofold degenerate, as even and odd fermion parity have the same energy. In general, the ground state of $m$ MBS pairs is $2^m$-fold degenerate, corresponding to every pair having even/odd fermion parity. This ground state degeneracy is a requirement for the Majorana non-Abelian exchange properties.

To understan the origin of non-Abelian properties, we can start with the simple picture of moving MBSs around each other, originally conceived by D. Ivanov for vortices in superconductors~\cite{Ivanov_PRL2001}. Encircling a vortex implies a change of $2\pi$ in the phase, which is mathematically taken into account via the branchcut associated to every vortex. During the exchange, one of the MBSs will cross a brnachcut, while the other will not, resulting in the exchange
\begin{equation}
    \gamma_1\to-\gamma_2,\qquad\gamma_2\to\gamma_1\,.
\end{equation}
The physical operation that implements this exchange is given by
\begin{equation}
    U_{12}=\frac{1}{\sqrt{2}}\left(1+\gamma_1\gamma_2\right)=e^{-i\gamma_1\gamma_2\pi/4}\,.
\end{equation}
The consequences of braiding two MBSs become clearer when considering the fermion parity of a pair of MBSs, defined in Eq.~\eqref{Eq:MMBS_parity}. If we consider the two MBSs are in an initial parity state $\ket{n}$, where $n=0,1$ for even/odd state, braiding two MBSs transforms the state as
\begin{equation}
    U_{12}\ket{n}=e^{\pi/4(1-2n)}\ket{n}\,.
\end{equation}
this corresponds to a global phase added to the wavefunction of the system.

The brid operation becomes more interesting in the situation with more than a pair of MBSs. The simplest non-trivial scenario is the one with two pairs of MBSs, whose state can be described by the wavefunction $\ket{n_{12}n_{34}}$. In this case, exchanging MBSs that belong to the same pair leads to global phases in the wavefunction,
\begin{eqnarray}
    U_{12}\ket{n_{12},n_{34}}=e^{\pi/4(1-2n_{12})}\ket{n_{12},n_{34}}\,,\\
    U_{34}\ket{n_{12},n_{34}}=e^{\pi/4(1-2n_{34})}\ket{n_{12},n_{34}}\,.
\end{eqnarray}
The implications of braiding become more evident when exchanging MBSs that do not belong to the same pair
\begin{eqnarray}
    U_{23}\ket{n_{12},n_{34}}=\frac{1}{\sqrt{2}}\left[\ket{n_{12},n_{34}}+i(-1)^{n_{12}}\ket{1-n_{12},1-n_{34}}\right]\,.
\end{eqnarray}
In this case, the exchange of two MBSs leads to a non-trivial rotation of the ground state. As expected, the total parity of the system is preserved. However, the local parity of the two MBS pairs can be reversed by either splitting or a Cooper pair, where the two electrons end up or recombine from the non-local state defined by the MBSs. Of course, the choice of two MBSs belonging or not to the same pair is arbitrary. Therefore, fact that a braid looks like a trivial phase or a non-trivial rotation of the state depends on the choice of the basis. This implies that the state $\ket{n_{12},n_{34}}$ is going to look as an equal superposition between the two states with the same total parity in the basis $\ket{n_{13},n_{24}}$ or $\ket{n_{14},n_{23}}$. It means that a system that is initialize in a given basis can look as a superposition between two states in another one. This exotic property is also known as Majorana fusion rules, and related to the possible outcomes when coupling an even number of MBSs, that depends solely on the system's state.

Protocols to demonstrate Majorana fusion and braid statistics are discussed in Sec.~\ref{Sec:fusion} and \ref{Sec:MajoranaBraid}, respectively.

\section{Top down approach toward topological superconductivity}
\label{sec:4}
In 2008 Fu and Kane put forward the conceptual breakthrough of effectively creating p-wave superconductivity and MBSs out of standard s-wave superconductors by virtue of the proximity effect acting onto the helical edge states of topological insulators (propagating edge states with spin-momentum locking)~\cite{Fu_PRL2008}. The possibility of combining different materials to engineer the topological superconducting state has spurred immense interest in the physics of Majorana states in hybrid systems. Fu and Kane's idea was soon extended to other materials with helical states, but different from topological insulators. In 2010 Lutchyn and Oreg described how to bring this proposal to reality by combining well-studied semiconducting materials \cite{Lutchyn_PRL2010,Oreg_PRL2010,AguadoReview,LutchynReview}. They proposed the use of hybrid superconducting-semiconducting nanowires subjected to an external magnetic field $B$. To induce a topological transition (namely, the closing and reopening of an energy gap), the idea is to exploit the competition among three different effects. First, the s-wave superconducting proximity effect, with an induced gap $\Delta_{\rm ind}$: the generation of spin-singlet Cooper pairs in a semiconductor. Second, the Zeeman energy
$E_Z=1/2g\mu_B B$, with $g$ the nanowire's Land\'e factor and $\mu_B$ the Bohr magneton, tends to break Cooper pairs by aligning their electron spins and closing the superconducting gap. Third, spin-orbit coupling, negates the external magnetic field by preventing the spins from reaching full alignment. The competition between the second and third effects creates regions in parameter space where the gap closes and reopens again.
Such phase transition occurs when
\begin{equation}
    E_Z= \sqrt{\mu^2+\Delta_{\rm ind}^2}\,,
    \label{Eq:topoBoundary}
\end{equation}
where $\mu$ is the chemical potential of the semiconductor taken from the bottom of the band. Interestingly, above this critical Zeeman field the proximitized semiconductor effectively behaves as a topological spinless $p$-wave superconductor, the conceptual model for one-dimensional p-wave superconductivity proposed by Kitaev \cite{Kitaev_2001}, see Sec.~\ref{sec:3}.

This idea has been extensively analyzed both theoretically and experimentally in the last two decades. In this platform, the semiconductor plays an important role as a booster of the Zeeman field in the semiconductor that allows crossing the topological phase transition without suppressing superconductivity (the Land\'e factor of the parent metallic superconductor is usually much smaller than the one of the semiconductor $g_S\ll g$). Also it is crucial to be able to gate the semiconducting nanowire to a low-density regime. Specifically, a small $\mu$ minimizes the value of the magnetic field needed to enter the topological regime, Eq.~\eqref{Eq:topoBoundary}, making it clear why a semiconductor with a low-carrier concentration is advantageous. Also, the Zeeman field cannot overcome the Chandrasekhar-Clogston limit in the superconductor, where superconductivity is suppressed, $\frac{1}{2}g_S\mu_B B<\Delta_S/\sqrt{2}$, where $\Delta_S$ is the gap of the parent superconductor. Therefore, the desirable situation involves a material combination with a very large g-factor imbalance, $|g/g_S|\gg1$ and small $\mu$. Although entering the topological regime is easier for small $\Delta_{\rm ind}$, the topological gap, which sets the topological protection, depends on this parameter. For this reason, it is desirable having a large induced gap. We note that the spin-orbit coupling does not appear explicitly in the topological criterion, Eq.~\eqref{Eq:topoBoundary}. The topological gap, however, depends on the value of the spin-orbit coupling, which has to be large to ensure a good energy separation between the MBSs and the continuum of states. 
\begin{svgraybox}
The profound connection between fermion parity crossings in QDs, which could be considered as the zero-dimensional limit of a proximitized semiconductor, and the above topological criterion is evident by just comparing Eq.~\eqref{Eq:topoBoundary} with Eq.~\eqref{parity-crossingQD}, which just expresses that parity crossings occur when the effective Zeeman effect in the QD overcomes the superconducting pairing induced by the proximity effect.
\end{svgraybox}
\subsection{The semiconductor-superconductor platform}
\label{subsec::experimentsMBS}
\begin{figure}[t]
\includegraphics[width=1\linewidth]{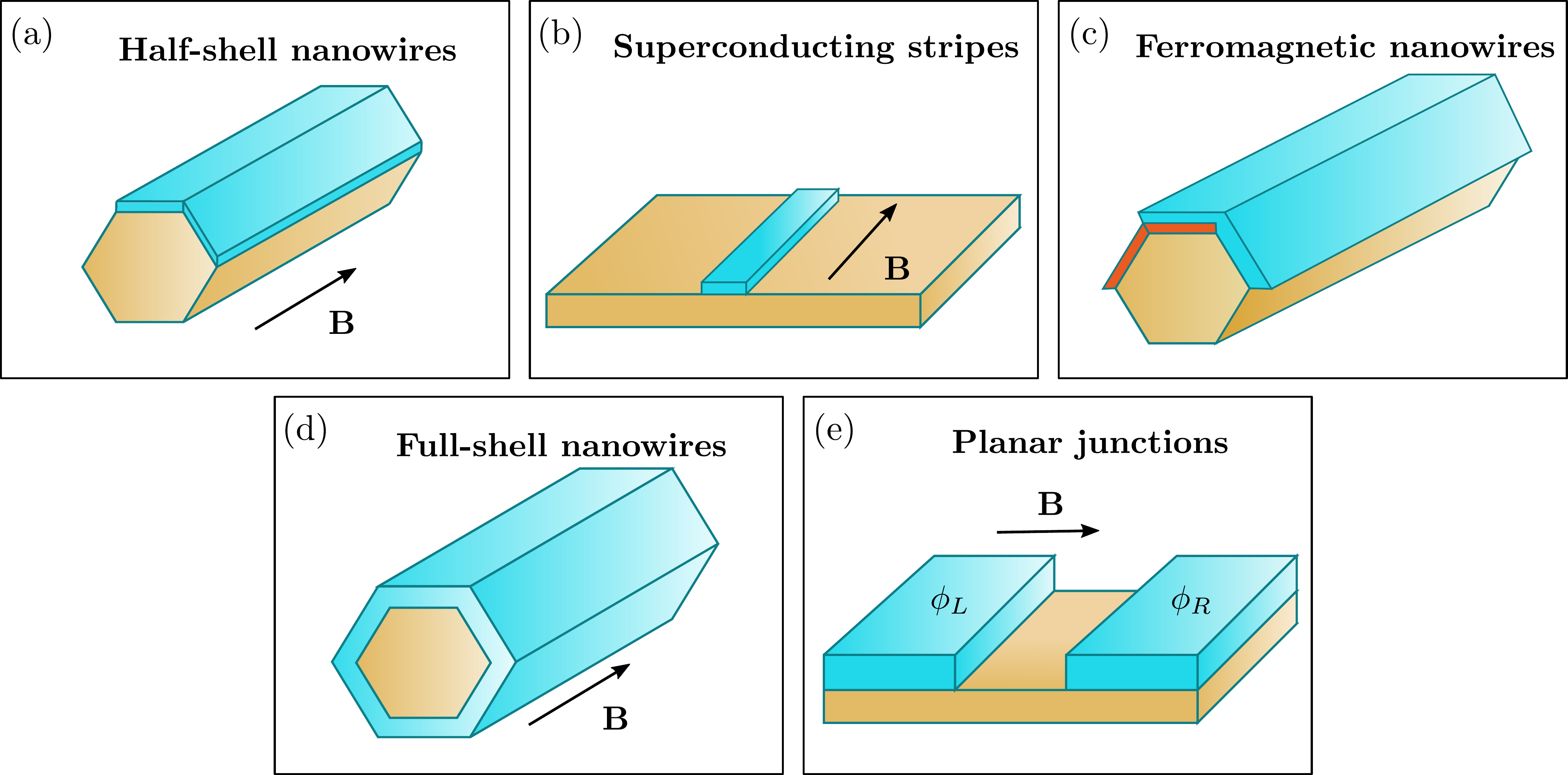}
\caption{Sketch of some of the most studied semiconductor-superconductor platforms for Majorana physics. (a) Semiconductor nanowire (orange) partially covered by a superconducting shell. (d) Illustration of a superconducting stripe on top of a semiconductor 2-dimensional electron gas. In this case, 1-dimensional channels can be achieved by electrostatic confinement. (c) Half-shell nanowire partially covered by a ferromagnetic insulator shell (red). (d) Full shell wire, where the superconductor wraps around the semiconductor nanowire. (e) Planar Josephson junction, where the phase difference between the superconductors ($\phi_L-\phi_R$) can tune the topological transition.}
\label{fig:ExperimentalPlatforms}
\end{figure}

From the experimental side, the Majorana field took over in 2012 with the first observation of robust zero-bias conductance peak in spectroscopy, signaling a zero-energy states at the end of semiconductor nanowires partially covered with a superconductor~\cite{Mourik_science2012}, see Fig.~\ref{fig:ExperimentalPlatforms}(a) for an sketch. This observation was followed by similar measurements obtained in other laboratories using different material combinations~\cite{Deng_NanoLett2012,Das_NatPhys2012,Rokhinson_NatPhys2012,Churchill_PRB2013}. A major breakthrough appeared with the advent of epitaxial semiconductor-superconductor heterostructures that allowed for clean interfaces that result in hard superconducting gaps, {\it i.e.} without undesired states inside the gap~\cite{Krogstrup_NatMat2015,Chang_NatNano2015,Sestoft_PRM2018,Gul_NatNano2018}. In parallel to these developments in half-shell wires, other semiconductor-superconductor platforms have been studied, see Fig.~\ref{fig:ExperimentalPlatforms} for sketches of the most well studied and Refs.~\cite{flensberg2021engineered,Yazdani_Science2023} for reviews. In the following we mention the most important ones.

Two dimensional electron gases are versatile platforms for topological superconductivity as they allow to define superconducting 1-dimensional channels in different orientations of the plane. This can be done by electrostatic confinement using gate electrodes. A superconducting stripe can be used to induce the proximity pairing needed to get MBSs, see Fig.~\ref{fig:ExperimentalPlatforms}(b) for a sketch. The superconductor screen the electric field coming from the gates, allowing a regime where a quasi 1-dimensional channel develops below the superconductor. Epitaxial growth has been also developed for this platform~\cite{Shabani_PRB2016,Kjaergaard_PRA2017}. Recent experiments have explored local and non-local spectroscopy has been reported in this platform~\cite{Poschl_PRB2022,Poschl_PRB2022_2,Danilenko_PRB2023}. Some of these experiments reported low-energy states appearing under a parallel magnetic field to the superconducting stripe, consistent with theory predictions for Majorana states~\cite{Nichele_PRL2017,Farrell_PRL2018,Aghaee_PRB2023}.

Orbital effects can destroy superconductivity well-below the theoretical limit imposed by the Chandrasekhar-Clogston limit. Orbital effects can be minimized by using thin superconductors and magnetic fields that align to their smallest cross section, imposing strong constrains on the geometries that can be studied using nanowires. Ferromagnetic materials offer the possibility to overcome these two issues. In order to avoid undesired quasiparticles in the system, the ferromagnet should be insulating. The platform combining semiconductor-superconductor-ferromagnet in epitaxial growth was developed In Ref.~\cite{Liu_NanoLett2020}, see the sketch of the system in Fig.~\ref{fig:ExperimentalPlatforms}(c). Early experiments in these systems reported low-energy states at the end of the wire~\cite{Vaitiekenas_NatPhys2021} that are spin-polarized~\cite{Vaitiekenas_PRB2022} and supercurrent reversal due to a large exchange field~\cite{Razmadze_PRB2023}. The theory of these devices showed that the exchange field induced in the semiconductor is essential to reach the topological regime~\cite{Woods_PRB2021,Escribano_PRB2021,Liu_PRB2021,Khindanov_PRB2021,Maiani_PRB2021,Poyhonen_SciPost2021}. Recently, the platform has been extended to two-dimension electron gas~\cite{Liu_PRB2022_ferro,Escribano_NPJ2022}.

Full-shell nanowires, {\it i.e.} semiconducting nanowires entirely surrounded by a superconductor, are another way to decrease the required external field to reach the topological regime. The narrow superconducting cylinder feature the so-called Little-Parks effect~\cite{Little_PRL1962}: a set of superconducting lobes that close when the magnetic field is half a flux quantum across the nanowire section, with a maximal gap when the flux is an integer number of the flux quantum~\cite{Vaitiekenas_Science2020,Vekris_SciRep2021,Ibabe_NatCom2023,Razmadze_PRB2024}. Signatures of low-energy states were found in the first lobe, where the phase of the superconducting parameter winds $2\pi$ and describes a fluxoid~\cite{Vaitiekenas_Science2020}, consistent with theory predictions~\cite{Penaranda_PRR2020,Paya_arxiv2023}. Later experiments found that near-zero energy subgap states of non-topological origin could explain the measured low-energy features in the data~\cite{Valentini_Science2021,Valentini_Nature2022}. Apart from the low-energy state, additional subgap states appear in the first lobe, analogous to the Caroli-de Gennes-Matricon states observed in type II superconductors~\cite{SanJose_PRB2023}.

The superconducting phase difference between superconductors offers a way to engineer topological phases without large magnetic fields, see Ref.~\cite{Lesser_JoPD2022} for a review. The simplest proposal in terms of superconductors is composed by a planar semiconductor proximitized by two superconductors in a Josephson junction geometry~\cite{Pientka_PRX2017,Hell_PRL2017,Laeven_PRL2020,Ikegaya_PRB2020,Paudel_PRB2021,Sardinero_arxiv2024}. This proposal exploits the gap closure in transparent bulk Josephson junctions close to phase difference of $\pi$. At this point, any small Zeeman field can induce a topological phase transition. Planar Josephson junctions have been also investigated experimentally in Refs.~\cite{Ren_Nature2019,Fornieri_Nature2019,Shabani_PRL2021,Banerjee_PRL2023,Banerjee_PRL2023_2}, finding low-energy states.
\begin{question}{Have Majoranas been detected?}
Several experiments revealed signatures consistent with the presence of MBSs in semiconductor-superconductor heterostructures. Apart from the aforementioned zero-bias conductance peak reproduced in different platforms, researchers studied the temperature dependence of the peak height, finding a behavior that is consistent with the theory prediction for MBSs~\cite{Nichele_PRL2017}. Other experiments studied the hybridization of the low-energy states in semiconductor-superconductor heterostructures with a QD~\cite{Deng_Science2016,deng2018nonlocality,Poschl_PRB2022}, that provides information about the localization of Majorana modes~\cite{Prada_PRB2017,Clarke_PRB2017}.

Simultaneous transport measurements at both sides of the nanowire provides additional information about the system, including the presence of subgap states at both ends. Additionally, the non-local conductance, {\it i.e.} local conductance measurements after applying a voltage at the other side, contain additional information about the localization of states inside the gap and their BCS charge~\cite{Danon_PRL2020,Menard_PRL2020,Poschl_PRB2022_2,Maiani_PRB2022_conductance}. These signatures were compiled in a protocol to identify the topological regime based on the simultaneous detection of states at the two ends of the wire and the closing and reopening of the induced gap in the nanowire~\cite{Pikulin_arxiv2021}. Recently, researchers at Microsoft claimed that some nanowires have passed their topological criterion~\cite{Aghaee_PRB2023}.

Majorana bound states can also appear in floating wires, where the thin superconductor is not connected to a big reservoir of electrons. In this configuration, charging energy becomes important, providing an effective coupling between the two MBSs that splits the ground state degeneracy. This splitting is tunable using an external gate that controls the charge occupation of the island. Early proposals studied electron transport through Majorana islands, demonstrating long-range coherent transport between the two ends of the wire, so-called electron teleportation \cite{Fu_PRL2010,Hutzen_PRL2012}. Experiments in semiconductor-superconductor islands revealed an exponential trend of the energy splitting with the length of the island~\cite{Whiticar_NatCom2020,Vaitiekenas_Science2020}. Also, the predicted periodicity after the addition of 1 electron in Majorana islands has been demonstrated~\cite{Shen_NatComm2018,Carrad_AdvMat2020,vanZanten_NatPhys2020}, including interferometry features~\cite{Whiticar_NatCom2020}. Finally, cotunneling features in the transport through the island allowed to determine the ``spinless'' nature of low energy states in ferromagnetic wires~\cite{Vaitiekenas_PRB2022}.

Despite all the evidence, a definitive demonstration of the topological origin of the measured zero-energy states would require measuring Majorana's non-local properties, Sec.~\ref{Sec:Sec:coherent_MBSs} for details. The reason is that robust zero-energy states of non-topological origin can appear at zero energy in the presence of disorder or smooth confinement potentials in semiconductor-superconductor wires, as discussed in, for example, Refs.~\cite{Prada_PRB2012,Kells_PRB12,Moore_PRB18,Vuik_SciPost19,Awoga_PRL2019,Pan_PRR20,Avila_ComPhys2019,Legg_PRL2023}. In fact, spatially overlapping MBSs at one of the of the nanowire, sometimes referred to quasi- or pseudo-Majoranas~\cite{Vuik_SciPost19}, can mimic the local properties of topological MBSs. Interestingly, this Andreev versus Majorana controversy, and why topologically trivial zero modes often mimic Majorana properties, can be clarified when framed in the language of non-Hermitian topology. This change of paradigm allows one to understand topological transitions and the emergence of zero modes in more general systems than can be described by band topology. This is achieved by studying exceptional point bifurcations in the complex spectrum of the system’s non-Hermitian Hamiltonian. Within this broader topological classification, Majoranas from both conventional band topology and a large subset of Andreev levels at zero energy are in fact topologically equivalent, which explains why they cannot be distinguished \cite{Avila_ComPhys2019}. 

Finally, and very importantly, unintentional QD formation in tunneling experiments is a common source of zero bias peaks in conductance owing to the YSR physics that we discussed in subsection \ref{InteractingQDS}. Specifically, as the charge state, and thereby the fermion parity, of the QD is tuned against some external parameter (e.g gate voltage or magnetic field), the ground state switches between the singlet and doublet states \cite{Lee-NatureNano14,PhysRevB.95.180502,Valentini_Science2021}.  Such parity crossings appear in tunneling conductance as zero-bias peaks \cite{PhysRevB.91.045441}.
\end{question}
\begin{svgraybox}
The presence of nontopological subgap states in semiconductor-superconductor hybrid devices is often considered a drawback that complicates the interpretation of experiments. However, as we now comment, this view is perhaps \emph{too short-sighted}: recent advances in this field are, in fact, strongly driven by the recent demonstration that topologically trivial ABS can also be used to tune artificial Kitaev chains in which topological MBSs can be designed, see Sec.~\ref{sec:5}. In addition, qubits based on Andreev and YSR states offer another interesting alternative in which hybrid qubit research is progressing very rapidly, see Sec.~\ref{Sec:qubits}. In the following sections, we review all these fascinating advances based on subgap states.
\end{svgraybox}

\section{Bottom-up approach: minimal Kitaev chains}
\label{sec:5}
The Kitaev chain, based on spinless fermions with p-wave pairing is the simplest toy model to understand topological superconductivity, see in Sec.~\ref{sec:4} for details~\cite{Kitaev_2001}. The model predicts a topological phase transition characterized by the onset of Majorana states at the ends of the system, which are non-local states with non-Abelian exchange properties. The Kitaev chain has a very simple solution when the nearest-neighbor hoping ($t$) and the p-wave pair amplitude ($\Delta$) are equal, and the chemical potentials of the sites ($\varepsilon_i=0$) are set to zero, see Sec.~\ref{sec:3}. For these parameters, the chain hosts two MBSs perfectly localized at the two ends of the system. The perfect localization implies that the Majorana states should persist independently from the chain's length. This is true even for minimal chains, composed by two sites. As explained below, Majorana states in minimal Kitaev chains are not topologically protected, as deviations from the ideal parameters will hybridize them, lifting the ground state degeneracy. For this reason, they are usually referred to as poor man's Majoranas (PMMs). Nevertheless, PMMs share all the properties with their topological counterparts counterparts, including non-Abelian statistics, as we shall discuss in Sec.~\ref{sec:PMM_coherent}.

The recent controversy on the topological nature of the measure subgap states in different platforms has set PMMs as one of the most promising routes to demonstrate non-Abelian quasiparticles in condensed matter physics. In this respect, the robustness of the platform against disorder and the ability of tuning the system in and out of the regime where PMMs appear might be seen as an advantage rather than a disadvantage.

\subsection{Majorana states in double quantum dots}
\label{seq:6.1}

\begin{figure}[t]
\includegraphics[width=1\linewidth]{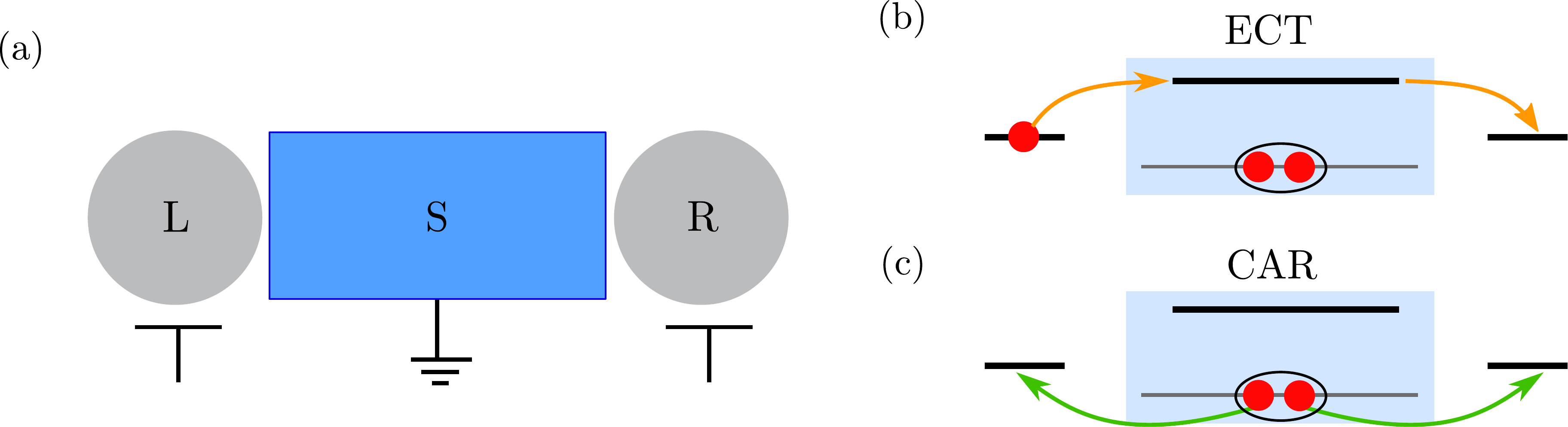}
\caption{(a) Sketch of a Poor man's Majorana (PMM) setup, composed by a narrow superconductor that mediates the coupling between two QDs. The superconducor is grounded and can host subgap states that allow for an efficient tuning of the coupling between the QDs~\cite{Liu_PRL2022}. (b) Schematic representations for the elastic cotunneling (ECT) process, where an electron (red circle) is transferred between the left and right dots occupying a virtual excited state in the superconductor (thick line in the middle segment). (c) Crossed Andreev reflection, where a Cooper pair splits (recombines) with one electron going (coming) from each QD.}
\label{fig:6.1.1}
\end{figure}

The first proposal for a minimal Kitaev chain appeared in 2012 by two groups independently, Refs.~\cite{Leijnse_PRB2012,Sau_NatComm2012}. Their setup is similar to the one used in Cooper pair splitter geometries, see for example Refs.~\cite{Recher_PRB2001,Hofstetter_Nature2009,Herrmann_PRL2010,Fulop_PRL2015}, with two QDs coupled at the two sides of a superconducting segment, Fig.~\ref{fig:6.1.1}. The QDs' properties can be tuned using electrostatic gates, including their energies and the tunnel coupling to the central superconductor. The superconductor should be smaller than the superconducting coherence length to allow for crossed Andreev reflection (CAR) processes, {\it i.e.} the splitting of a Cooper pair, where each one of the electrons ends up in a different QD. CAR plays the role of the p-wave pairing, $\Delta$, in the Kitaev chain. Elastic cotunneling (ECT) allows for an electron to jump between the dots, occupying a virtual excited state at the superconductor, which plays the role of the nearest-neighbor hoping, $t$, in the Kitaev model.

To get Majorana states in the minimal setup sketched in Fig.~\ref{fig:6.1.1}, the modulus of $t$ and $\Delta$ has to be the same and the system subject to a strong magnetic field to polarize the QDs. In the absence of spin-mixing terms, either $\Delta$ or $t$ are zero, depending on whether equal or different spin QD orbitals align. Therefore, spin-mixing terms are required to engineer MBSs, which can be introduced by different mechanisms, including spin-orbit coupling or different magnetization angles between the QDs. Irrespectively from the mechanism, the minimal model for the PMM system is given by~\cite{Leijnse_PRB2012}
\begin{equation}
	H=\sum_{\alpha=L,R}\varepsilon_\alpha d^{\dagger}_\alpha d_\alpha + \left(t d^{\dagger}_L d_R + \Delta d_L d_R +{\rm H.c.}\right),
    \label{eq:PMM_Ham_simpleModel}
\end{equation}
where $d_\alpha$ ($d^{\dagger}_\alpha$) creates (annihilates) an electron in dot $\alpha=L,R$, $\varepsilon_\alpha$ is the dot's energy measured from the superconductor's chemical potential, and $t$ and $\Delta$ the ECT and CAR amplitudes. In this equation, we have integrated the effect of the central superconductor and assumed that the system is subject to a strong magnetic field that polarizes the QDs. Therefore, one can ignore one of the spin-spices of the QDs. Extensions of this model are described below.

\begin{example}{The single particle formulation}
In the single particle basis, the Hamiltonian of the system is given by
\begin{equation}
H=\Psi^\dagger\begin{pmatrix}
\varepsilon_L & t & 0 & \Delta\\
t & \varepsilon_R & -\Delta & 0\\
0 & -\Delta & -\varepsilon_L & -t\\
\Delta & 0 & -t & -\varepsilon_R
\end{pmatrix}\Psi\,,
\end{equation}
where we are using the Nambu spinor $\Psi=(d_L,d_R,d^{\dagger}_L,d^{\dagger}_R)$ and $t$ and $\Delta$ are chosen real. The model has a sweet spot for $t=\Delta$ and $\varepsilon_L=\varepsilon_R=0$, where the system has two eigenstates with energy $E=0$ and wavefucntions given by
\begin{equation}
    \psi_1=\frac{1}{\sqrt{2}}(1,0,1,0)^T\,,\qquad
    \psi_2=\frac{i}{\sqrt{2}}(0,1,0,-1)^T\,.
    \label{eq:PMM_wavefunctions}
\end{equation}
In second quantization, the operators associated with these two states are given by $\gamma_1=(d_L+d^{\dagger}_L)/\sqrt{2}$ and $\gamma_2=i(d_R-d^{\dagger}_R)/\sqrt{2}$, which are Hermitian operators [$(\gamma_{1,2})^\dagger=\gamma_{1,2}$] and describe Majorana states completely localized in the dots. The same result can be found in the many-body basis, which reveals the even and odd fermion parity structure, see discussion around Eq.~\eqref{eq:H_PMM_1body} and Ref.~\cite{Leijnse_PRB2012}.

\begin{figure}[t]
\includegraphics[width=1\linewidth]{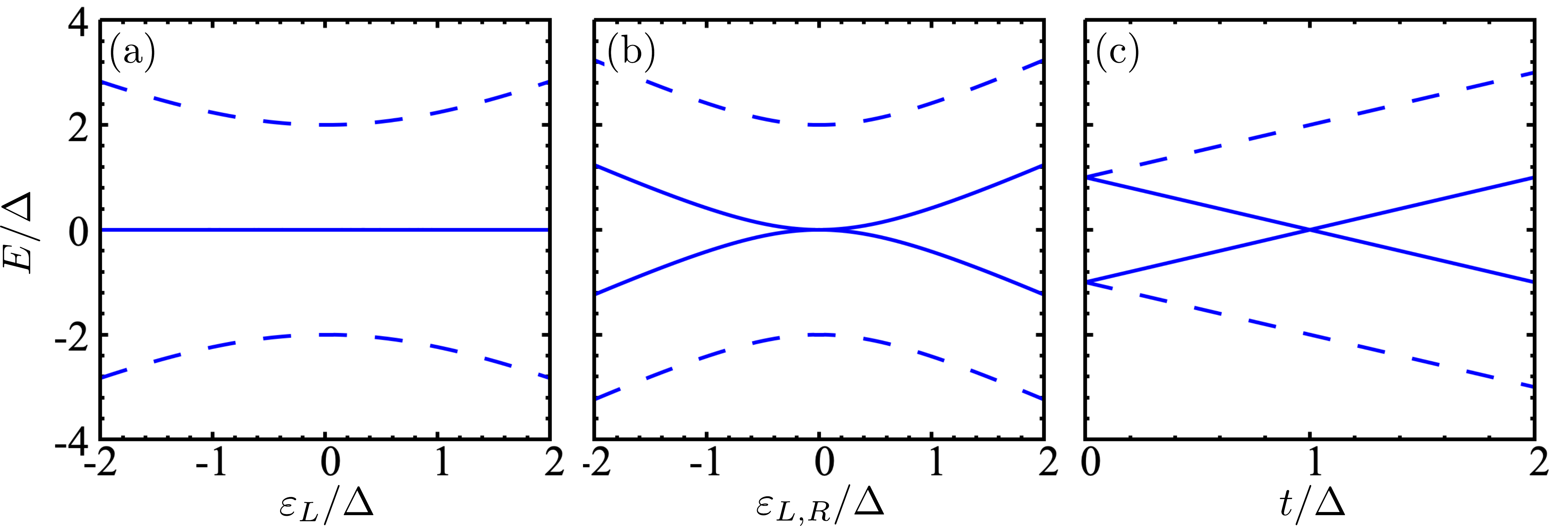}
\caption{Energy states of the PMM Hamiltonian~\eqref{eq:PMM_Ham_simpleModel}, where the two states closer to 0 are represented with solid lines. (a) energy levels for $t=\Delta$ and $\varepsilon_R=0$ when detuning $\varepsilon_L$. Two states remain degenerate, regardless of the $\varepsilon_L$ value. (b) Energy levels for $\varepsilon_L=\varepsilon_R$ and $t=\Delta$. (c) Energy levels for $\varepsilon_L=\varepsilon_R$ and different values of $t$. In all the panels, we used $\Delta=1$.}
\label{fig:6.1.2} 
\end{figure}

PMMs are not topologically protected, in the sense that detuning the system away from the ideal sweet spot makes the Majorana overlap and, eventually, split away from zero energy. For instance, if one QD is detuned away from 0 energy, the Majorana that was originally localized in the dot licks into the opposite QD. For example, if we take $\varepsilon_L\neq0$, the two wavefunctions are modified as
\begin{equation}
    \psi_1=\frac{1}{A\sqrt{2}}(1,\delta,1,\delta)^T\,,\qquad
    \psi_2=\frac{i}{\sqrt{2}}(0,1,0,-1)^T\,.
\end{equation}
with $\delta=-\varepsilon_L/2t$ and $A=\sqrt{1+\delta^2}$. In this case, both wavefunctions are still Majorana. While $\psi_2$ is still localized at the right QD, $\psi_1$ has some weight in both dots, differently from the sweet spot result, Eq.~\eqref{eq:PMM_wavefunctions}. To the lowest order in the dots onsite energies, the splitting between the two ground states for $t=\Delta$ is given by
\begin{equation}
    E_{1,2}=\pm\frac{\varepsilon_L\varepsilon_R}{2t}
\end{equation}
up to second order corrections in $\varepsilon_{L,R}$. This expression makes it clear that shifting the energy of only of the QDs does not split the energy degeneracy, eventhough two Majorana states can overlap in the other dot, see Fig.~\ref{fig:6.1.2}(a). Therefore, the energy degeneracy is linearly protected to local perturbations on the dots energies, Fig.~\ref{fig:6.1.2}(b). In contrast, deviations on the ECT and CAR amplitudes result in a linear splitting on the ground state energy, Fig.~\ref{fig:6.1.2}(c)
\begin{equation}
    E_{1,2}=\pm(|\Delta|-|t|)\,.
\end{equation}
Therefore, controlling the relative amplitude between CAR and ECT is a crucial condition to realize Majorana states in dot systems.
\end{example}

\begin{example}{Many-body formulation}
The many-body basis provides a complementary point of view to the problem, allowing also to introduce interactions between the QDs. This description accounts for the fermion occupation of each of the two QDs. In the simple description of Eq.~\eqref{eq:PMM_Ham_simpleModel}, where one of the spin spices is disregarded (limit of infinite exchange field in the dots) and the central superconductor is not described explicitly (it only mediates CAR and ECT processes), the four possible states are $\ket{00}$, $\ket{11}$ and $\ket{01}$, $\ket{10}$ for the even and odd subspaces. Here, $0$ and $1$ denote the fermion occupation of the left/right QDs. In this model, the Hamiltonian becomes block-diagonal, as there are no terms mixing the two sectors with different fermion parity
\begin{equation}
H=\begin{pmatrix}
0 & \Delta & 0 & 0\\
\Delta & \varepsilon_R+\varepsilon_L & 0 & 0\\
0 & 0 & \varepsilon_R & t\\
0 & 0 & t & -\varepsilon_L
\end{pmatrix}\,.
\label{eq:H_PMM_1body}
\end{equation}
\begin{svgraybox}
Interestingly, the above equation written in terms of the two parity sectors is nothing but a non-local version of Eq. \eqref{matrix-SC-QD}, where the spin in a single QD is replaced by a pseudospin living in the double QD system (the fermion occupation of the left/right QDs) and the hopping $t$ acting as a pseudospin mixing term. This analogy is even more evident when explicitly including a Coulomb interaction between the QDs, see Eq. \eqref{PMMfiniteU}.
\end{svgraybox}

The Hamiltonian in Eq.~\eqref{eq:H_PMM_1body} has two even and two odd fermion parity eigenvalues, given by
\begin{eqnarray}
    E^{\pm}_e&=&\frac{\varepsilon_L+\varepsilon_R\pm\sqrt{(\varepsilon_L+\varepsilon_R)^2+4\Delta^2}}{2}\,,\nonumber\\
    E^{\pm}_o&=&\frac{\varepsilon_L+\varepsilon_R\pm\sqrt{(\varepsilon_L-\varepsilon_R)^2+4t^2}}{2}\,,
    \label{eq:PMM_energies}
\end{eqnarray}
These expressions imply that the system ground state can be tuned from a total even to a total odd parity by either tuning the values of $t$ and $\Delta$, or by shifting the energy of the QDs, $\varepsilon_L$ and $\varepsilon_R$. First experiments on minimal Kitaev chains have demonstrated that the ground state change is a good way to determine the position in parameter space of Majorana sweet spots~\cite{Dvir2023}, see discussion in Sec.~\ref{sec:PMM_experiments}. The ground state degeneracy occurs for 

\begin{figure}[t]
\includegraphics[width=1\linewidth]{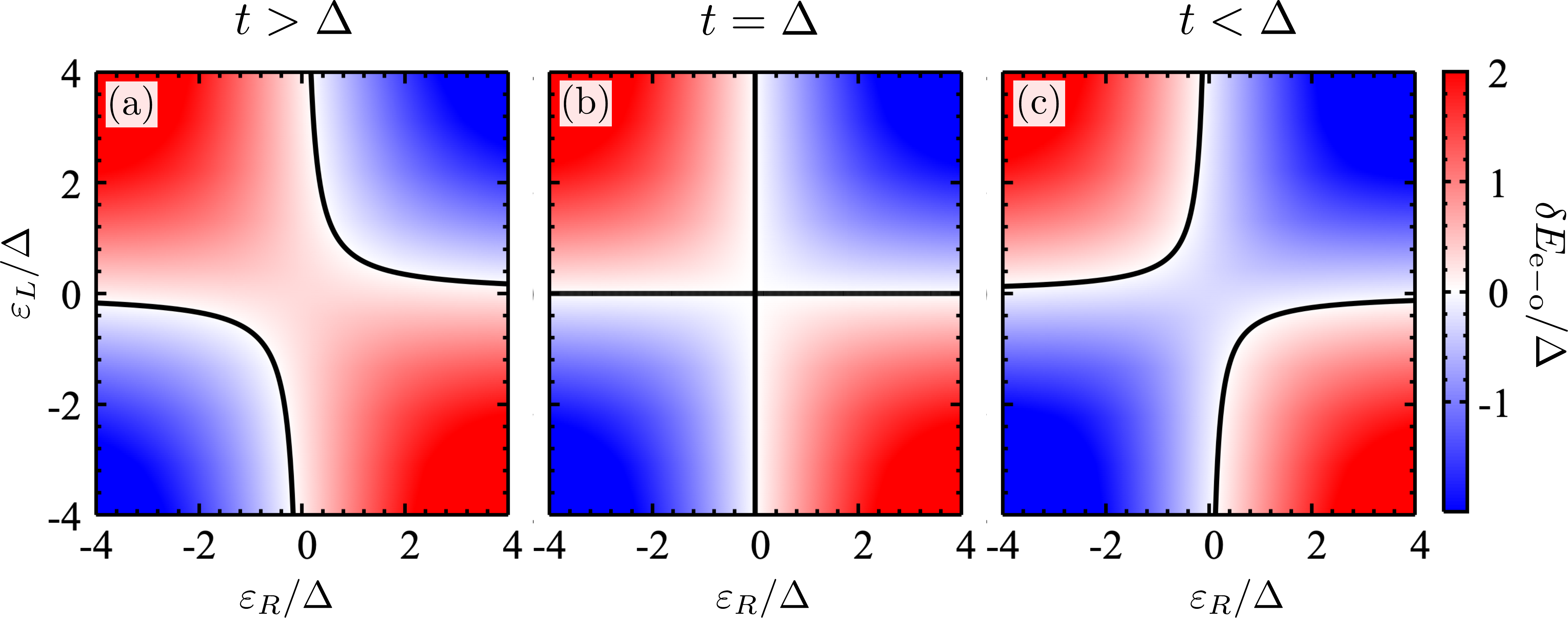}
\caption{Energy splitting between the even and odd ground states in a 
PMM system, $\delta E_{\rm e-o}=E_{\rm o}^{-}-E_{\rm e}^{-}$, given in Eq.~\eqref{eq:PMM_energies}. We show results as a function of the energy of the PMM dots, $\varepsilon_L$ and $\varepsilon_R$. The solid lines are the points with even--odd degeneracy, given by Eq.~\eqref{eq:PMM_even-odd_degeneracy}. (a) ECT-dominated regime ($t=1.3\Delta$), (b) sweet spot regime ($t=\Delta$), and (c) CAR-dominated regime ($t=0.7\Delta$). The direction of the avoided crossing can serve to distinguish between the different regimes and to find sweet spots.}
\label{fig:6.1.3} 
\end{figure}

\begin{equation}
    \varepsilon_L\,\varepsilon_R=t^2-\Delta^2\,.
    \label{eq:PMM_even-odd_degeneracy}
\end{equation}
For $t=\Delta$, this equation implies that the ground state is degenerate when either $\varepsilon_L=0$ or $\varepsilon_R=0$, see Fig.~\ref{fig:6.1.3}(b). When both are zero, the system features a Majorana sweet spot with wavefunctions for the ground state (considering $t=\Delta>0$) that are given by
\begin{eqnarray}
    \psi^{\pm}_e=\frac{1}{\sqrt{2}}\left(\ket{00}-\ket{11}\right)\,,\nonumber\\
    \psi^{\pm}_o=\frac{1}{\sqrt{2}}\left(\ket{01}-\ket{10}\right)\,.
    \label{eq:PMM_gs_simpleModel}
\end{eqnarray}
It is straightforward to proof that the local Majorana operators $\gamma_L=(c_L+c^{\dagger}_L)$ and $\gamma_R=i(c_R-c^{\dagger}_R)$ induce transitions between the two opposite parity ground states, up to an irrelevant complex phase in the wavefunction. In the Majorana language, the two states in Eqs.~\eqref{eq:PMM_gs_simpleModel} are eigenstates of the number operator $n=f^\dagger f$, corresponding to the non-local fermionic operator $f=(\gamma_L-i\gamma_R)/2$ that includes both Majorana states.

For $t\neq\Delta$, Eq.~\eqref{eq:PMM_even-odd_degeneracy} describes two hyperbolas in the $(\varepsilon_L,\,\varepsilon_R)$ plane, as shown in Figs.~\ref{fig:6.1.3}(a,c). Depending on the sign of $\Delta^2-t^2$, either CAR or ECT processes dominate, changing the orientation of the avoided crossing between the two degeneracy lines. This provides an experimental way to determine which of the two processes dominate in the system and to tune it into the sweet spot configuration. In Sec.~\ref{sec:PMM_transport} we discuss the transport signatures in the three described regime.

We conclude the section by discussing the role of non-local Coulomb interaction between the PMM dots. This interaction can be described by adding an additional term $H_U=U_{LR}\,n_L\,n_R$ to the Hamiltonian in Eq.~\eqref{eq:PMM_Ham_simpleModel}, with $n_\alpha=d^{\dagger}_\alpha d_\alpha$ being the number operator in the two dots. Therefore, the Hamiltonian matrix is given by
\begin{equation}
H=\begin{pmatrix}
0 & \Delta & 0 & 0\\
\Delta & \varepsilon_R+\varepsilon_L+U & 0 & 0\\
0 & 0 & \varepsilon_R & t\\
0 & 0 & t & -\varepsilon_L
\end{pmatrix}\,
\label{PMMfiniteU},
\end{equation}
from where it is clear that the Coulomb repulsion only affects to the states in the even parity sector, that now become
\begin{equation}
    E^{\pm}_e=\frac{\varepsilon_L+\varepsilon_R+U_{LR}\pm\sqrt{(\varepsilon_L+\varepsilon_R+U_{LR})^2+4\Delta^2}}{2}\,,
    \label{eq:PMM_energies_U}
\end{equation}
while the odd-parity energies are described in Eq.~\eqref{eq:PMM_energies}. As shown in Eq.~\eqref{eq:PMM_energies_U}, the non-local charging energy only renormalizes the effective energy of the PMM QDs. Therefore, the energy splitting between the even and odd fermion parity sectors can be compensated by gating the QDs. The system shows even-odd degeneracy for 
\begin{equation}
    U_{LR}=\frac{\left(\Delta^2-t^2+\varepsilon_L\varepsilon_R\right)\left(\varepsilon_L+\varepsilon_R+2\sqrt{t^2+(\varepsilon_L-\varepsilon_R)^2}\right)}{2\left(t^2-\varepsilon_L\varepsilon_R\right)}\,.
\end{equation}
Even if we can find sweet spots with ground state degeneracy, these points do not correspond to well-localized Majorana states in the two dots, loosing the property of non-locality. This means that local measurements in one of the dots can determine the state of the system and the system becomes sensitive to local perturbations, in contrast to the quadratic protection described in the ideal scenario. Therefore, non-local interactions between the dots are detrimental for the realization of PMMs. This is in principle not an issue, as non-local interactions are screened by the central superconductor. In Sec.~\ref{sec:PMM_realistiModels} we comment on the role of local Coulomb interactions, an important in QDs that has been disregarded up to this point.

The local superconducting pairing is another important ingredient that has been ignored when considering a large Zeeman field, so one spin component can be disregarded. Adding local BCS pairing in the dots will hybridize the different spins of the dot, mixing them and ruining the Majorana properties. However, it is possible to exploit a local BCS pairing to induce PMM states~\cite{Miles_arXiv2023,Samuelsson_arXiv2023}. The simplest geometry to understand this problem are two QDs where one of them is strongly coupled to a grounded superconductor and features an ABS. In this case, the equivalent to ECT is a direct tunneling between the QDs, {\it i.e.} a first-order tunneling. In contrast, the equivalent to CAR is a third-order tunnel process, where the a Cooper pair splits from the superconductor into the two dots. In order to make these two processes equal, one can align opposite spins in the two QDs, leading to a potential sweet spot. The price to pay is a suppression of the CAR-like tunnel amplitude, that results in a lower superconducting gap at the sweet spot.
\end{example}

\subsubsection{Microscopic model for ECT and CAR}
\label{sec:PMM_CAR_ECT}
Tuning the energy of individual dots is relatively easy and can be done using external electrostatic gates that control the charge occupation of each dot, schematically shown in Fig.~\ref{fig:6.1.1}, and the tunnel amplitudes to the superconductor. However, tuning the relative amplitude of CAR and ECT is more challenging, as they scale in a similar way with external gates controlling the dots' properties. This issue was theoretically circumvented in Ref.~\cite{Leijnse_PRB2012} by considering non-colinear Zeeman fields in the dots, that may arise from an anisotropic Landé g-factor tensor in the dots. In this way, ECT and CAR amplitudes scale as $t=t_0\cos{\theta/2}$ and $\Delta=\Delta_0\sin{\theta/2}$, with $\theta$ being the angle between the spins and $t_0$ ($\Delta$) is the maximum ECT (CAR) amplitudes for parallel (anti-parallel) spin configurations in the dot.

\begin{figure}[t]
\includegraphics[width=1\linewidth]{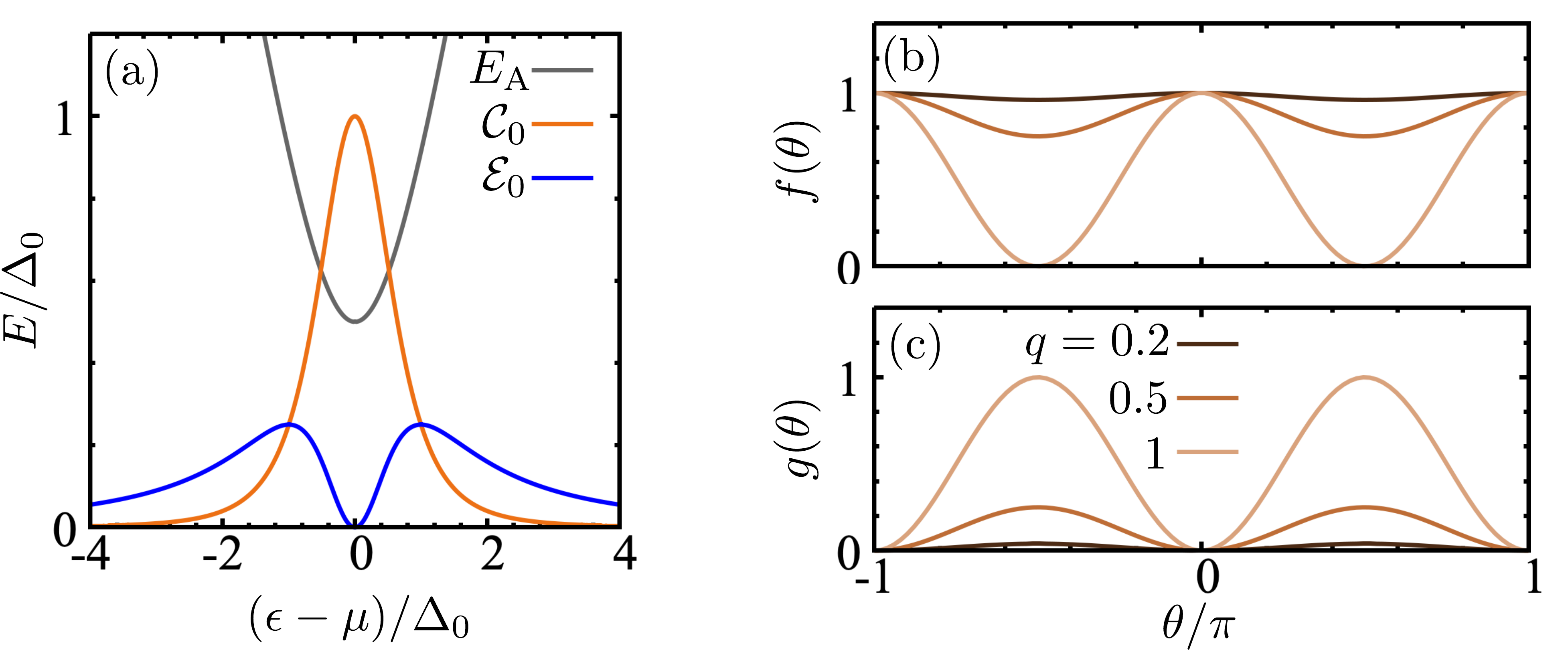}
\caption{CAR and ECT amplitudes, calculated using the model in Ref.~\cite{Liu_PRL2022}. (a) Gray line: energy of a subgap state, considering a parabolic model given in Eq.~\eqref{eq:parabolicABS}, with $\Delta_0=1$. The curve has been shifted down by $\Delta_0/2$. The orange and blue line show the amplitude for CAR and ECT amplitudes in Eq.~\eqref{eq:CAR_ECT_Wimmer}. (b) and (c) show the angle dependence of the $f$ and $g$ functions in Eq.~\eqref{eq:CAR_ECT_Wimmer} for different values of the spin precession, parameterized with $q$.}
\label{fig:6.2.1} 
\end{figure}

The spin-orbit coupling in semiconductor-superconductor and heavy compound superconductors provides a spin-mixing term for tunneling electrons. The spin-mixing term can lead to finite ECT and CAR amplitudes for  spin-polarized QDs when the spin-orbit and magnetic fields are non-colinear. The tunability of the relative amplitudes between the two processes can be achieved by using a subgap state in the middle region~\cite{Liu_PRL2022}. This subgap state provides a low-excitation energy for ECT and CAR, therefore dominating the coupling between the two QDs. In the case where the central superconductor screens the external magnetic field and up to the lowest order in the tunnel amplitudes
\begin{eqnarray}
    t&=&f(\theta)\,\mathcal{E}_0=f(\theta)\frac{t_Lt_R}{\Delta_0}\left(\frac{2uv}{E/\Delta_0}\right)^2\,,\nonumber\\
    \Delta&=&g(\theta)\,\mathcal{C}_0=g(\theta)\frac{t_Lt_R}{\Delta_0}\left(\frac{u^2-v^2}{E/\Delta_0}\right)^2\,,
    \label{eq:CAR_ECT_Wimmer}
\end{eqnarray}
where
\begin{equation}
    E=\sqrt{(\epsilon-\mu)^2+\Delta^{2}_0}
    \label{eq:parabolicABS}
\end{equation}
is the energy of the central Andreev state, with BdG coefficients (see Sec.~\ref{sec2} for a discussion) given by $u^2=1-v^2=1/2+(\epsilon-\mu)/2E$. Here, $\Delta_0$ is the gap of the central superconductor and $\mu$ controls the energy of the Andreev state, which has a minimum for $\mu=\epsilon$, Fig.~\ref{fig:6.2.1}(a). For large values of $\mu$, the Andreev state is mostly an electron ($u\approx1$) or a hole ($v\approx1$) like-state, dispersing linearly with $\mu$. In this regime, the amplitude for ECT dominates, as CAR would depend on the product between $u$ and $v$. On the other hand, the Andreev state has a minimum for $\mu=0$, where $u=v=1/\sqrt{2}$. In this regime, CAR reaches its maximum value, while ECT has a dip coming destructive interference between second-order processes, see discussion in Ref.~\cite{Liu_PRL2022}. Therefore, the ECT and CAR amplitude intersects at two values as a function of the gate controlling the energy of the central superconductor, close to where Majorana sweet spots can appear. We note that the energy of the bound state in the middle can be also tuned using the superconducting phase if it embedded in a Josepshon junction~\cite{Torres_arXiv2024}.

In this picture, we have ignore the spin texture, described by the functions $f$ and $g$ in Eq.~\eqref{eq:CAR_ECT_Wimmer}, that account for the electron's spin-rotation due to the spin-orbit coupling. These functions depend on $\theta$, the relative angle between the spin-orbit field and the local magnetization in the QDs. They are given by $f(\theta)=p^2+q^2\cos^2\theta$ and $g(\theta)=q^2\sin^2\theta$ for equal spin polarization in the dots. The definitions are exchanged for dots with opposite magnetization. Here, $p=\cos(k_{\rm so}L)$ and $q=\sin(k_{\rm so}L)$ describe the spin precession probability for an electron traveling through the system, with $p^2+q^2=1$, $k_{\rm so}$ being the spin-orbit wavevector and $L$ the length of the system. When the exchange field aligns with the exchange field direction, $\theta=n\pi$, the electron spin remains invariant as it moves along the system. Therefore, $f=1$ and $g=0$, as illustrated in Fig.~\ref{fig:6.2.1}(b,c). This situation is detrimental for spin-triplet superconductivity and, therefore, to generate Majorana states in the system, as CAR will be suppressed, see Eq.~\eqref{eq:CAR_ECT_Wimmer}. On the other hand, $\theta=\pi/2$ boosts CAR. In this situation, ECT is only suppressed in the limit where $L=1/k_{\rm so}$. In a generic situation, there is a sweet spot where both amplitudes are equal and Majoranas are expected to appear in the system. A finite exchange field does not affect to the presence of crossings where CAR and ECT are equal, see Ref.~\cite{Liu_PRL2022} for a discussion.

\subsubsection{Microscopic models for the QDs}
\label{sec:PMM_realistiModels}
As shown before, the amplitudes for CAR and ECT are boosted if the two PMM QDs interact via a subgap Andreev states, which originates originates at the interface between a superconductor with other materials. To couple the two dots via the Andreev state, the length of the superconducting segment should be much smaller than the superconducting coherence length. In this regime, the state allows for an independent control of the ECT and CAR contributions via electrostatic gating of the superconducting segment between the QDs. In a realistic scenario, the situation will be a bit more complex to the one described previously, as (i) the QDs exhibit interactions; (ii) the magentic field that can be applied to the system, and therefore the Zeeman field in the QDs, is limited by the paramagnetic limit of the superconductor; and (iii) boosting the gap at the Majorana sweet spot requires to consider a non-perturbative coupling between the QDs and the island~\cite{Liu_arXiv2023}. In this regime, the notions of CAR and ECT are not well-defied, as high-order tunneling processes between the dots dominate the physics. For this reason, in the following we will use a different notation to refer to the coupling strength between the even and odd fermion parity subspaces: even and odd coupling.

\begin{figure}[t]
\sidecaption
\includegraphics[scale=.20]{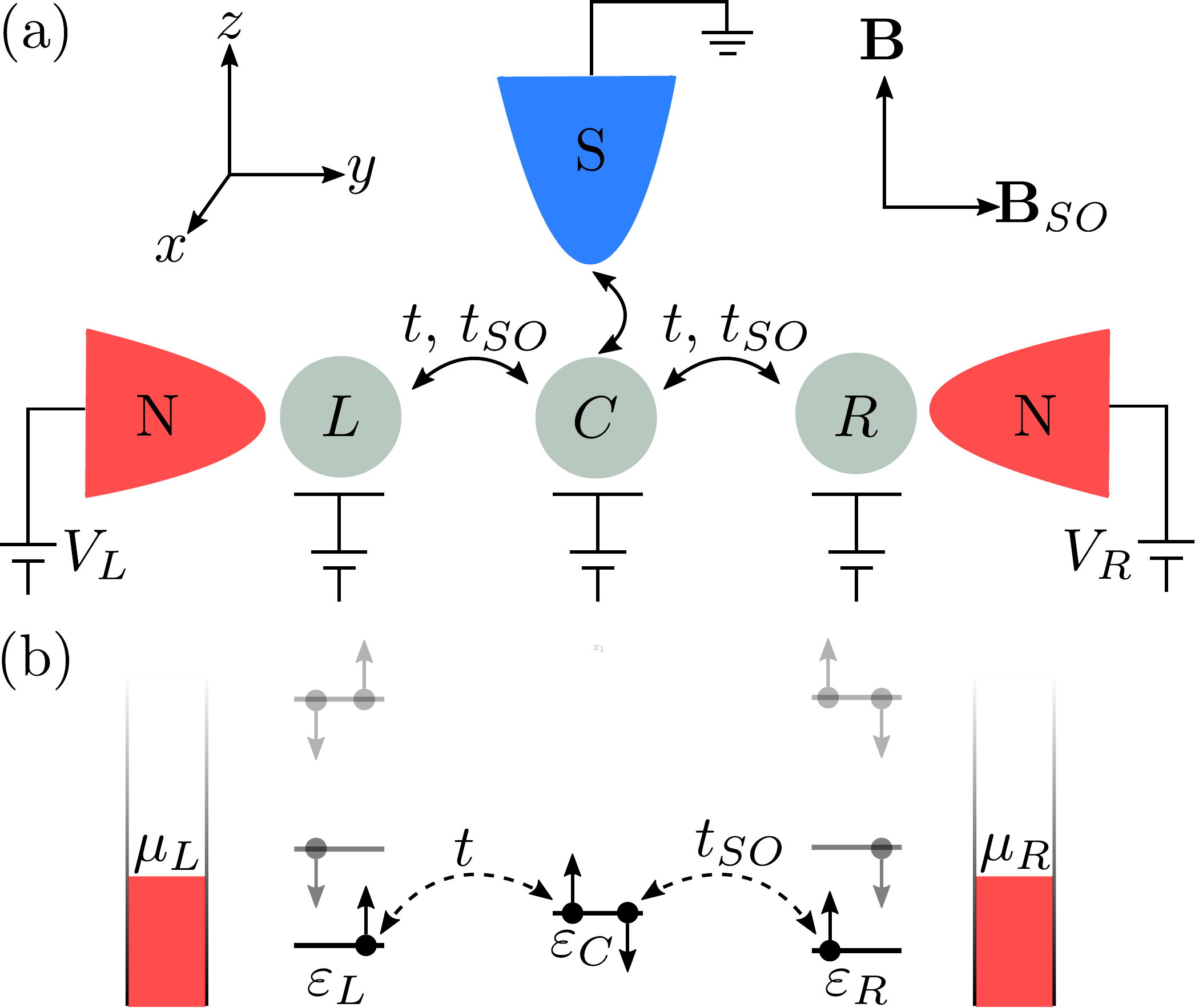}
\caption{(a) Sketch of the PMM system, containing three tunnel-coupled QDs (gray circles). The central QD is strongly coupled to a grounded superconductor (blue). We allow for spin preserving tunneling, with an amplitude $t$, and spin-flip tunneling due to spin-orbit coupling, $t^{\rm so}$. The system couples weakly to two metallic electrodes to measure the local density of states and the non-local conductance. Panel (b) shows schematically the relevant energy scales and tunneling processes due to $t$ and $t^{\rm so}$. The left and right QDs are spin-polarized due to a external magnetic field, while the middle one is in the singlet state. Reprinted from Ref.~\cite{Tsintzis2022} under CC-BY-4.0 license, \textcopyright~2022, The Author(s).}
\label{fig:6.4}
\end{figure}

The minimal microscopic model that describes all these effects is sketched in Fig.~\ref{fig:6.4} and described by the Hamiltonian

\begin{equation}
	H=H_{D}+H_{S}+H_{T},
    \label{eq:ThanosHam}
\end{equation}
where dot Hamiltonian is described by
\begin{equation}
	H_{D}=\sum_{\alpha=L,R}\left[\sum_\sigma (\varepsilon_\alpha+sE_{Z,\alpha}) d_{\alpha\sigma}^\dagger d_{\alpha\sigma}+U_\alpha n_{\alpha\uparrow} n_{\alpha\downarrow}\right]\,,
\end{equation}
with $\varepsilon_\alpha$ being the energy of the $\alpha=L,R$ QD, $d_\sigma$ the annihilation operator for an electron with spin $\sigma=\uparrow,\downarrow$, $E_Z$ the exchange field ($s=\pm1$ for $\uparrow,\downarrow$ spin), $U_\alpha$ the charging energy, and $n_\sigma=d_{\sigma}^\dagger d_{\sigma}$. The physics of the central region is dominated by the onset of the bound state. The origin of this state is not important for the described physics below. For this reason, we consider a simple realization, based on a QD coupled to a bulk superconductor that screens charging energy and exchange field in the central region. We consider a regime where the gap in the central QD dominates over the charging energy. In this limit, the infinite gap limit provides a simple description for the Andreev state, see the large gap limit in Sec.~\ref{InteractingQDS} for details, 
\begin{equation}
	\label{H_super_PMM}
	H_{S}=\sum_\sigma (\varepsilon_S+sE_{Z,S}) d_{S\sigma}^\dagger d_{S\sigma}+U_S n_{S\uparrow} n_{S\downarrow}+\Delta d_{S\uparrow}^\dagger d_{S\downarrow}^\dagger+{\rm H.c.}\,,
\end{equation} 
where $\varepsilon_S$ is the energy of the subgap state and $\Delta$ the induced pair amplitude in the central QD. The tunneling Hamiltonian is given by
\begin{equation}
	H_{T}=\sum_{\alpha\sigma}t_\alpha d^{\dagger}_{\alpha\sigma}d_{S\sigma}+t^{\rm so}_\alpha d^{\dagger}_{\alpha\sigma}d_{S\bar{\sigma}}+\mbox{H.c.}\,,
\end{equation} 
which describes the tunneling of electrons between the QDs and the superconductor, with $t$ and $t^{so}$ being the amplitudes for normal and spin-flip tunneling due to the spin-orbit coupling ($\bar{\sigma}$ denotes the opposite spin to $\sigma$). This model recovers the one described the simple PMM models, described in Sec.~\ref{seq:6.1} \cite{Leijnse_PRB2012}, in the limit $|t_\alpha|,\;|t^{\rm so}_\alpha|,\;|U_\alpha|,\;|E_{Z,\alpha}|\ll |\Delta|,\;|E_{Z,\alpha}|$. In  this limit, the outer QDs are occupied by a single electron  for $|\varepsilon_{\alpha}|\ll E_{Z,\alpha}$ and the central is in a superposition between 0 and 2 electrons. A second-order perturbation theory leads to an effective coupling between $L$ and $R$ QDs that preserves spin ($\propto t_L t_R$), ECT, and flips spin ($\propto t_L t^{\rm so}_R$, $\propto t_L t^{\rm so}_R$), CAR.

\begin{svgraybox}
In interacting Kitaev chains with finite magnetic field, there are no ideal Majorana sweet spots, as the presence of the other spin species introduces corrections to the simple picture described in Sec.~\ref{seq:6.1}. However, sweet spots with very good Majorana localization and (close to) degenerate ground states with even and odd parity can still appear in the system. The Majorana localization is quantified by the so-called Majorana polarization (MP), that describes how much the local wavefunction in the $L/R$ QDs looks like a Majorana state~\cite{Aksenov2020}. For real Hamiltonians, as the one described in Eq.~\eqref{eq:ThanosHam}, MP is defined as follows
\begin{equation}
    M_\alpha=\frac{ W_{\alpha}^2-Z_{\alpha}^2}{ W_{\alpha}^2+Z_{\alpha}^2}
    \label{eq:MP_definition}
\end{equation}
where,
\begin{eqnarray}
    W_{\alpha}=\sum_\sigma\left\langle O\left|d_{\alpha\sigma}+d^{\dagger}_{\alpha\sigma}\right|E\right\rangle\,,\nonumber\\
    Z_{\alpha}=\sum_\sigma\left\langle O\left|d_{\alpha\sigma}-d^{\dagger}_{\alpha\sigma}\right|E\right\rangle\,,
\end{eqnarray}
with $\ket{O(E)}$ being the ground state wavevector with total odd (even) fermion parity. In an ideal sweet spot where two Majoranas are perfectly localized at the left/right QD, leading to $ W_{\sigma}=1$, $Z_{\alpha}=0$ and $W_{\alpha}=0$, $Z_{\alpha}=1$ for the two sides. It means that $M_L=-M_R=\pm1$. The presence of additional Majorana components on the dot, due either to a non-perfect localization or a finite Zeeman field results on an absolute MP value smaller than 1.
\end{svgraybox}

\begin{figure}[t]
\includegraphics[width=1\linewidth]{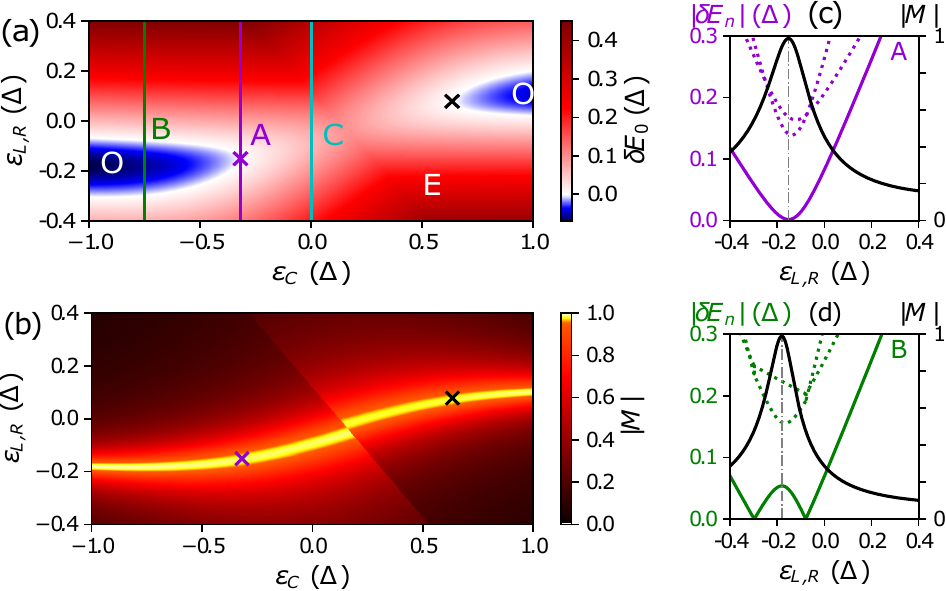}
\caption{(a) Energy spectrum as a function of the energy of the left and right QDs ($\varepsilon_{L,R}$) and the central QD ($\varepsilon_C$). Red (blue) colors indicate an even (odd) ground state, while the white stripes are regions with even--odd ground state degeneracy. (b) Absolute value of MP. The purple (black) cross indicates the spot where a ground state degeneracy coincides with a high MP values, appearing for $\varepsilon_{L,R}=-0.319\Delta\; (0.634\Delta)$ and $\varepsilon_C=-0.151\Delta\; (0.0785\Delta)$. (c) and (d) are cuts along lines A and B in panel (a). Color lines correspond to the excitation energies, where the solid line is the ground state. The black line represents the absolute value of MP and the vertical dashed-dotted line is a guide to the eye at the maximum MP value. The remaining parameters are $t=0.5\Delta$, $t^{\rm so}=0.2\Delta$, $U_{L,R}=5\Delta$, $U_C=0$, $E_{Z,L}=E_{Z,R}=1.5\Delta$, $E_{ZC}=0$. Reprinted and Adapted from Ref.~\cite{Tsintzis2022} under CC-BY-4.0 license, \textcopyright~2022, The Author(s).}
\label{fig:6.1.2.1}
\end{figure}

We note that MP does not contain information about the wavefunction localization, as it is a measurement that involves only local operators in one of the QDs. Therefore, having $|M_\alpha|\approx1$ does not imply a perfect Majorana localization on the outer $L/R$ QDs, as the Majorana wavefunctions can still overlap in the central region without splitting the ground state degeneracy. An ideal Majorana sweet spot would require having an almost unity value for $W$ and $Z$ in both dots. However, a state with a high MP still preserves the relevant Majorana properties, that implies that non-local experiments, including fusion and braiding, will also work in the setup, as we describe below in Sec.~\ref{sec:PMM_coherent}. Therefore, three conditions are required to define high-quality Majorana sweet spots: (i) degenerate ground states with even and odd fermion parity; (ii) localized MBSs of high quality, characterized by a high MP value in both QDs; and (iii) substantial gap to the excited states.

Figure \ref{fig:6.1.2.1}(a) shows results for the energy spectrum as a function of the energy of the three QDs, that can be controlled with external gates. For sufficiently large exchange field in the external PMM dots, the system can exhibit a ground state with a total odd fermion parity (blue color in the figure). Around this region, the system has an even--odd degeneracy, needed to fulfill the energy degeneracy (i). The absolute value of MP in one of the QDs is shown in Fig.~\ref{fig:6.1.2.1}(b). It peaks for some discrete values in gate configuration. For generic parameters, there are two spots at which conditions (i) and (ii) are simultaneously fulfilled, illustrated by the cuts shown in Fig.~\ref{fig:6.1.2.1}(c) and (d), where also the energy spectrum is shown.

\begin{figure}[t]
\includegraphics[width=0.9\linewidth]{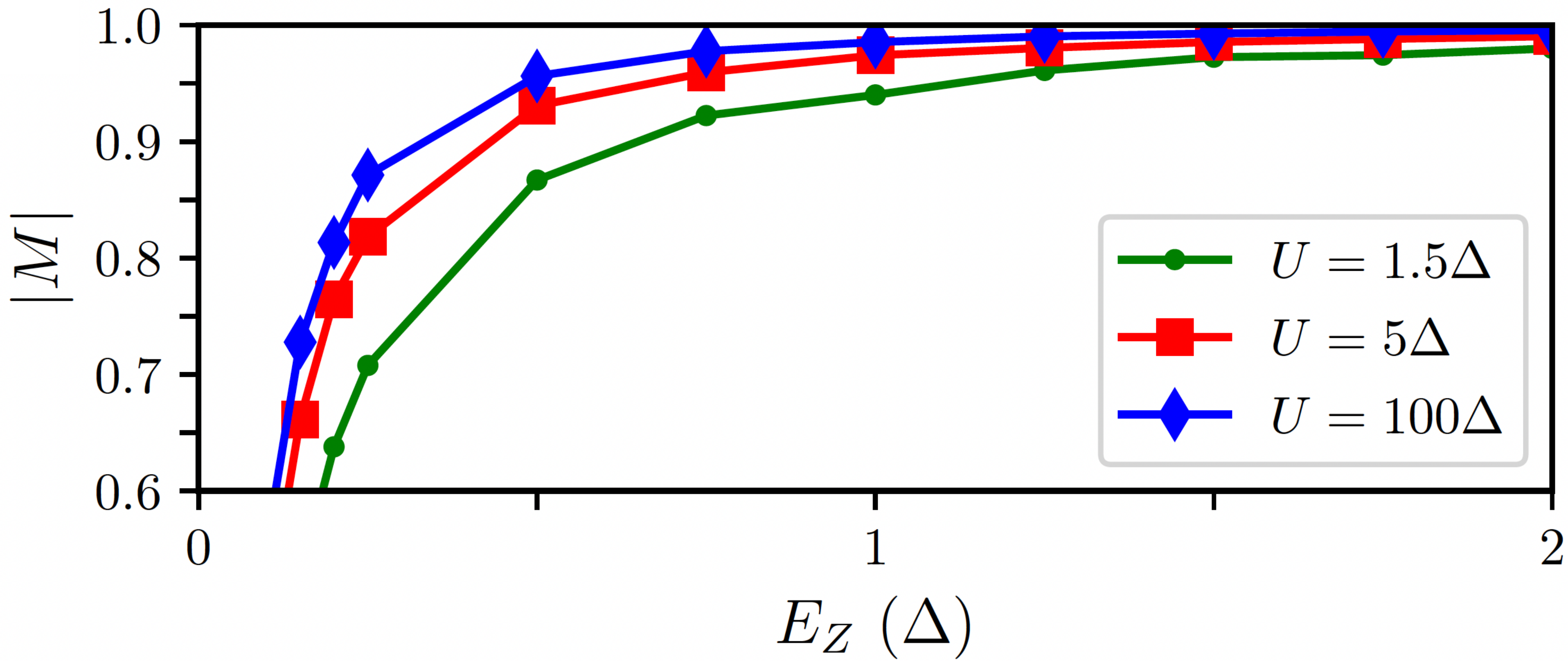}
\caption{Maximum MP values for sweet spots found with increasing exchange field and local charging energy in the left and right dots, $U$ ($U_S=0$). The sweet spots are energy crossings found by tuning $\varepsilon_L$, $\varepsilon_C$, and $\varepsilon_R$, and maximizing MP. Reprinted from Ref.~\cite{Tsintzis2022} under CC-BY-4.0 license, \textcopyright~2022, The Author(s).}
\label{fig:6.1.2.2}
\end{figure}
The system features low-MP sweet spot when either the QD energies shift from the sweet spot or the Zeeman field is reduced. Lowering the external magnetic field will re-introduce the second spin spices in the PMM dots, leading to overlapping Majorana states. Differently from the low-MP sweet spots found when varying the ratio between $t$ and $\Delta$ from unity (or, equivalently, $\varepsilon_C$ in Fig.~\ref{fig:6.1.2.1}), low-MP sweet spots can appear as (almost) zero excitation energy crossings in $\varepsilon_L$ and $\varepsilon_R$. In this regime, the local charging energy in the outer PMM dots help at increasing the MP, as illustrated in Fig.~\ref{fig:6.1.2.2}. At the mean-field level, the effect of the local interactions is to increase the local exchange field.

\subsubsection{Transport characterization}
\label{sec:PMM_transport}
Electron transport is one of the most broadly used methods to characterized the electronic properties nanoscopic systems. In this section, we introduce three different experimental measurements that can be used to characterize the PMM system and to identify Majorana sweet spots with high MP. These measurements involve coupling the PMM system to normal leads, that can be used to measure the local spectrum of the system.

\begin{figure}[t]
\includegraphics[width=1\linewidth]{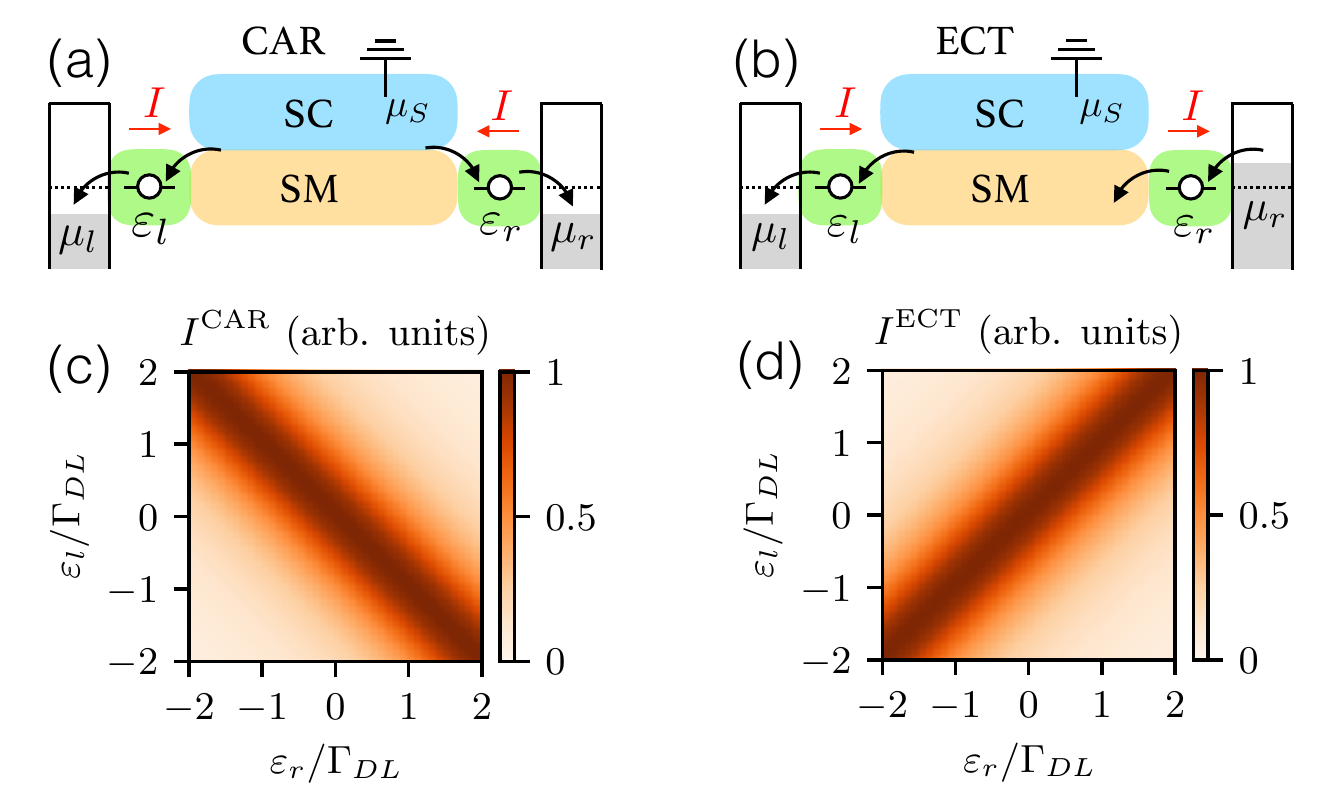}
\caption{Current measurement of CAR and ECT in the weak coupling regime. Top panels show the gate configurations used measure these two processes. The bottom panels show numerical calculations using rate equations on the model introduced in Sec.~\ref{sec:PMM_CAR_ECT}. Reprinted from Ref.~\cite{Liu_PRL2022} with permission, \textcopyright~2022, American Physical Society. All rights reserved.}
\label{fig:6.1.3.1}
\end{figure}

The first experimental challenge consists on finding regimes where the coupling between the even and odd fermion parity subspaces (or CAR and ECT in the weak superconductor-dot tunneling regime) are equal. The experimental setup to measure these two amplitude is shown schematically in the top panels of Fig.\ref{fig:6.1.3.1}, taken from Ref.~\cite{Liu_PRL2022}. In the setup, two metallic leads couple to the outer PMM dots, allowing for current to flow between the three different terminals (the two metallic electrodes and the grounded superconductor, blue rectangle in the figure). 

For gate configurations where the dots have opposite energies, $\varepsilon_L=-\varepsilon_R$, processes involving the splitting of Cooper pairs are resonant in energy, Fig.~\ref{fig:6.1.3.1}(a). The split electrons can flow out from the QDs to the metallic leads if their chemical potential is smaller than the QDs' energies, $\mu_L<\varepsilon_L$ and $\mu_R<\varepsilon_R$. Under these conditions, a net current will cross the system, flowing from the superconductor and dividing equally to the left and right metallic leads. There is a corresponding process where electrons recombine into Cooper pairs. This process is also resonant for $\varepsilon_L=-\varepsilon_R$ and dominates for a large enough bias voltage in the leads, $\mu_L>\varepsilon_L$ and $\mu_R>\varepsilon_R$, resulting in a net current flowing from the metallic electrodes into the central PMM superconductor.

In contrast, the elastic transference of electrons between the dots is resonant when the orbital energy of the two dots is the same, $\varepsilon_L=\varepsilon_R$, Fig.~\ref{fig:6.1.3.1}(b). In this case, a finite current can flow between the leads if the bias is applied asymmetrically, so that electrons can tunnel on the right dot ($\mu_R>\varepsilon_R$) and leave to the left electrode ($\mu_L<\varepsilon_L$). This generates a current between the metallic leads, equal in magnitude but with opposite sign. The current direction is changes if the biasing conditions are reversed. In this picture, the central superconductor only mediates the coupling between the dots and does not add additional charges into the system. This is only true if the energy of the two dots is small enough (smaller than the lowest state in the superconductor) to avoid the tunneling of quasiparticles between the superconductor and the QDs.

\begin{example}{Measuring ECT and CAR amplitudes from current}
Simple expressions for the current in the two configurations described above can be found using the model described in Sec.~\ref{sec:PMM_CAR_ECT} that considers very large exchange field (ignoring one of the spin-spices in the dots) and small tunneling coupling between the dots and the central superconductor. In the regime where also the coupling to the metallic leads is weak, perturbation theory provides an accurate way to obtain the current through the system. For the two chemical potential configurations described above, the current through the left lead is given by
\begin{eqnarray}
    I^{CAR}&=&\frac{e}{\hbar}\frac{\Gamma}{(\varepsilon_L+\varepsilon_R)^2+\Gamma^2}\left|\Delta\right|^2\,,\qquad {\rm for}\;\mu_S>\mu_{L,R}\,,\\
    I^{ECT}&=&\frac{e}{\hbar}\frac{\Gamma}{(\varepsilon_L-\varepsilon_R)^2+\Gamma^2}\left|t\right|^2\,,\qquad {\rm for}\;\mu_L<\mu_S<\mu_{L}\,,.
    \label{eq:PMM_Icar_Iect}
\end{eqnarray}
Results for these current contributions as a function of the energy levels of both dots are shown in Figs.~\ref{fig:6.1.3.1}(c,d). From Eqs.~\eqref{eq:PMM_Icar_Iect}, it is easy to see that the relation between the maximum current in both configurations provides a way to measure the relative strength between CAR and ECT. In fact, $I^{CAR}/I^{ECT}=|\Delta/t|^2$ for $\varepsilon_L=\varepsilon_R=0|$, providing a way to determine the effective CAR and ECT amplitudes.
\end{example}

\begin{example}{Local and non-local conductance}
\begin{figure}[t]
\includegraphics[width=1\linewidth]{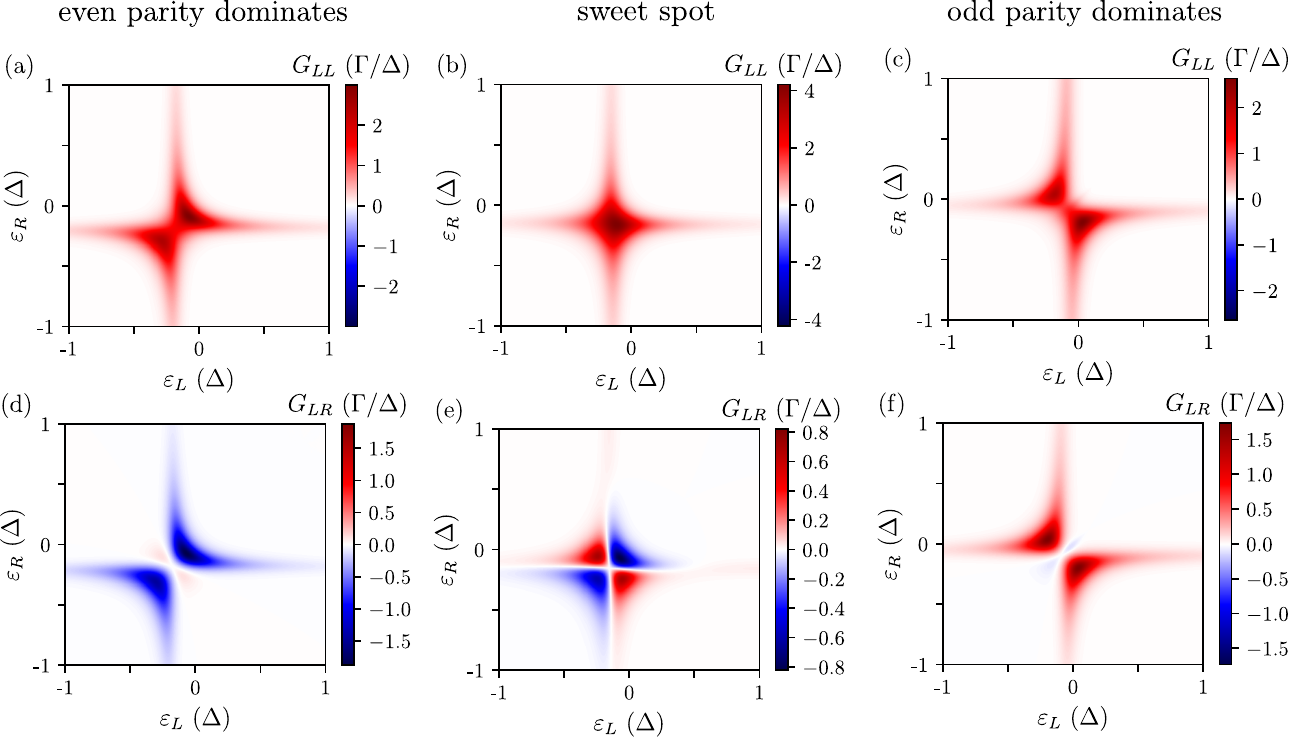}
\caption{Zero-bias conductance through a PMM system, as a function of the energy of the outer dots, which are attached to metallic electrodes, see Fig.~\ref{fig:6.4}. Top panels show local conductance at the left lead, where the bias is applied symmetrically in the two leads. Lower panels show the non-local conductance, where bias is applied to right lead and current measured in the left one. From left to right, we show results for even parity-dominated regime (the coupling between the even states is stronger), the sweet spot (equal coupling within the two fermion parity sectors), and the odd parity-dominated regime, corresponding to the $\varepsilon_C$ values highlighted in Fig.~\ref{fig:6.1.2.2}. Reprinted from Ref.~\cite{Tsintzis2022} under CC-BY-4.0 license, \textcopyright~2022, The Author(s). Similar theoretical and experimental results were reported in Ref.~\cite{Dvir2023}, see also Sec.~\ref{sec:PMM_experiments}}
\label{fig:6.1.3.2}
\end{figure}

The leads attached to the system can be also used to identify sweet spots with potentially high MP. In local spectroscopy, the low-bias transport and, therefore the zero-bias conductance, is enhanced when the full system has a degenerate ground state. This means that electrons at the Fermi level of the superconductor can resonantly tunnel to the system. This means that the local spectroscopy will result in conductance peaks at the resonant lines. This is illustrated in the top panels of Fig.~\ref{fig:6.1.3.2}, where the red lines correspond to the degeneracy between states with even and odd fermion parity (see also Fig.~\ref{fig:6.1.3}). Depending on the whether the coupling between even or odd states dominates, we expect to see an avoided crossing between the zero-bias conductance lines that go through the anti-diagonal (diagonal). In contrast, the zero-bias conductance lines cross at the sweet spot, located at the maximum of the local conductance value.

However, these features might be hard to resolve experimentally, mostly close to the sweet spot where the avoided crossing between the conductance lines might be small. The local conductance in Fig.~\ref{fig:6.1.2.1} in the upper panels can serve as an illustration. For this reason, it is very convenient to study the non-local transport through the system, that provides additional information about the nature of the crossing. To measure the non-local current, a bias voltage is applied asymmetrically to the system, while current is measured in the other side. Therefore, the non-local conductance is defined as
\begin{equation}
    G_{\mu\nu}=\frac{\partial I_\mu}{\partial V_\nu}\,,
\end{equation}
with $\mu,\nu=L,R$. The lower panels of Fig.~\ref{fig:6.1.3.2} show results for the non-local conductance, $G_{LR}$, through a PMM system around the Majorana sweet spot and at zero bias. The three cases correspond to the line cuts shown in Fig.~\ref{fig:6.1.2.1}. 

In the regime where the even parity states couple more strongly than the odd ones, transport is dominated by CAR-like processes, changing simultaneously the occupation of both dots. For $\mu_R>\mu_S$ it means that electrons in both dots will recombine in Cooper pairs at the central superconductor, leading to a current flowing in from the left lead, Fig.~\ref{fig:6.1.2.1}(d). In the opposite situation where the coupling between the odd parity states dominate, transport will be dominated by the injected electrons to the right dot will cotunnel to the left one. Therefore, current will mostly flow from right to left, Fig.~\ref{fig:6.1.2.1}(f). At the sweet spot, both processes are equally important and the non-local conductance shows a symmetric pattern of positive and negative conductance, Fig.~\ref{fig:6.1.2.1}(e). The Majorana sweet spot locates at the center of this pattern. In general, sweet spots with low-MP are characterized by a deformed non-local conductance pattern with respect to the one shown in Fig.~\ref{fig:6.1.2.1}(e), see Ref.~\cite{Tsintzis2022}. 

However, the differences between high and low-MP in the non-local conductance are sometimes faint, which makes it hard to distinguish between the two. For this reason, it is important to develop ways to identify high-MP sweet spots before proceeding to non-local experiments like braiding, Sec.~\ref{sec:PMM_coherent}.

The Majorana localization on one side of the PMM system can be proven by coupling the system to an additional dot, as theoretically proposed and analyzed in Refs.~\cite{Prada_PRB2017,Clarke_PRB2017} for Majorana wires. These proposals inspired experiments that showed similar patterns to the ones predicted by theory~\cite{Deng_Science2016,deng2018nonlocality}. In these references, they showed that a QD cannot split the ground state degeneracy when the dot couples to only one of the Majorana component. In contrast, the coupling between the dot and two Majoranas split the degeneracy. The energy splitting shows a ``diamond-like'' shape as a function of the energy of the additional dot, becoming maximal when one of the additional dot's energy levels align with the chemical potential of the superconductor. The situation is reversed if the two Majoranas overlap but only one couple to the additional dot. In this case, the ground state degeneracy is split, except when one of the additional dots level align with the superconductor's chemical potential, leading to a ``diamond-like'' pattern. This idea has been recently extended to PMM systems, showing that an additional QD can be used to identify high-MP sweet spots~\cite{Souto_PRR2023}.
\end{example}

\subsubsection{Experiments in minimal Kitaev chains}
\label{sec:PMM_experiments}
Early experiments studied CAR processes in a superconducting heterostructures, with the aim of generating a source for entangled electrons, so-called Cooper pair splitters~\cite{Recher_PRB2001,Beckmann_PRL2004,Russo_PRL2005,Hofstetter_Nature2009,Herrmann_PRL2010,Schindele_PRL2012,Das_NatCom2012,Fulop_PRL2015}. The experimental realization of these splitters require electron channels separated by a distance smaller than the superconducting coherence length, that is usually of the order of tens of nanometers to a micron for conventional BCS superconductors. In this context, QDs offer a fundamental advantage as their charging energy suppresses local Andreev reflections for a sufficiently small coupling to the central superconductor~\cite{Recher_PRB2001,Schindele_PRL2012}: processes where a Cooper pair tunnels to the same terminal.
\begin{figure}[h]
\includegraphics[width=1\textwidth]{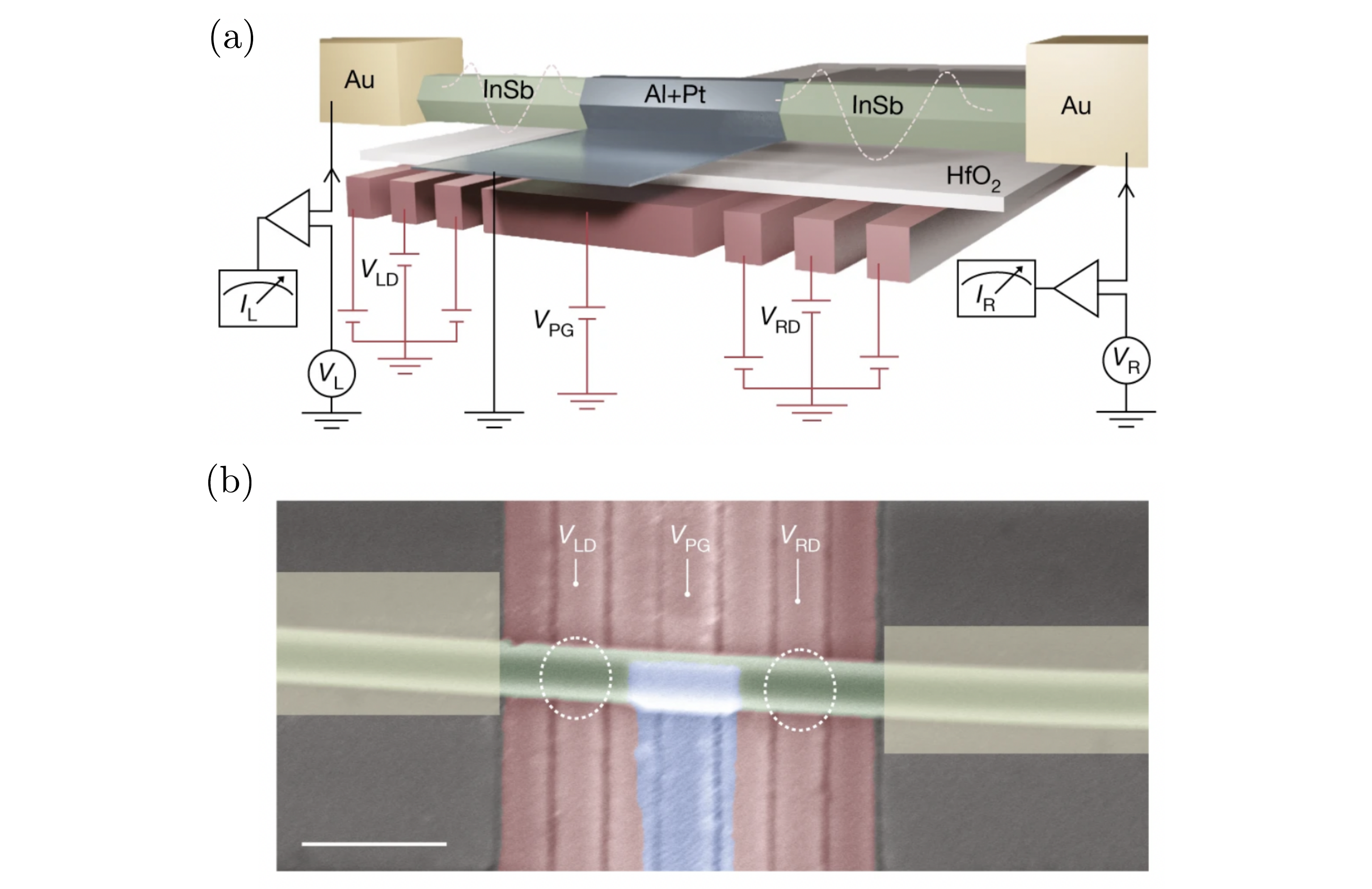}
\caption{Illustration of the PMM system used in Ref.~\cite{Dvir2023}. Two QDs are defined in an InSb nanowire and coupled via a central region that has proximity-induced superconductivity coming from a grounded superconductor. The outer Au leads are used for spectroscopy. The gates at the bottom (red) allow to tune the device into the desired configuration (b) False-colored scanning electron microscopy image of the device, before the fabrication of the N leads (schematically shown in yellow). The scale bar is 300 nm. Reprinted from Ref.~\cite{Dvir2023} with permission, \textcopyright~2023, The Author(s), under exclusive licence to Springer Nature Limited. All rights reserved.}
\label{fig:6.6}
\end{figure}
\begin{figure}[t]
\includegraphics[width=1\linewidth]{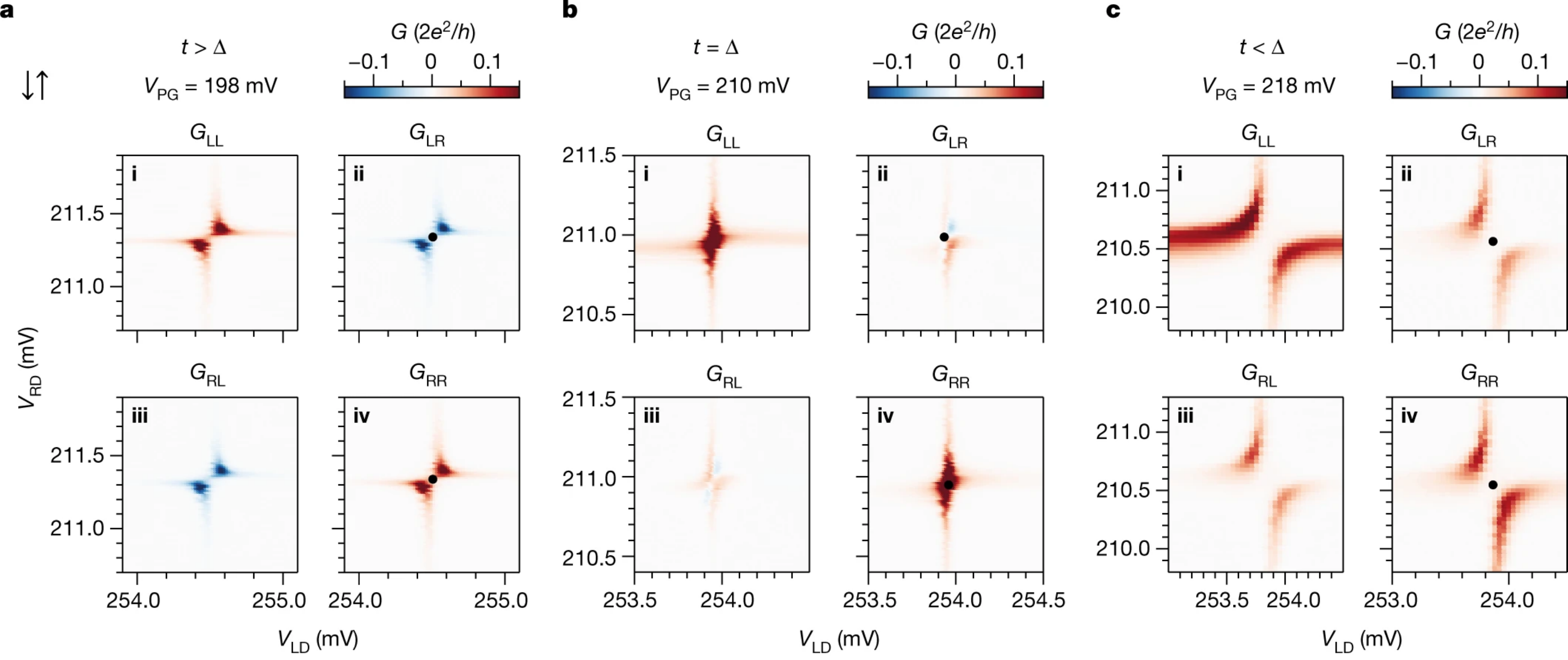}
\caption{Conductance matrices measured for different values of the gate controlling the semiconductor segment underneath the superconductor  ($V_{\rm PG}=$198, 210 and 218 mV, for panels (a-c). The three panels are representative measurements of the three possible regimes (ECT-dominated regime, close to the sweet spot with ECT and CAR amplitudes similar, and CAR-dominated regime) and in qualitative agreement with theory, see Fig.~\ref{fig:6.1.3.2}. Reprinted from Ref.~\cite{Dvir2023} with permission, \textcopyright~2023, The Author(s), under exclusive licence to Springer Nature Limited. All rights reserved.}
\label{fig:6.3.2}
\end{figure}
As mention in Sec.~\ref{sec:PMM_CAR_ECT} (see also Ref.~\cite{Liu_PRL2022}), an Andreev bound state in the superconducting region can mediate the coupling between the electrons. These bound states, with energies below the superconducting gap, offer the dominant contribution for Cooper pairs tunneling to the dots due to their reduced energies. Therefore, they offer advantages, including the increase the Cooper pair splitting efficiency and the possibility of tuning the relative amplitudes between CAR and ECT by modulating the energy of the bound state, that can be achieved through electrostatic gates acting on the central superconductor. For this task, superconductor-semiconductor hybrid structures are advantageous, as they can exhibit subgap states that are gate tunable. Experiments have demonstrated the possibility of hybridizing a subgap state with a QD~\cite{Deng_Science2016,deng2018nonlocality,Poschl_PRB2022} and tuning the particle/hole component, {\it i.e.} the BCS charge of a subgap state, using electrostatic gates~\cite{Poschl_PRB2022_2}. These are required ingredients to realize PMM systems and tune them to the sweet spots where localized Majoranas appear.

Independent measurements on semiconductor-superconductor hybrid devices demonstrated CAR processes of Cooper pairs with the same spin in nanowires~\cite{Wang_Nat2022,Bordoloi_Nat2022,Bordin_PRX2023} and two-dimensional electron gases~\cite{Wang_Natcom2023}. This observation points towards the presence of equal-spin-triplet Cooper pairs due to the interplay between spin-orbit coupling and an external magnetic field. The existence of these Cooper pairs is an essential requirement to realize topological superconductivity. In particular, Refs.~\cite{Bordoloi_Nat2022,Wang_Natcom2023} determined the amplitudes for CAR and ECT using transport measurements across the device, as described in Sec.~\ref{sec:PMM_transport}. They demonstrated a fine-tuned regime of parameters where the amplitude of both processes is the same and where PMMs can appear. Additionally, a Cooper pair splitting efficiency around 90\% was demonstrated, that is over the threshold required for a Bell test experiment~\cite{Schindele_PRL2012}.

Additional local and non-local spectroscopic measurements in a PMM system were reported in Refs.~\cite{Dvir2023,Haaf_arXiv2023,Zatelli_arXiv2023}. The device in Ref.~\cite{Dvir2023}, sketched in Fig.~\ref{fig:6.6}(a), is composed by a InSb nanowire fabricated using the shadow-wall lithography technique~\cite{Heedt_NatCom2021,Borsoi_AdvMat2021}. The bottom gates (red) control the energy and the coupling of the QDs to the central superconductor (blue) and the metallic leads used for spectroscopy (yellow rectangle). The QDs are spin-polarized using an external magnetic field, boosted by the high Land\'e g-factor measured in InSb. The grounded superconductor at the center mediates CAR and ECT processes between the spin-polarized QDs. The central region hosts an Andreev bound state that mediates both processes and allows to tune their relative amplitude, as explained in Sec.~\ref{sec:PMM_CAR_ECT}.

Measurements for the local and non-local conductance are shown in Fig.~\ref{fig:6.3.2} \cite{Dvir2023}. The figure shows representative results for the three important regimes: $t>\Delta$ (left panels), $t<\Delta$ (right panels), and $t\approx\Delta$ (middle panel). Local spectroscopy (subpanels panels i and iv in every panel) reveals an avoided crossing between two levels. The direction of the avoided crossing along the diagonal/anti-diagonal reveals the dominance of CAR/ECT processes. In contrast, for $t\approx\Delta$, the levels seem to cross, showing a high conductance peak at the sweet spot. We also note a change on the sign of the non-local conductance between the left and right panels (subpanels ii and iii), while at the sweet spot (middle panels) there seems to be an alternation between positive and negative features, although with a dominance of the positive ones. 

In the simplest model proposed by Kitaev, the protection to deviations from the ideal sweet spot improves when adding additional dots to the chain. In particular, the truly topological phase is recovered in the limit of chain being much larger than the localization of the Majorana states. In this limit, deviations from the sweet spot does not affect to the topological properties of the system. This vision motivated recent experiments on Kitaev chains with three sites, {\it i.e.} three QDs that couple via two superconducting segments~\cite{Bordin_PRL2024,Bordin_arXiv2024}.

\section{Hybrid semiconductor-based superconducting Qubits}
\label{Sec:qubits}
Superconductors constitute one of the most promising platforms for quantum technologies. Unlike typical quantum systems that are observable only at the microscopic level, superconductors allow quantum effects to be seen and manipulated at scales much larger than individual atoms or particles. At the heart of this phenomenon is the Cooper pair: two electrons that move through a superconductor without resistance. These pairs behave as a single quantum entity, displaying characteristic quantum properties like tunneling and entanglement over macroscopic distances. One of the most striking demonstrations of this is the Josephson effect, where an electric current flows between two superconductors separated by a thin barrier, without any voltage applied. 

Transmon qubits are paradigmatic examples of the application of superconductivity for the implementation of quantum technologies. They are pivotal in the advancement of quantum computing, showcasing how macroscopic quantum phenomena can be harnessed for computational purposes. Originating from the Cooper pair box design, transmon qubits have been modified to increase their stability and coherence. Their key advantage lies in their diminished sensitivity to charge noise, which plagued earlier superconducting qubit designs. This is achieved through a combination of a Josephson junction and a large shunting capacitor, resulting in a system with distinct energy level separations~\cite{Koch_PRA2007}. Quantum information in transmon qubits is encoded in the energy states of the junction and manipulated using microwave pulses. This allows transmon qubits to maintain quantum states, like superpositions and entanglements, for extended durations, essential for effective quantum computation. Additionally, their relatively straightforward design and manufacturability facilitate integration into larger, more complex quantum circuits. Transmon qubits thus represent the realization of scalable and practical quantum computers, poised to tackle computational challenges far beyond the reach of classical computing~\cite{Arute_Nature2019}. 

The semiconductor-superconductor platform allows to tune the critical current of Josephson junctions using external gate electrons. This capability allowed to design the so-called gatemon: a transmon qubit with a tunable critical current~\cite{Larsen_PRL2015,Lange_PRL2015,Casparis_NatNano2018,Zheng_arXiv2023}. This gate-tunability provides a key advantage: it allows for a more flexible and precise control over the qubit's frequency, enabling to tune the qubit into and out of resonance with other qubits or microwave cavities. It facilitates the qubit coupling and entanglement operations. Moreover, gatemon qubits retain the charge noise insensitivity characteristic of transmons, making them a promising architecture for scalable quantum computing.

Even though transmon qubits are protected against charge noise, they are still sensitive to other sources of noise that limit their performance. Different proposals based on Josephson junctions have been designed to improve the performance of qubits based on Josephson junctions, including the so-called ``$0-\pi$'' qubit~\cite{Brooks_PRA2013,Groszkowski_NJP2018,Gyenis_PRXQ2021}, parity protected qubits~\cite{Smith_NPJQI2020,Larsen_PRL2020,Schrade_PRXQ2022,Smith_PRX2022,Maiani_PRXQ2022}, and the fluxonium qubit~\cite{Bao_PRL2022}. These proposals are compatible with macroscopic superconductors featuring Josephson junctions. On the other hand, microscopic superconductors offer the possibility of manipulating superconductivity at the nanoscale, as we have illustrated in the previous sections, providing a new realm for quantum technologies. In this context, superconductor-semiconductor platforms offer many advantages. In the following of this chapter, we present qubits based on nanoscopic superconductors that have been analyzed both, from the theory and the experimental side, in the recent years.

\subsection{Qubits based on a single Andreev state}
Andreev states, appearing inside the superconducting gap, are ideal candidates to storage and process quantum information. Spin-degenerate Andreev states can host up to two quasiparticle excitations. Therefore, the system has four states in total, two with even and two with odd fermion parity. Assuming that electrons in the superconductor form Cooper pairs, the fermion parity is a good quantum number and the even and odd fermion parity subspaces are disconnected. In the ground state, all the electrons are assumed to form Cooper pairs, occupying the states with energies below the superconductor's Fermi level ($\ket{g}$). The lowest-excited state with the same parity has two excited quasiparticles in the Andreev bound state
\begin{figure}[t]
\includegraphics[width=1\linewidth]{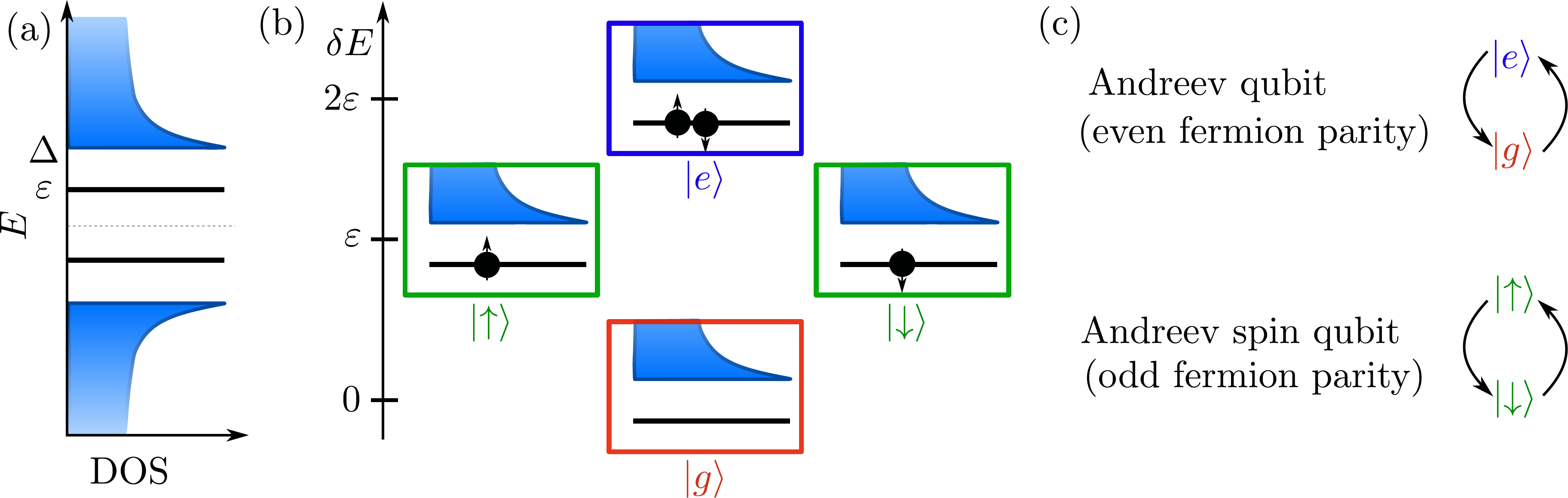}
\caption{Qubits based on single Andreev levels. (a) Energy spectrum of a superconducting system featuring one ABS at energy $\varepsilon$. (b) States of the spin-degenerate Andreev levels. Red (ground state) and blue (excited state with two opposite spin quasiparticles in the ABS) represent the even fermion parity states. The green squares show the odd fermion parity states. (c) Basis for the Andreev qubit and the Andreev spin qubit.}
\label{fig:7.1.1}
\end{figure}
\begin{equation}
    \ket{e}=\gamma^{\rm ABS}_{\uparrow}\gamma^{\rm ABS}_{\downarrow}\ket{g}\,,
\end{equation}
with $\gamma^{\rm ABS}_{\sigma}$ being the BdG operator of the Andreev state. The state $\ket{e}$ has an energy $2\varepsilon$ higher than the ground state, as depicted in Fig.~\ref{fig:7.1.1}(b). There are additional excited states with even fermion parity that involve the excitation of quasiparticles above the gap. However, quasiparticles in the continuum can diffuse fast and scape the system. For this reason, these states are avoided for quantum applications.

On the other hand, the odd fermion parity subspace has two states that correspond to excited quasiparticles with opposite spins
\begin{equation}
    \ket{\sigma}=\gamma^{\rm ABS}_{\sigma}\ket{g}\,.
\end{equation}
These states are degenerate in the absence of spin-breaking symmetry terms and split in the presence of magnetic field. Therefore a spinful subgap state defines two, in principle, disconnected subspaces where information can be encoded and processed.

\begin{center}
\begin{figure}[t]
\includegraphics[width=0.85\linewidth]{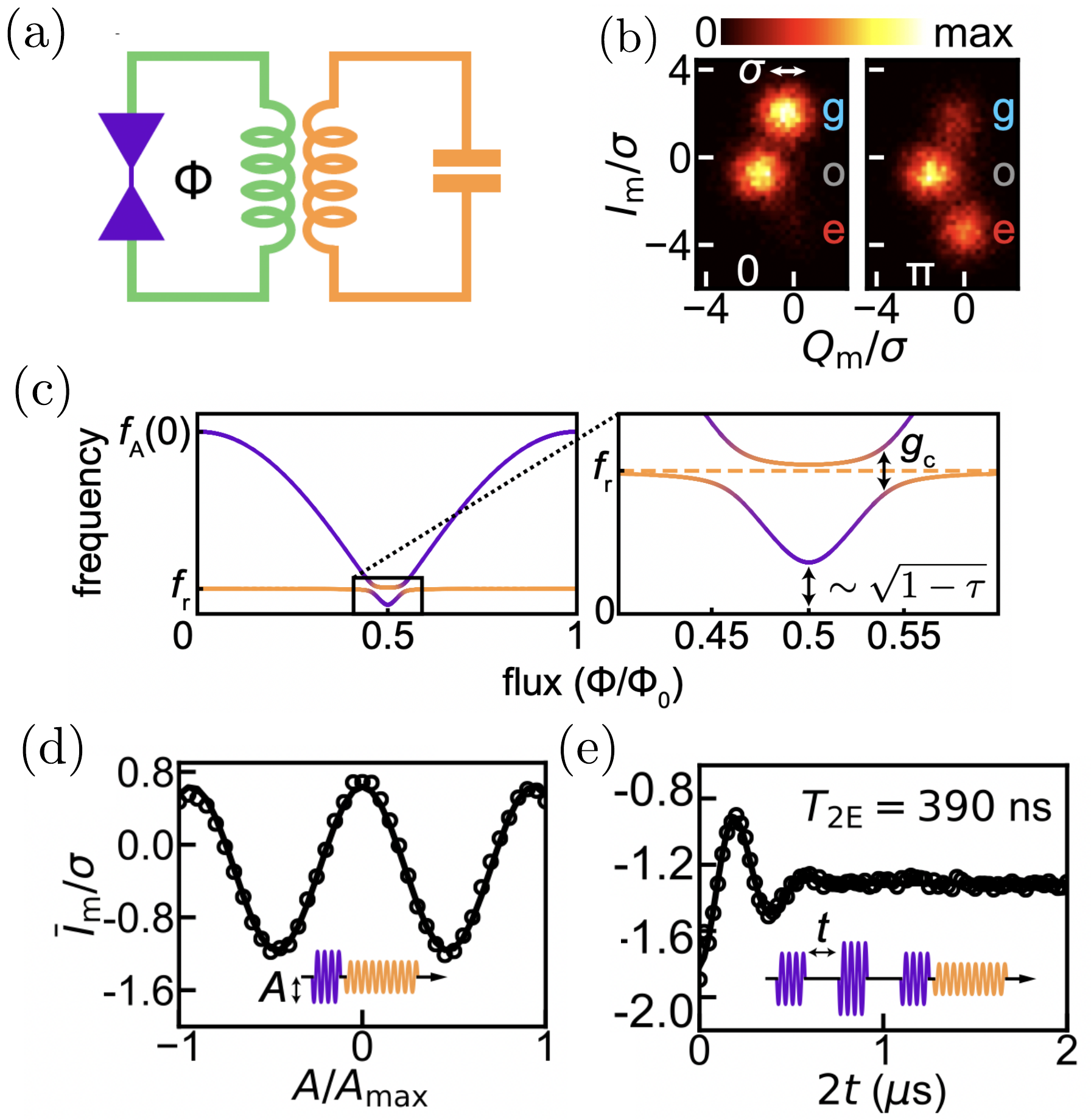}
\caption{(a) Sketch of the used setup to measure the coherent properties of the Andreev qubit, where a superconducting junction (purple) is embedded into a loop (green) and inductively coupled to a resonator (yellow). (b) Histogram of the Im and Qm quadratures of the readout tone with no driving (left) and after a $\pi$ pulse (right). (c) Theory illustration of  the spectrum of the system. (d) Rabi oscillations. (e). Coherence of the qubit measured using a Hahn-echo pulse sequence. Reprinted and Adapted from Ref. \cite{Hays_PRL2018} with permission, \textcopyright~2018, American Physical Society. All rights reserved.}
\label{Fig:ABS_qubit}
\end{figure}
\end{center}

\begin{example}{The Andreev level qubit}
The first experimental demonstration of a qubit based on Andreev states focused on the even fermion parity state, therefore using the states $\ket{g}$ and $\ket{e}$ as qubit states, see Fig.~\ref{Fig:ABS_qubit}. These experiments exploited the discrete Andreev states appearing in Josephson junctions between two superconductors. Andreev qubits have been realized in superconducting quantum point contacts~\cite{Janvier_Science2015} and semiconductor-superconductor nanowires~\cite{Hays_PRL2018}. To measure and manipulate the qubit, it is advantageous to embed the Josephson junction into a superconducting loop. The current through the loop depends on the state of the qubit, having different directions between the $\ket{g}$ and $\ket{e}$ states, due to their opposite curvature with phase, originated from particle-hole symmetry, see for example Fig.~\ref{fig:1.1}(b). This means that the qubit can be measured by inductively coupling it to a microwave LC resonator, whose properties get renormalized depending on the qubit state. 

Specifically, the full Hamiltonian for the LC resonator+qubit system is a Jaynes Cummings Hamiltonian of the form
\begin{equation}
H_{JC}=\hbar\omega_ra^\dagger a+H_A+\hbar g_c(a^\dagger\sigma_-+a\sigma_+),
\end{equation}
with $a^\dagger (a)$ the creation (anhilation) operators for the LC resonator with frequency $\omega_r=1/\sqrt{LC}$ and $H_A=\frac{\hbar\omega_A}{2}\sigma_z$ being the Hamiltonian defining the Andreev qubit, writen in terms of the Pauli matrix $\sigma_z=\ket{e}\bra{e}-\ket{g}\bra{g}$ and $\omega_A=2E_A(\phi)$ from Eq. (\ref{ABS}). The Jaynes Cummings coupling term of strength $g_c$ is writen in terms of the corresponding Pauli matrices $\sigma_+$ and $\sigma_-$.

For this particular implementation, the Jaynes Cumming coupling is achieved through the inductive coupling. Specifically, one can write a direct coupling beween the resonator and the supercurrent operator of the Andreev levels
\begin{equation}
\hat{I}_A(\phi)=\frac{2\pi}{\Phi_0}\frac{\partial H_A}{\partial\phi},  
\end{equation}
with $\Phi_0=h/2e$ being the superconducting flux quantum. The calculation of the supercurrent operator involves some subtleties since, apart from the obvious $\phi$ dependence of the qubit energy, one needs to consider the phase dependence of the Andreev level wavefunctions (since $\sigma_z$ is written in the basis of the Andreev level). Putting everything together results in a supercurrent operator that involves off-diagonal elements (which in turn allow to drive in qubit transitions):
\begin{equation}
\hat{I}_A(\phi)=\frac{2\pi}{\Phi_0}\frac{\partial E_A(\phi)}{\partial\phi}[\sigma_z(\phi)+\sqrt{1-\tau} \tan(\phi/2)\sigma_x(\phi)].   
\end{equation}
The direct coupling appears when Taylor-expanding the total flux-dependent hamiltonian $H=H_A+\Phi\frac{\partial H_A}{\partial\Phi}+...$, which results in a Jaynes Cummings hamiltonian
\begin{equation}
H_c=\Phi\hat{I}_A(\phi)=\Phi_r(a^\dagger+a)\hat{I}_A(\phi)\approx g_c(\phi)(a^\dagger\sigma_-+a\sigma_+),
\end{equation}    
with the operator describing vacuum quantum phase fluctuations of the LC resonator (strength $\Phi_r=\sqrt{\frac{L\hbar\omega_r}{2}}$) defined as $\Phi=\Phi_r(a^\dagger+a)$. In the last step, the coupling is defined as
$g_c(\phi)=2\pi\frac{\Phi_r}{\Phi_0}\sqrt{1-\tau}\frac{\partial E(\phi)}{\partial\phi} \tan(\phi/2)$, where we have dropped the diagonal part of the current operator and performed a rotating
wave approximation.

\begin{svgraybox}
In circuit QED, experiments can be performed such that the frequency difference between the resonator and the qubit is much bigger than the coupling between them $g_c\ll |\omega_r-\omega_A|$, which is known as the dispersive regime. In such dispersive limit, the hamiltonian can be written as:
\begin{equation}
H_{dis}=\hbar(\omega_r+\chi\sigma_z) a^\dagger a+\frac{\hbar\omega_A}{2}\sigma_z,
\end{equation}
with 
\begin{equation}
\label{dispersive}
\chi=\frac{g_c^2}{\omega_A-\omega_r},
\end{equation}
which is known as the dispersive shift.
This equation has a profound meaning since it means that the Jaynes Cummings coupling in the dispersive regime results is a \emph{qubit state dependent} shift of the resonator frequency by $\pm\chi$. By monitoring the resonator response to a microwave readout tone, the quantum state of the qubit can be resolved.
\end{svgraybox}

Interestingly, $g_c$ is maximum at $\phi=\pi$ for the Andreev qubit.
This is very convenient, since at half flux, the dispersive shift $\chi$  is maximized because the numerator in Eq. (\ref{dispersive})
becomes big while the denominator becomes small. This results in a strong avoided crossing near $\phi=\pi$  as illustrated in Fig. \ref{Fig:ABS_qubit}c.

In the same way, resonant pulses, with frequency $\omega_A=2\pi f_A$, induce Rabi oscillations, changing the relative population of $\ket{g}$ and $\ket{e}$ states, Fig. \ref{Fig:ABS_qubit}d. The decoherence times of such qubit can be extracted by pulse sequences such as Hahn-echoes, Fig. \ref{Fig:ABS_qubit}e.
Finally, its is important to mention the recent achievement of a very long range coupling of two Andreev qubits mediated by a microwave photon in a superconducting cavity coupler \cite{cheung2023photonmediated}. Instead of using inductive coupling like in the previous cases, this experiment uses a coupling capacitor designed between two resonators at the respective voltage antinodes. This allows to perform fast readout of each qubit using the strongly coupled mode, while the weakly coupled mode is utilized to mediate a very long range coupling between the qubits (over a distance of six millimeters).
\end{example}
\begin{figure}[t]
\includegraphics[width=1\linewidth]{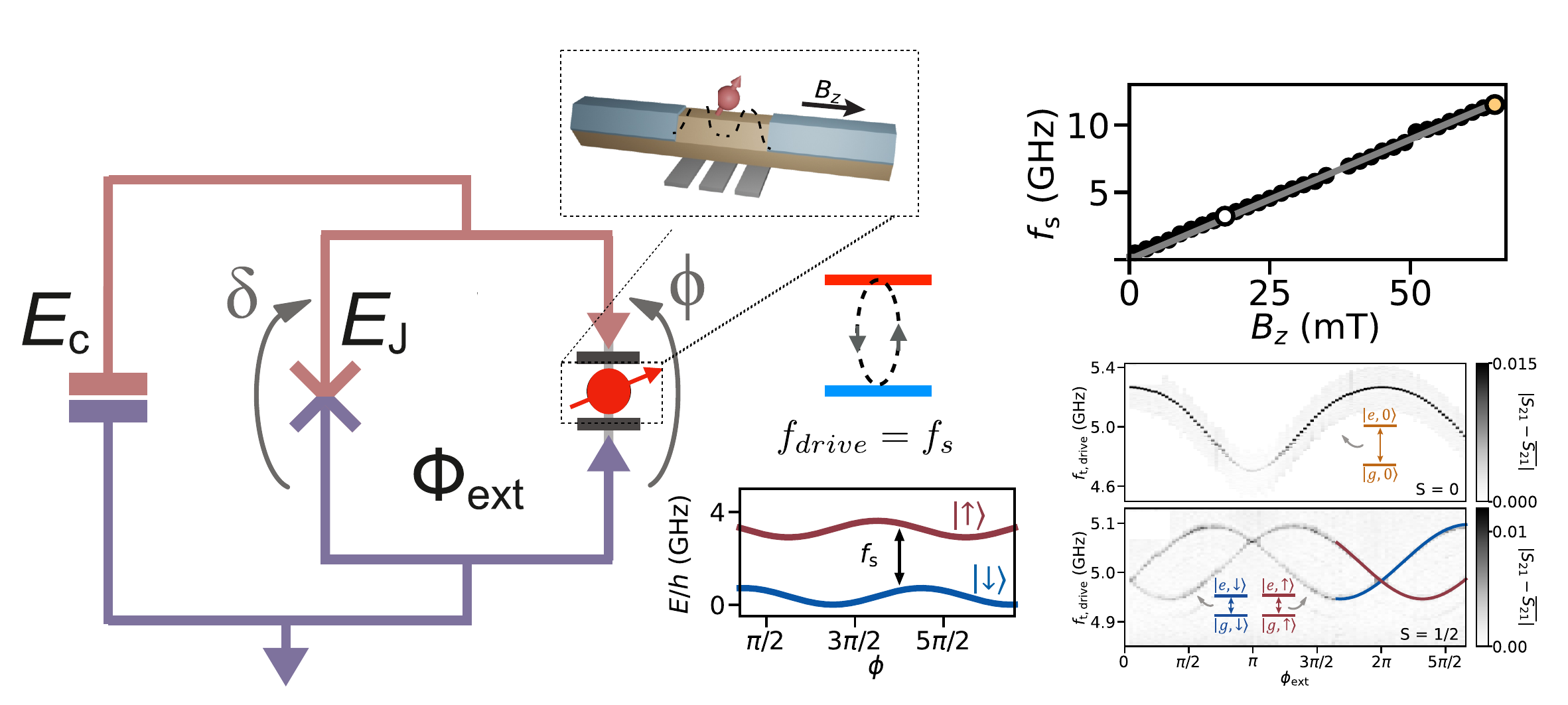}
\caption{Left: Circuit model of the Andreev spin qubit embedded in a transmon circuit. The transmon island with charging energy $E_C$ is connected to ground by a SQUID formed by the parallel combination of the QD Josephson junction and a reference Josephson junction of energy $E_J$. 
$\phi$ and $\delta$ denote the superconducting phase difference across the quantum dot and reference junctions respectively, they can be tuned by means of an externally applied magnetic flux through the SQUID loop $\Phi_{\textrm{ext}}$. 
A microwave drive of frequency $f_{\textrm{drive}}$ applied to a central gate electrode (not shown) in the QD region (red circle with arrow) allows to manipulate the spin state. The dashed inset shows a sketch of the QD region with the Andreev spin qubit confined in a hybrid semiconductor-superconductore nanowire. Right Top: An applied magnetic field $B_z$ along the nanowire axis allows to tune the frequency of the qubit $f_s$. Right bottom:  Comparison of singlet and spin-split doublet ground states in transmon two-tone spectroscopy.  Transmitted microwave signal versus external flux, $=\phi_{\textrm{ext}}=(2e/\hbar)\Phi_{\textrm{ext}}$, and transmon drive frequency, $f_{\textrm{t,drive}}$, for the quantum dot junction in the singlet state (top) and for the doublet state (bottom), revealing the spin- splitting of the doublet state. Reprinted and Adapted from Refs. \cite{PhysRevLett.131.097001} and \cite{Pita-Vidal_NatPhys2023} with permission, \textcopyright~, American Physical Society. \textcopyright~2023, The Author(s), under exclusive licence to Springer Nature Limited. All rights reserved.}
\label{Fig:ABS_spin_qubit}
\end{figure}

\begin{example}{The Andreev spin qubit}
The odd parity subspace is formed by two states characterized by a single BdG quasiparticle trapped in the ABSs. This quasiparticle have a well-defined spin, that can be decomposed into up/down components of a given basis, leading to an effective spin 1/2 degree of freedom where information can be encoded. This qubit \cite{PhysRevLett.90.226806} is the superconducting counterpart of conventional spin qubits, realized in quantum dots~\cite{Burkard_RMP2023}, and is hence dubbed Andreev spin qubit. Single qubit operations, like transitions between $\ket{\uparrow}$ and $\ket{\downarrow}$, require transitions between opposite spin states in the Andreev level. These transitions are allowed by terms that mix the two spins, like spin-orbit coupling and time-reversal breaking terms, like magnetic fields or a finite phase difference. 
Two implementations of Andreev spin qubits in semiconducting nanowire junctions have been studied. The first  was experimentally realized using spin-split ABSs of a long Al/InAs semiconducting Josephson junction \cite{Hays_Science2021}. In these experimental implementation, the qubit was incorporated into the inductor of a lumped element resonator. By doing so, the intrinsic spin-supercurrent coupling of the superconducting spin qubit could be exploited to monitor the spin state of the qubit using circuit quantum electrodynamics techniques \cite{HaysNaturePhysics2020}. This first Andreev spin qubit implementation based on a long Josephson junction, however, presents a fundamental challenge: the ground state of the system is a  singlet, making the two qubit states an excited manifold. Decay from the qubit states to the ground state, consequently, leads to leakage and constrains the qubit’s coherence times. In this implementation, the singlet and doublet switching rates were found to be of the same order of magnitude, with spin lifetimes of tens of  $\mu$s \cite{Hays_Science2021,HaysNaturePhysics2020}.
\begin{figure}[t]
\includegraphics[width=\linewidth]{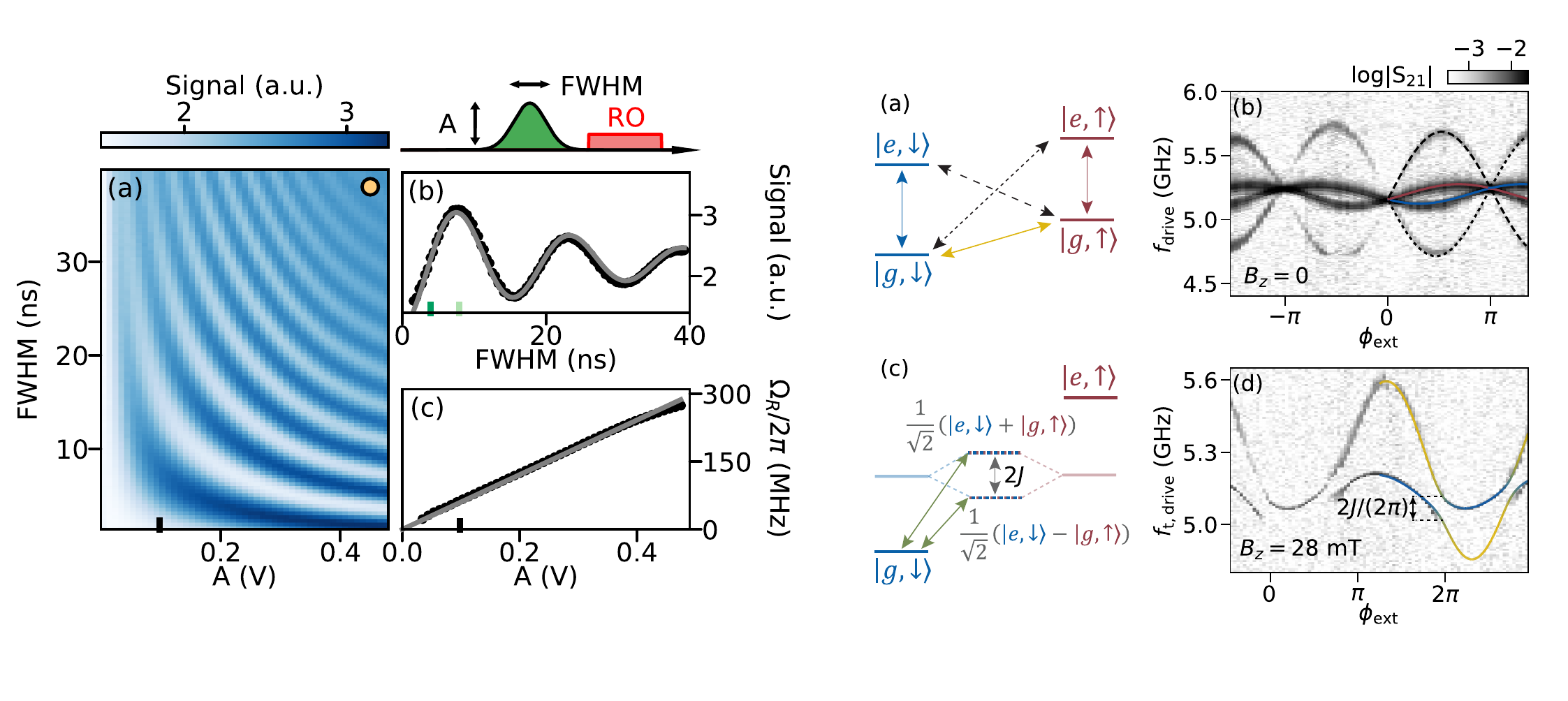}
\caption{Coherent manipulation of the Andreev spin qubit. Left: (a) Rabi oscillations for a range of Gaussian pulses characterized by their amplitude $A$ at the waveform generator output and their full width at half maximum (FWHM). (b) Rabi oscillation corresponding to $A = 0.1V$. (c) Extracted Rabi frequencies versus pulse amplitude. Right: Coherent ASQ-transmon coupling. (a) Frequency
diagram of the joint Andreev spin qubit-transmon system at
large detuning between ASQ and transmon qubit energy levels. In addition to the two spin-conserving transmon transitions (solid red and blue) and the transmon-conserving spin qubit transition (solid yellow), two additional swap transitions $\ket{g,\downarrow}\leftrightarrow \ket{e,\uparrow}$ and $\ket{g,\uparrow}\leftrightarrow \ket{e,\downarrow}$ involving both qubits can take place in the presence of coherent coupling between them (dashed and dotted black). (b) Two tone spectroscopy of the joint two-qubit system at Bz = 0 showing the additional swap transitions. (c)  In the presence of coherent coupling,
the two qubits hybridize into states $1/\sqrt{2}(\ket{e,\downarrow}-\ket{g,\uparrow})$ and $1/\sqrt{2}(\ket{e,\downarrow}+\ket{g,\uparrow})$ with a frequency splitting
of $2J$. Green arrows denote the transitions from ground to
the two hybridized states. (d) Experimental demonstration of strong coupling between both qubits with Two-tone spectroscopy. Reprinted and Adapted from Ref. \cite{Pita-Vidal_NatPhys2023} with permission, \textcopyright~2023, The Author(s), under exclusive licence to Springer Nature Limited. All rights reserved..}
\label{Fig:ABS_spin_qubit2}
\end{figure}
An alternative approach to implementing Andreev spin qubits in semiconducting nanowires, involves exploiting the finite charging energy within the Josephson junction when the semiconducting segment contains a QD. Importantly, this method enables the operation of the qubit's state in a regime where they are the lowest energy states of the system. As discussed in subsection \ref{InteractingQDS}, such QD-based Josephson junctions can be tuned into a regime where the doublet states constitute the lowest energy manifold of the system, see Fig. 1(d). The first step towards this QD-based Andreev spin qubit was demonstrated in Ref. \cite{PRXQuantum.3.030311} which presented microwave spectroscopy experiments in a hybrid superconductor-semiconductor transmon device in which the Josephson effect was controlled by a gate-defined quantum dot in an In As-Al nanowire. In such setup, the microwave spectrum of the transmon depends on the ground-state parity of the quantum dot. This allows to measure the parity phase diagram as a function of different experimental tuning knobs, such as gate voltages and external magnetic flux, in very good agreement with that predicted by a single-impurity Anderson model with superconducting leads. Furthermore, continuous-time monitoring of the circuit allows to resolve the quasiparticle dynamics which, owing to the QD in the Josephson junction, result in a notable increase in the doublet lifetime up to the order of 1 to 10 milliseconds. Furthermore, the application of a magnetic field not only increases the spin splitting due to the Zeeman effect $f_s=E_\uparrow-E_\downarrow$, but, in combination with spin-orbit interaction, also allows for direct driving of the spin-flip transition \cite{PhysRevLett.131.097001}, as opposed to Ref. \cite{Hays_Science2021}.

Tunability of the doublet ground state \cite{PRXQuantum.3.030311} together with direct  spin-flip transition \cite{PhysRevLett.131.097001} allowed to demonstrate an Andreev spin qubit  based on QDs in Ref.~\cite{Pita-Vidal_NatPhys2023}. Such configuration allows to 
to tune the qubit frequency by means of an external magnetic field over a frequency range of 10 GHz (Fig.\ref{Fig:ABS_spin_qubit} top right) and to investigate the qubit performance using direct spin manipulation. 
Specifically, the spin-split doublet phase (Fig.\ref{Fig:ABS_spin_qubit} bottom right) can be described by a minimal extension of the superconducting Anderson model of subsection \ref{InteractingQDS} in terms of a Josephson potential
\begin{equation}
U(\phi) = E_0\cos\left(\phi\right) - E_{\rm SO}\, \vec{\sigma} \cdot \vec{n}\,\sin\left(\phi\right) +\frac{1}{2} \vec{E}_{\rm Z} \cdot\vec{\sigma}\,.
\label{eq:ESOpotential}
\end{equation}
Here, $\vec{\sigma}$ is the spin operator, $\vec{n}$ is a unit vector along the spin-polarization direction set by the spin-orbit interaction, and $E_{\rm SO}$ and $E_0$ are spin-dependent and spin-independent Cooper pair tunneling rates.
Note that the $E_0$ term proportional to $\cos\left(\phi\right)$ has a minimum at $\phi=\pi$, describing the doublet phase. Finally, $\vec{E}_{\rm Z}$ is the external Zeeman field.

Interestingly, using an all-electric microwave drive, Rabi frequencies exceeding 200 MHz can be achieved. Furthermore, the Andreev spin qubit can be embedded in a superconducting transmon qubit, which allows to demonstrate strong coherent qubit-qubit coupling~\cite{Pita-Vidal_NatPhys2023}. Specifically, the transmon circuit consists of a capacitor, with charging energy $E_C$, shunting a superconducting quantum interference device (SQUID) formed by the parallel combination of a gate-tunable Josephson junction with Josephson energy $E_J$, and the quantum dot
Josephson junction hosting the Andreev spin qubit, Fig.\ref{Fig:ABS_spin_qubit} top schematics. By operating in a regime where the ratio $E_J/\sqrt{E_0^2+E_{SO}^2}$ is large, the phase 
 phase difference $\phi$ across the quantum dot Josephson junction can be controlled through the magnetic flux through the SQUID loop $\Phi_{ext}= \phi_{ext}\Phi_0/(2\pi)$. Due to the presence of the $E_{SO}\sin\left(\phi\right)$ term, the transmon frequency becomes spin-dependent. This can be exploited to readout the Andreev spin qubit state by capacitively coupling the transmon circuit to a readout resonator. Finally, due to the transmon-resonator dispersive coupling, the resonator frequency in turn becomes spin-dependent. Therefore, spectroscopy of the spinful Andreev levels can be performed using standard two-tone circuit QED techniques. The spin-flipping qubit transition can be directly driven, while maintaining the transmon in its ground state, by applying a microwave tone on the central quantum dot gate.
Since there is an intrinsic coupling between the spin degree of freedom and the supercurrent across the quantum dot Josephson junction, it is possible to tune the system such that a coherent coupling induces transitions that involve both qubits, in addition to the single-qubit transitions. Such transitions (Fig. \ref{Fig:ABS_spin_qubit2} top right), $\ket{g,\downarrow}\leftrightarrow \ket{e,\uparrow}$ and $\ket{g,\uparrow}\leftrightarrow \ket{e,\downarrow}$, with $\ket{g}$ and $\ket{e}$ denoting the ground and first excited transmon states, respectively, are swap-like and could be used to construct entanglement and two qubit gates between the two different qubit platforms. 

Finally, by performing a Taylor expansion of the phase-dependent SO term in the transmon+spin qubit hamitlonian:
\begin{equation}
H=-4E_C\partial_\phi^2-E_Jcos(\phi)+U(\phi-\phi_{ext}),
\end{equation}
we can obtain an interesting coupling term of the form
\begin{equation}
H_{c}\approx E_{\rm SO}[\sin(\phi_{ext})-\phi \cos(\phi_{ext})](\cos \theta\sigma_{\bar{x}}+\sin\theta\sigma_{\bar{z}}).
\end{equation}
This term of the Hamiltonian couples the Andreev spin qubit to the transmon via the phase operator $\phi=\phi_{zpf}(a\dagger+a)$, , where $\phi_{zpf}=(2E_C/E_J)^\frac{1}{4}$ is the magnitude of zero-point fluctuation of the transmon phase,  and is, thus, reminiscent of a dipole coupling. The coupling contains both longitudinal and transverse terms in the spin eigenbasis, with $\theta$ being the angle between the Zeeman field and the spin-orbit direction. By varying the external flux $\phi_{ext}$ such that the Andreev spin qubit frequency crosses the transmon frequency, one can obtain superpositions $1/\sqrt{2}(\ket{e,\downarrow}-\ket{g,\uparrow})$ and $1/\sqrt{2}(\ket{e,\downarrow}+\ket{g,\uparrow})$ separated by a frequency splitting $2J/2\pi\approx 100MHz$ (Fig. \ref{Fig:ABS_spin_qubit2} bottom right), greatly exceeding the decoherence rates of both qubits and hence demonstrating the first realization of a direct strong coupling between a spin qubit and a superconducting qubit~\cite{Pita-Vidal_NatPhys2023}. 
The analytical estimation $J\approx E_{\rm SO}\phi_{zpf}\sin(\theta)$ suggests that by gating the system (or choosing a different material) to reach a larger $E_{\rm SO}$  and by aligning the magnetic field perpendicular to the spin-orbit direction, coupling rates of hundreds of MHz could be achieved, which would enable rapid two-qubit gates between the transmon and the Andreev spin qubit.

We finish this part by mentioning another experimental breakthrough in the field: a supercurrent-mediated
coupling between two distant Andreev spin qubits \cite{Pita-Vidal_arXiv2023}. This qubit-qubit interaction is induced by a a shared Josephson inductance by using an early theoretical idea \cite{PhysRevB.81.144519} that large spin-dependent supercurrents can lead to strong long-range and tunable spin-spin coupling. In the experiments, in particular, this qubit-qubit interaction is longitudinal type and tunable by bith gate voltages and external fluxes.

Despite all this amazing progress, experiments with Andreev spin qubits still report rather small $T_2$ coherence times, of the order of tens of ns. The interaction to the surrounding spin bath of the nuclei is one of the dominant dephasing mechanisms. For this reason, the recent realization of hard-gap superconductivity in materials without a net nuclear spin, like Germanium~\cite{Tosato_ComMat2023,Valentini_NatCom2024}, sets a promising route to improve the coherence times of the Andreev Spin qubits. First steps towards hybrid qubits using a two-dimensional hole gas based on Germanium quantum wells are already being demonstrated with the realization of a gate tunable transmon qubit in planar Germanium \cite{Katsaros-private}.
\end{example}

\begin{svgraybox}
The spaces for the Andreev and Andreev spin qubits are disconnected as they have different fermion parity. However, random quasiparticle tunneling events can drive transitions between the two subspaces, bringing the system outside form the computational space. Therefore, the timescale of the quasiparticle poisoning events set an upper limit for operations of these systems. Fortunately, poisoning events ($T_1$ times in qubit language) of $\sim100\mu$s have been reported in superconductor-semiconductor systems, see for example Refs.~\cite{Pita-Vidal_NatPhys2023,Janvier_Science2015,Hays_PRL2018,Hays_Science2021}. 
\end{svgraybox}

\subsection{Beyond single Andreev states}
\begin{figure}[t]
\includegraphics[width=1\linewidth]{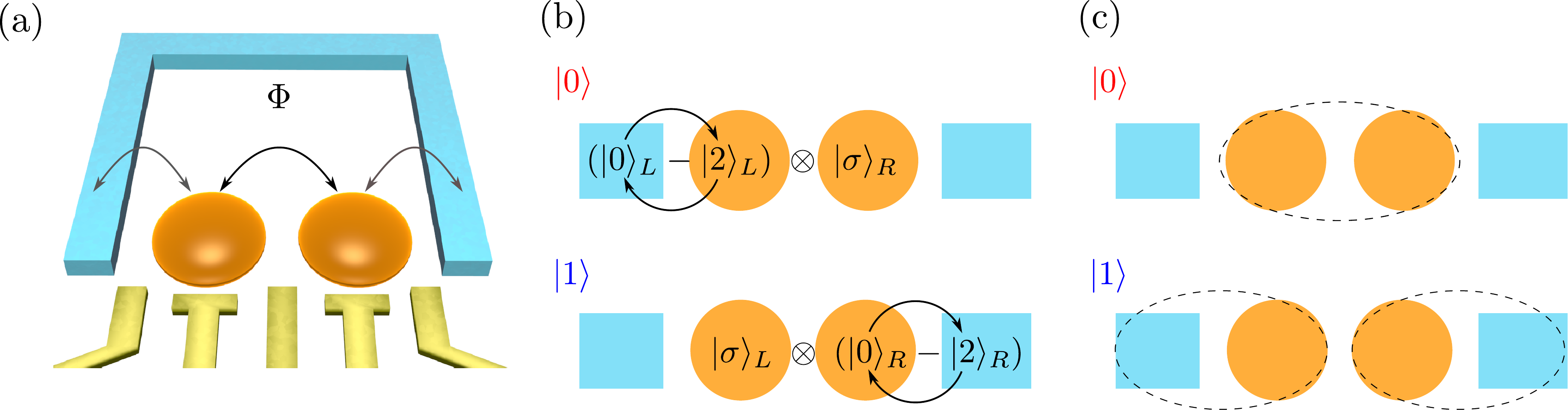}
\caption{Qubits based on coupled double dots. (a) Qubit sketch composed by two tunnel-coupled QDs. The two QDs are embedded into a superconducting loop that allows for phase control. Panel (b) and (c) shows two possible qubits that can be defined in the total odd  and even subspaces, respectively. (b) Two states with a single quasiparticle excited in either left/right QD, while the other QD is in the even parity state. (c) Singlet formed between the dots or between the dots and the superconductors.}
\label{fig:7.2.1}
\end{figure}

The protection of superconductor qubits can be improved in systems of coupled Andreev states. These systems that can be engineered in, for example, sets of quantum dots coupled to superconductors. We have introduced in Sec.~\ref{sec:5} a way to improve the coherent properties of qubits based on nanoscopic superconductors by making an artificial Kitaev chain. In this section, we focus on proposals that couple two Andreev states appearing, for example, in double QD systems coupled to superconductors~\cite{Mishra_PRXQ2021,Malinowski_arxiv2023,Geier_arXiv2023}. These proposals exploit the superconducting properties and the tunability of QDs to find sweet spots where the qubit features some robustness to certain sources of noise. In this subsection, we discuss two different proposals of qubits based on coupled ABSs, engineered in a superconductor-QD-QD-superconductor device, as sketched in Fig.~\ref{fig:7.2.1}. Similarly to the previously discussed qubits based on single ABSs, one can define different qubits, depending on the total parity states.

\begin{example}{The parity qubit}
The first version we discuss is based on the total odd fermion parity state and introduced in Ref.~\cite{Geier_arXiv2023}. In this qubit, information is encoded in a single quasiparticle occupying either the left or the right superconductor-QD subsystem, dubbed parity qubit, Fig.~\ref{fig:7.2.1}(b). Therefore, the two half of the system have different fermion parity. The qubit can be seen as the superconducting version of the conventional charge qubit, studied in semiconductor QDs. In conventional charge qubits, charge noise coming from the coupling to the electrostatic environment provides the main source of decoherence. It has been recently shown that the use of charge neutral environments using, for example noble gases, can greatly enlarge the qubit's coherence times~\cite{Zhou_NatPhys2023}. 

The parity qubit uses a different strategy to enlarge the coherence times, exploiting the exotic properties of subgap states in superconductor. Andreev bound states, appearing for example in QD-superconductor systems, have sweet spots where the ground and the excited states have the same total charge and the system is insensitive to charge fluctuations. Mathematically, this means that the Bogoliubov-de Gennes coefficients describing the subgap state fulfill $u=v$. In superconductor-QD systems, this situation is achieved at discrete points in parameter space, so-called sweet spots. To achieve good protection, the parity qubit requires that the two half of the system are tuned to their respective sweet spots, where it is linearly insensitive to charge noise acting on the level energy of the QDs. The parity qubit is, however, still linearly sensitive to noise on the tunnel coupling between the QDs. This issue can be improved by increasing the tunnel coupling, at the cost of reducing the relaxation times. Quantum operations between the QDs can be performed by driving the tunnel coupling at the resonant frequency, that makes the excited quasiparticle jump across the junction. Finally, charge measurements in the QDs away from the sweet spot allows to determine the state of the system.
\end{example}

\begin{example}{The YSR bond qubit}
Similarly to the single Andreev qubits, one can also define a qubit in the total even subspace. One possibility is to define a qubit in the regime where both QDs have a single charge~\cite{Steffensen_arXiv2024}. In this regime, there are two possibilities for screening the spin of the escess electron in the dots: by forming a singlet between the two QDs or by forming two YSR states with the two coupled superconductors. These two possibilities, illustrated in Fig.~\ref{fig:7.2.1}, have in general different energies, that are determined by the tunneling amplitudes, representing a two level system where one can encode quantum information. Importantly, the two states have the same charge and total spin. Therefore, the qubit is insensitive to small fluctuations of the level energies of the dots and magnetic fields. Quantum operations between the two qubit states require a quasiparticle being transfer through the full system, recombining/splitting a Cooper pair. These operations can be performed by driving the level energy of one of the QDs.
\end{example}

\section{Coherent experiments with Majorana states}
\label{Sec:Sec:coherent_MBSs}
In Section~\ref{sec:3}, we introduced the exotic properties of non-local Majorana states, which are predicted to emerge at the ends of one-dimensional topological superconductors. These superconductors can be engineered in semiconductor-superconductor heterostructures under strong magnetic fields. Section~\ref{sec:4} reviewed experimental progress and identified signatures consistent with the presence of topological MBSs. However, the existence of low-energy trivial states complicates the picture as they can mimic most of the properties of MBSs. Therefore, demonstrating the topological origin of the measured subgap states would require probing the non-local properties of MBSs, related to their non-abelian statistics. Achieving this goal requires coherent control on the properties and the state of several MBSs.

In this section, we review the theoretical advancements and theoretical proposals for operating MBSs. We begin the section with a discussion on qubit designs based on topological superconductors. Following this, we examine experimental designs aimed at demonstrating the non-Abelian properties of MBSs and, consequently, their topological origin. The section includes proposals for fusion rules and braiding experiments. The section concludes with a discussion about operating on the state encoded in PMMs, realized in minimal Kitaev chains.

\subsection{Topological qubits}
\label{Sec:topoQubits}
Majorana bound states, appearing at the ends of 1-dimensional topological superconductors, are a promising route for the implementation of quantum technologies. The key feature that makes Majorana qubits apart from alternatives is their non-local nature: the information in a Majorana qubit is stored not in a local quasiparticle, but in the collective state of a pair of spatially separated Majorana states. This non-local encoding of information offers inherent protection against local perturbations, a significant advantage for quantum computing. In Majorana qubits, information is encoded in the fermion occupation of Majorana pairs, that can be in either the even or the odd fermion parity state. A single Majorana pair cannot encode all the possible states of a qubit. The reason is that transitions between even and odd states would require changing the overall fermion parity, considered to be a good quantum number in superconductors, except for undesired poisoning events. Therefore, four MBSs are needed to encode all the states of a qubit. The Majorana qubit states can be defined as $\ket{0}=\ket{00}$, $\ket{1}=\ket{11}$ for the total even parity subspace, where $0/1$ denotes the even/odd fermion occupation of a given Majorana pair. These two states have the same electron number and energy for decoupled MBSs, and differ from one Cooper pair that can split occupying the two non-local fermion modes defined by the Majoranas. We note there is another identical computational subspace with total odd fermion parity ($\ket{01}$, $\ket{10}$) that can be also used as a basis for computation.

The most intriguing property of Majorana qubits is their potential for topological quantum computation. In a topological quantum computer, information is stored and manipulated through the exchange of Majorana fermions. When these fermions are exchanged or ``braided'' around each other, the quantum state of the system changes in a way that depends only on which MBSs are exchanged, becoming independently from other details like the speed and the time of the exchange (if everything is done adiabatic)~\cite{Kitaev_AoP2003,Sarma_NPJQ2015}. The topological nature of Majorana qubits makes them inherently resistant to certain types of errors and decoherence, a major challenge in conventional quantum computing systems. This resilience to local noise and perturbations is due to the fact that the quantum information is stored globally in a non-local state rather than in a local variable like an electron's spin or charge. As a result, disturbances or changes in the system do not easily disrupt the stored information, unless the gap is closed, offering a promising route to more stable and reliable quantum computation.

Topologically protected operations, like braiding, can be used to implement quantum gates, including the Hadamard and control-Z. However, the set of operations allowed by braiding do not form a universal gate set~\cite{NayakReview}. For this reason, braid operations need to be complemented with gates that are not topologically protected to achieve universal quantum computation. The $\pi/8$ rotation, also known as T gate or magic gate is an operation that can complement the previous two. Despite that this operation is not topologically protected, some schemes have been develop to cancel errors that may arise during unprotected gate operations~\cite{Karzig_PRB2019}.

Several qubit devices have been proposed, exploiting the Majorana degree of freedom to encode quantum information. In the following, we discuss some of these proposals.

\begin{figure}[t]
\includegraphics[width=1\linewidth]{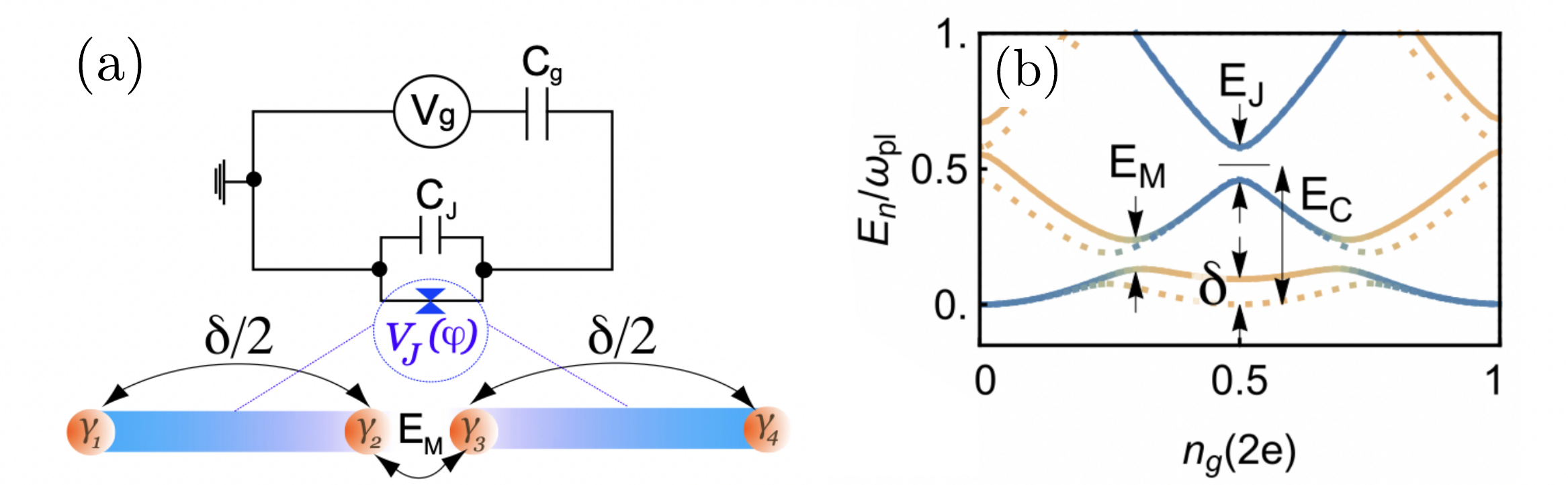}
\caption{(a) Sketch of a transmon Majorana qubit, where the connection between the two superconductors is made via two coupled Majorana wires, bottom panel. (b) Energy spectrum of the system in the charging-energy dominated regime, known as Cooper pair box regime ($E_J/E_C=0.5$, $E_M/E_C\sim0.12$), and close to the topological phase transition for the wires. Blue/orange
colors denote fermionic even/odd parities of the inner MBSs. 
Reprinted and Adapted
from Ref.~\cite{Avila_PRR2020} under CC-BY-4.0 license, \textcopyright~2022, The Author(s).}
\label{fig:Majorana_transmon}
\end{figure}

\begin{example}{The Majorana transmon qubit}
The Majorana transmon is an example of a qubit that exploits the extra degree of freedom introduced by four Majoranas, see Fig.~\ref{fig:Majorana_transmon}. This qubit is an extension of the celebrated transmon qubit, where the Josephson coupling between the superconductors includes a term that is originated from the coupling between two 
MBSs~\cite{Avila_PRR2020,Ginossar14,Yavilberg15,Li2018,Hassler_NJP2011,Hyart_PRB2013,Keselman19,Avila_PRB2020,Smith2020,Lupo2022}. The Hamiltonian of the system can be written as 
\begin{equation}
    H=H_C+H_J+H_M\,,
    \label{Eq:transmonMBS}
\end{equation}
where $H_C=4E_C(n-n_g)^2$, with $n$ being the charge, $n_g$ a charge offset, and $E_C$ the charging energy of the island. The coupling between the Cooper pair box and the superconductor has two terms. The first one, $H_J=E_J \cos(\phi)$, describes the contribution from all the superconductors' states, except MBSs, with $\phi$ being the phase operator, conjugated variable to the number operator $n$. The sum of $H_C+H_J$ corresponds to the conventional transmon qubit. The Majorana physics appears in the last term of Eq.~\eqref{Eq:transmonMBS} that describes the coupling between two MBSs,
\begin{equation}
    H_M=iE_M \gamma_2\gamma_3\,,.
\end{equation}
This coupling splits the ground state degeneracy, allowing to encode information in the degree of freedom of the coupled MBSs. The qubit can be manipulated and readout using microwave pulses, similar to what is currently done in transmon qubits.
\end{example}

\begin{example}{The Majorana box qubit}
The Majorana box qubit is another qubit geometry, where two topological superconductors couple via a trivial superconducting backbone, see Fig.~\ref{fig:Majorana_box_qubit}. Therefore, the two wires and the superconductor form a floating island where charging energy dominates, fixing the total fermion parity~\cite{Plugge_NJP2017}. For a given total fermion number, the system has two degenerate states ground states, corresponding to the two possible occupations of the four MBSs. For instance, for a total even parity, the two relevant quantum states are $\ket{0}=\ket{00}\otimes\ket{N_C}$ and $\ket{1}=\ket{11}\otimes\ket{N_C-1}$, where $N_C$ denotes the number of Cooper pairs in the ground state. Therefore, the two states differ by one Cooper pair that splits, where the two electrons occupy the low-energy state defined by the MBSs. The measurement of the qubit can be done via charge sensing of two of the MBSs, coupling them to a nearby QD~\cite{Munk_PRR2020,Steiner_PRR2020,Schulenborg_PRB2021,Smith_PRXQ2020}, or by measuring transport in an interferometry setup. The qubit initialization can be done using the current blockade through the system that fixes the parity of the system to a well-defined value~\cite{Nitsch_PRB2022}. The idea of fixing the parity of coupled Majorana wires inspired more complicated geometries that include several coupled wires~\cite{Karzig_PRB2017} and networks of wires for Majorana color codes~\cite{Litinski_PRB2018}.

The ground state degeneracy in the blockaded superconducting island plays the role of an effective spin. Early theory works proposed to use this fact to demonstrate the so-called topological Kondo effect, that can be used as a demonstration of the non-locality of the Majorana degrees of freedom~\cite{Beri_PRL2012}. 

\begin{figure}
\includegraphics[width=1\linewidth]{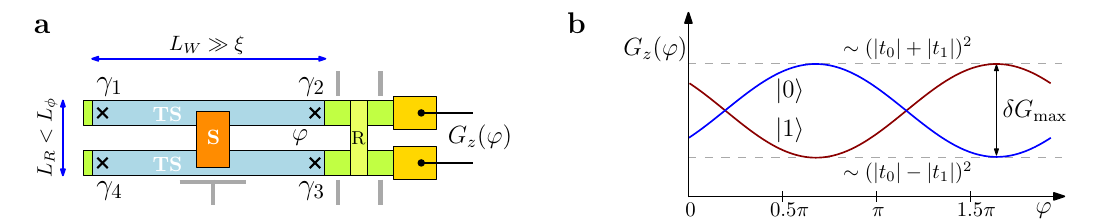}
\caption{(a) Sketch of the Majorana box qubit, where two topological superconductors (light blue) couple via a trivial superconductor (dark orange). The qubit readout can be done using conductance measurement via the leads attached to one end (yellow). (b) Conductance as a function of the flux.  Reprinted  from Ref.~\cite{Plugge_NJP2017} 
under CC-BY-3.0 license, \textcopyright~2022, The Author(s).}
\label{fig:Majorana_box_qubit}
\end{figure}

\end{example}

\subsection{Majorana fusion rules}
\label{Sec:fusion}
Majorana fusion rules are a key concept in understanding the non-Abelian nature of Majorana states. These rules describe how pairs of MBSs can combine or ``fuse'' to form different quantum states. The fusion of MBSs is not only a fundamental theoretical aspect but also has profound implications for demonstrating their non-Abelian properties, which are essential for topological quantum computation.

The basic idea of Majorana fusion rules is that, when two MBSs come together, they hybridize, resulting in a trivial (or vacuum) state, or they can fuse into a non-trivial fermionic state. This behavior is mathematically represented by the fusion rule
\begin{equation}
    \sigma\times\sigma=I+\phi\,,
    \label{Eq:fusion}
\end{equation}
where $\sigma$ represents a Majorana fermion, $I$ the vacuum state, and $\phi$ a fermionic state. The outcome of this fusion process depends only on the collective topological state of the system and is not determined by local properties. This ambiguity in the fusion outcome is a manifestation of the Majorana fermions' non-abelian properties.

A less formal but equivalent way to understand Eq.~\eqref{Eq:fusion} is that that the fusion (coupling and measurement) of two MBSs has two possible outcomes corresponding to the occupation of the fermion spanned by the measured MBS pair: empty ($I$) and full ($\phi$). The probability for each outcome depends on the joint state of the two MBSs. The experimental demonstration of fusion rules requires the initialization of the system in a given basis and the measurement in a different one. It means that, pairs of Majoranas are initialized on a well-defined fermion parity state, while measurements are performed on different using a different MBS pair combination. The result of the fusion is topologically protected and insensitive to the details of the measurement. Different protocols have been proposed to demonstrate the MBS fusion rules~\cite{Bishara_PRB2009,Ruhman_PRL2015,Aasen_PRX2016,Rowell_PRA2016,Hell_PRB2016,Clarke_PRB2017,Barkeshli_PRB2019,Zhou_PRL2020,Zhou_NatCom2022,Souto_SciPost2022,Nitsch_arXiv2022}.

\subsection{Majorana braiding}
\label{Sec:MajoranaBraid}
The ground state of a system composed by $N$ Majoranas has a degeneracy $2^{N/2}$, corresponding to every pair of MBSs being in the even/odd parity state. Majorana bound states are non-Abelian anyons: the exchange of a set of Majorana transforms the state of the system between the different ground states that are locally indistinguishable. This operation is topologically protected and does not depend on the details on how it is performed if MBSs remain well-separated and the exchange is adiabatic. Different set of Majoranas exchange correspond to distinct unitary operations that cannot be transformed into the other. This opens up the possibility of performing topologically-protected operations by just moving MBSs around. In addition, the demonstration of the non-trivial outcome after the Majorana exchange would constitute a direct proof of the topological origin of the measured zero-energy states, see Sec.~\ref{subsec::experimentsMBS}. In this subsection, we describe different Majorana braiding protocols introduced in the literature, see also Ref.~\cite{BeenakkerReview_20} for a review.

The topological phase of a semiconductor-superconductor nanowire can be controlled using external gates. That means that the topological properties of the system can be controlled locally. When a segment of a nanowires changes phase, a MBS moves accordingly to the boundary between the trivial and the topological phase. It also implies that MBSs could be moved preserving coherence if a set of gate electrodes are controlled adiabatically between different values, but sufficiently fast to avoid poisoning events. Majorana braiding would require extending the wire to a second dimension, forming a T or a Y shape, using another ancillary wire. In this configuration, two Majoranas can be exchanged, using the ancillary wire to place one of the Majoranas to avoid overlap, see Ref.~\cite{Alicea_NatPhys2011}. This braiding protocol would require full control over tens of gates~\cite{Harper_PRR2019} in a timescale much smaller than the typical quasiparticle poisoning time. For this reason, the proposal is considered to be challenging and another simplified schemes have been develop, where Majoranas are braid in parameter space instead.

\begin{figure}[t]
\includegraphics[width=1\linewidth]{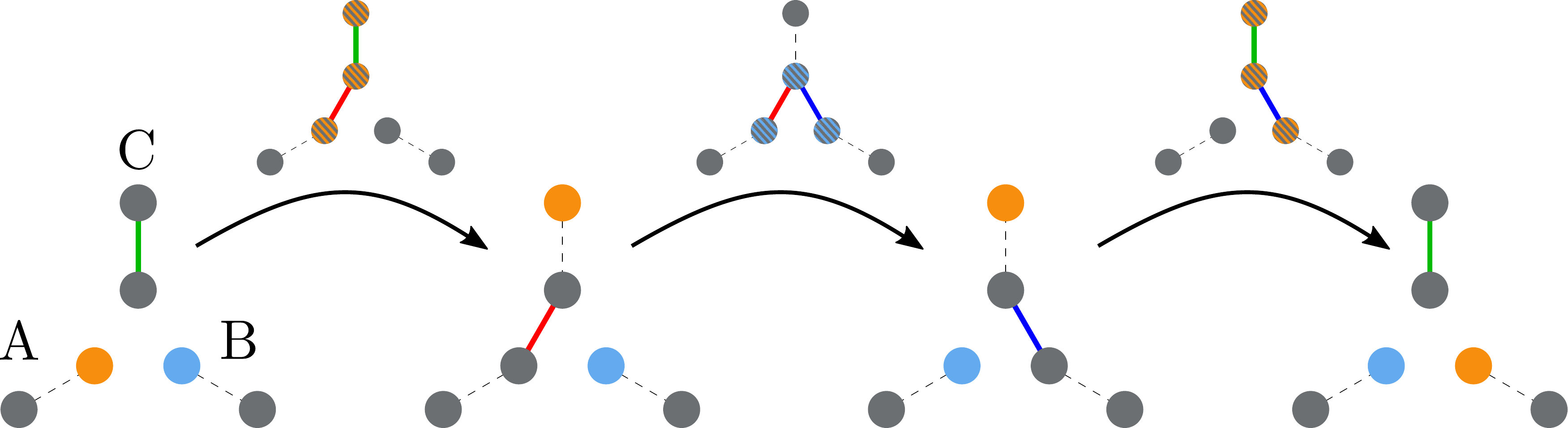}
\caption{Sketch of the hybridization-induced braiding. The system consists on 3 pairs of Majoranas (circles) that are initialize pairwise. By switching on and off the coupling between MBSs (thick lines) two of them are effectively braid, highlighted with orange and blue colors.}
\label{fig:starBraid} 
\end{figure}

\begin{example}{Hybridization-based braiding}
The approach that conceptually approximates the most to spatial braiding is the hybridization-induced braiding protocol, also referred to as the ``three-point turn braid''. For a visual representation, refer to Fig.~\ref{fig:starBraid}, which schematically illustrates both the device structure and the braiding protocol. In this proposal, there is a central MBS capable of coupling with three additional MBSs. The key to this proposal is the controlled hybridization between these MBSs, as represented by solid lines in the figure. The hybridization, needs to be dynamically switched on and off to allow for effectively braid two MBSs. When two MBSs hybridize, their ground state degeneracy is lifted, leading to a low energy state with well-defined fermion parity. This is however not the case when 3 MBSs couple, where 2 of them can be grouped into a regular fermion and we remain with a single uncoupled MBS. This fact is used in the intermediate steps, where three MBSs hybridize, leading to a single MBS delocalized along the three places. Removing one of the connections adiabatically localizes the MBS again, leading to an effective move of the topological mode. For the demonstration of Majorana fermions' non-Abelian statistics, it is crucial to initialize MBSs in pairs. For this reason, we included the additional MBSs in subsystems A and B. Various versions of this protocol have been proposed and discussed in the literature~\cite{Aasen_PRX2016, Hell_PRB2016, Clarke_PRB2017, Clarke_PRB2011, van_Heck_NJP2012,Burrello_PRA2013, Karzig_PRB2015, Hell_PRB2017, Liu_SCP2021}.
\end{example}

\begin{example}{Charge-transfer based braiding}
Another way to test non-Abelian properties consist on shuttling single charges to a system composed by several MBSs~\cite{Flensberg_PRL2011,Krojer_PRB2022}. This charge can be provided by a QD whose energy is controllable using local gates. When the energy of the QD is adiabatically swept from below to above the system's Fermi level, one electron will be transferred between the QD and the system. In a successful protocol, the electron will be transferred to the zero-energy MBSs~\cite{Souto_PRB2020}. If the system couples to only one MBS (denoted by 1), the operation can be understood as acting with the operator $C_1=\gamma_1$ to the state defined by the Majoranas. If the QD instead couples to 2 MBSs, the operator is written as $C_{1,2}=(\gamma_1+\gamma_2)/\sqrt{2}$. Therefore, the difference between applying these two operations in different orders, $C_1C_{1,2}=(1+\gamma_1\gamma_2)/\sqrt{2}$ and $C_{1,2}C_1=(1+\gamma_2\gamma_1)/\sqrt{2}$, is due to the non-Abelian properties of MBSs: the final wavefunction is the same after the two sequencies if Majorana operators are replaced by fermion operators. In this picture, we have ignored the dynamical phase that arise from the unprotected nature of these operations. A protocol has been recently developed to echo away this phase~\cite{Krojer_PRB2022}.
\end{example}

\begin{example}{Measurement-based braiding}
\begin{figure}[t]
\includegraphics[width=1\linewidth]{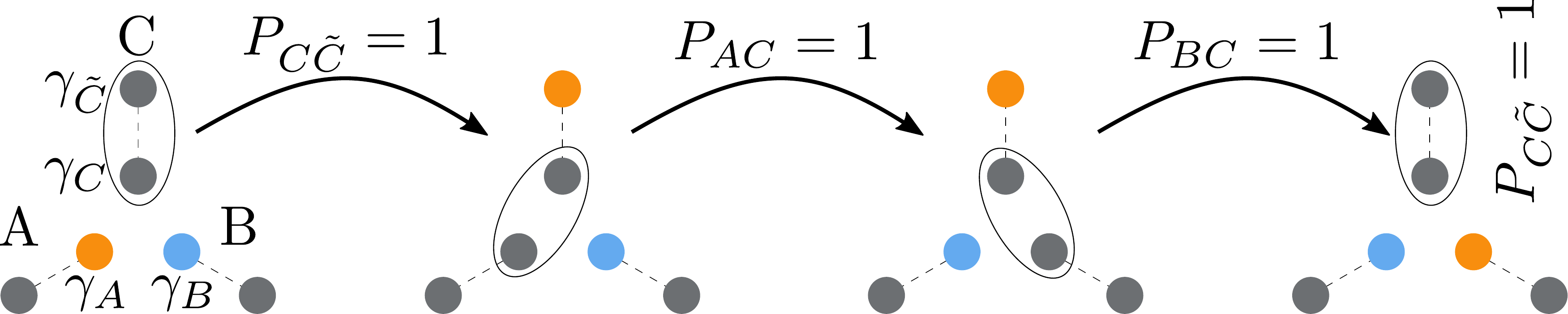}
\caption{Sketch of the measurement-based braiding. The fermion parity of two nearby MBSs is projected and read at every step, $P_{jk}=i\gamma_J\gamma_k$, schematically represented by the oval. The next step of the protocol is done only if the measured parity is 1. Otherwise, the protocol has to restart. After 4 successful measurements of the fermion parity, two MBSs are effectively braid.}
\label{fig:measurementBraid} 
\end{figure}

The measurement of the joint parity of two Majorana states provides a way to effectively braid MBSs~\cite{Bonderson_PRL2008,Karzig_PRB2017,Plugge_NJP2017}. The protocol is shown in Fig.~\ref{fig:measurementBraid}. The setup is in essence similar to the one used for the hybridization-based braiding: it consists of 4 Majoranas, plus another two that only used for initialization and readout (lower two MBSs in systems A and B). The protocol consists on measuring alternatively the parity of two MBSs at every step, enclose by an ellipse in Fig.~\ref{fig:measurementBraid}. We post select for a given outcome, in this case even parity. If the outcome is odd at any of the steps the protocol has to be restarted. Mathematically, the measurement operation can be written as $M_{jk}=(1+P_{jk})/2$, with $P_{jk}=i\gamma_j\gamma_k$. It can be shown mathematically that the braiding operation for $\gamma_A$ (orange) and $\gamma_B$ (blue) is equivalent to the set of measurements $B_{AB}\propto M_{C\tilde{C}}M_{AC}M_{BC}M_{C\tilde{C}}$.
\end{example}

In the proposals mentioned above, the operations to demonstrate Majorana braiding have to be adiabatic, ensuring that the system remains in the ground state during the protocol. On the other side, the full protocol has to be faster than the characteristic time for single quasiparticles to tunnel into the system, so-called quasiparticle poisoning time. In semiconductor-superconductor systems, poisoning times in the range of $\mu$s to ms~\cite{Albrecht_PRL2017,Menard_PRB2019,Nguyen_PRB2023,Hinderling_arXiv2023} have been reported, also under relatively strong magnetic field~\cite{Aghaee_arXiv2024}. 

\begin{svgraybox}
We want to conclude this part by commenting on the initialization and readout of the state encoded in a set of MBSs. As mentioned above, these states are non-local and local perturbations cannot affect the state of the system. It means that local probes coupled to only one MBS cannot project the state of the system. Therefore, the detection schemes for the state encoded in a pair of Majoranas rely on measuring the properties of the system after the two (or more) MBSs couple. These measurements include the charge, the energy, or the transport properties through the system. On the other hand, two coupled MBSs have a single ground state with well-defined parity. This can be used to initialize the system into a given state.
\end{svgraybox}

\subsection{Coherent experiments with PMMs}
\label{sec:PMM_coherent}
In the previous subsection, we have introduced protocols to operate MBSs in topological superconducting wires, focusing specially in protocols to demonstrate Majorana non-Abelian statistics. However, the measurements reported on superconductor-semiconductor nanowires cannot determine unambiguously the topological origin of the measured low-energy states. For this reason, minimal Kitaev chains have emerged as interesting alternatives to demonstrate the topological properties of MBSs, so-called poor man's Majoranas (PMMs) due to their lack of topological protection, see Sec.~\ref{sec:5} for a description of the system. In this subsection, we introduce a set of experiments that have been proposed in PMM systems to demonstrate the non-local properties of Majorana states. Before jumping into the proposals, we first discuss on the initialization and readout of the non-local fermionic state encoded in a pairs of PMM.

\subsubsection{Initialization and readout}
\label{sec:PMM_readout}
The lack of topological protection of Majorana defined in minimal Kitaev chains turns into an advantage for initialization and readout. The ground state of the system is in general non-degenerate, except for some fine-tuned situations, including the sweet spot characterized by well-separated Majorana states. Therefore, the system can be initialize shifting the energy of the orbitals of the two QDs away from the sweet spot. The energy shift splits the ground state degeneracy, leading to a ground state with a well-defined fermion parity. If the splitting is much larger than the temperature, the system will relax to the ground state after a sufficiently long enough time, that requires releasing the excess energy to the environment. A symmetric shift of the dots energy can be used to initialize the system in an even parity state, while shifting the orbitals of the dots in different directions can result on an initial odd fermion parity. 

Regarding readout, the conceptually easiest way to readout the state of the system is through charge detection in one dot. This measurement can be achieved by using a capacitively-coupled system, for example a quantum point contact or an additional QD that senses changes on the local charge of one or both dots. At the Majorana sweet spot, a charge detector coupled to one of the will not be able to measure, and therefore, project the PMM's state. This is a direct consequence of the inherent non-local character of well-separated PMMs: the local charge in each of the QDs is the same, independently from the state of the pair of PMMS. However, as the QDs are detuned from the sweet spot, there is a charge difference between the lowest even and odd parity states that is measurable by means of local charge detectors. This parity-to-charge conversion is similar to spin-to-charge conversion used for single-shot readout of spin qubits in double QDs~\cite{Barthel_PRL2009}.

An alternative way to measure the state of the system is based on measuring the quantum capacitance at the dots, see for example, Refs.~\cite{Petersson2010,Lambert2016,Vigneau2023}. At the sweet spot, the quantum capacitance measured of the even and odd ground states are the same, becoming indistinguishable for local measurements. Similar to the parity-to-charge conversion, a detectable difference between the even and the odd ground states signals appear when the system detunes form the sweet spot. This is also true at low-MP sweet spots. This fact can be also use to identify high-MP sweet spots using local charge measurements, see Ref~\cite{tsintzis2023roadmap} for a discussion.

In both cases, it is only possible to measure the system state away from the Majorana sweet spot. Non-local detectors couple to both dots simultaneously that can measure the state of the system at the Majorana sweet spot. This would require adding an extra indirect coupling between the dots, see for example proposals in Refs.~\cite{Leijnse_PRB2012,Liu2022_PMMfusion}. However, adding detectors capable of readout at the sweet spot not only complicates the device design and measurements (requiring either measurements sensitive to charge fluctuations or capacitive coupling to both QDs) but also introduces a decoherence mechanism at the sweet spot. These decoherence mechanisms might become relevant in experiments aiming at study the quantum coherent properties of PMMs.

The fermion parity degree of freedom is a good quantum number in superconducting systems, where electrons are paired together forming Cooper pairs, except from potentially unpaired electrons occupying low-energy states, like Majorana states. However, quasiparticle poisoning events can change the total femionic parity in the PMM system in an uncontrolled way, undermining coherence in the system, and being an obstacle toward quantum coherent experiments. Independent charge or capacitance measurements on both dots can help determining the poisoning timescale at  and away from the Majorana sweet spot, limiting timescale for future experiments.

\subsubsection{Coherent operations between two PMMs}
\label{sec:PM_coherentCoupling}
\begin{figure}[t]
\includegraphics[width=1\linewidth]{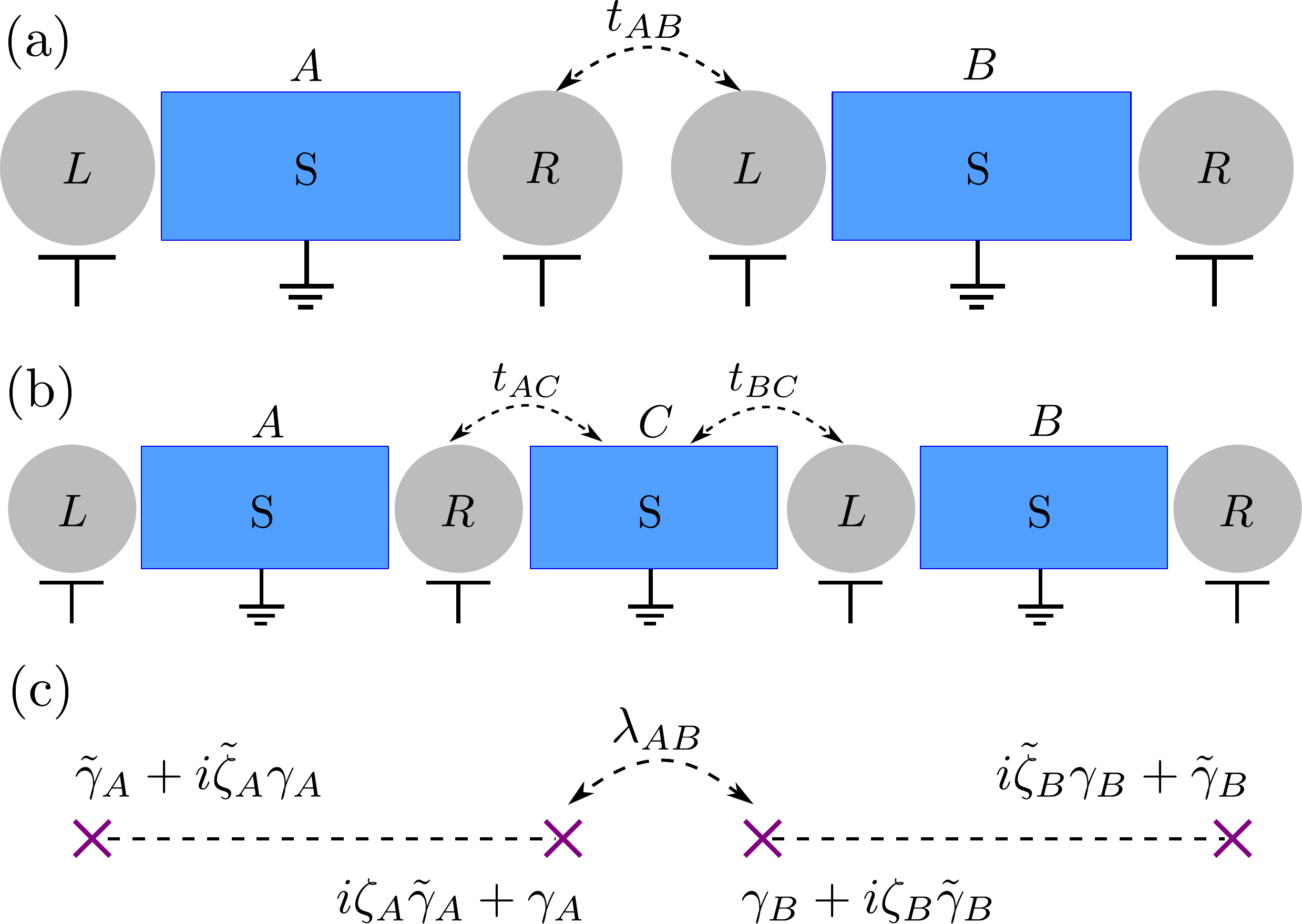}
\caption{Setup to demonstrate the coherent coupling between two PMM systems. Panels (a) and (b) show two different ways of coupling the two system, through direct tunneling and via a central superconductor. In both cases, the tunnel coupling results in a hybridization of the inner Majoranas. Panel (c) shows a minimal model, composed by four Majoranas, $\gamma_j$ and $\tilde{\gamma}j$ belonging to sub-system $j=A,B$, to describe the coupling between the PMM systems.}
\label{fig:6.3.2.1}
\end{figure}

Fermion parity is a good quantum number in PMM systems, meaning that the even and odd fermion parity states do not couple to each other. To couple both states and, therefore, perform coherent oscillations in the fermion parity degree of freedom, we need to couple the PMM system to another fermionic system, like a QD or another PMM system. This last situation is specially interesting, as this would be the corresponding poor man's version of the topological transmon qubit, introduced in Sec.~\ref{Sec:topoQubits}. The system is sketched in Fig.~\ref{fig:6.3.2.1}, where the two inner dots can couple directly (direct hoping) or via an intermediate superconductor that mediates crossed Andreev reflection and elastic cotunneling. Using the minimal model described in Sec.~\ref{sec:PMM_realistiModels}, the Hamiltonian can be written as
\begin{equation}
    H=H_{\rm PMM}^{A}+H_{\rm PMM}^{B}+H_T+H_U\,,
    \label{eq:double_PMM_fullModel}
\end{equation}
where $H_{\rm PMM}^{A,B}$ describe the left/right PMM system, given by Eq.~\eqref{eq:ThanosHam}. The tunneling is described by $H_T$, which depends on the details of the central region connecting the two PMM systems. For simplicity, we will focus on the case where the systems couple via normal tunneling, where the Hamiltonian can be written as
\begin{equation}
    H_T=\sum_{\sigma}\left[t_{AB}d^{\dagger}_{AR\sigma}d_{BL\sigma}+t_{AB}^{\rm so}d^{\dagger}_{AR\sigma}d_{BL\bar{\sigma}}+{\rm H.c.}\right]\,,
\end{equation}
where $t$ and $t^{\rm so}$ describes the spin conserving and the spin-flip tunneling terms, respectively. Finally, we have included an extra term in Eq.~\eqref{eq:double_PMM_fullModel} that accounts for non-local Coulomb interactions between the nearby QDs
\begin{equation}
    H_U=U_{AB}n_{L}^{B}n_{R}^{A}\,,
\end{equation}
with $n_{\alpha}^{j}$ being the charge of the $\alpha=L,R$ QD of the $j$ PMM. A finite $U_{AB}$ will induce correlations between the charge on the innermost QDs and become an obstacle to find Majorana sweet spots with well-localized MBSs, {\it i.e.} with high MP, and zero energy. This contribution can be suppressed by placing a a grounded metallic gate between the dots, that can also tune tunneling amplitude. Another way to suppress the non-local Coulomb interaction is to couple the two PMM systems via a grounded superconductor, forming a 4-dot Kitaev chain. In the following, we are going to ignore the effect of $U_{AB}$. Numerical results for the low-energy states can be found in Ref.~\cite{tsintzis2023roadmap}.

\begin{example}{Low-energy model for Majorana qubits}

It is instructive to introduce a low-energy model describing the Majorana states an their couplings, that qualitatively captures the dominant physics in the PMM system. For the two PMM systems, the model is given by
\begin{equation}
    H_{\rm Low-E}^{A,B}=\frac{i}{2}\sum_{j=A,B}\xi_{j}\gamma_j\tilde{\gamma}_j+\frac{i}{4}\left[\lambda_{AB}\left(\gamma_A-i\zeta_A\tilde{\gamma}_A\right)\left(\gamma_B-i\zeta_B\tilde{\gamma}_B\right)-{}\rm H.c.\right]\,,
    \label{eq:double_PMM}
\end{equation}
where the first term describes the overlap between PMMs belonging to the same system, defined as $\gamma=d+d^\dagger$ and $\tilde{\gamma}=i(d-d^\dagger)$, causing an energy splitting of the ground states $\xi$. As we have shown previously in the chapter, two PMMs can overlap without splitting the ground state degeneracy ($\xi=0$). They result in a zero-energy crossing characterized by a low MP. The model accounts for this scenario by allowing the tunneling to the second Majorana, given by $\zeta$, with $\lambda$ being the tunneling amplitude. The case $\zeta=0$ describes the ideal sweet spot scenario with well-localized PMMs in each dot. In contrast, the limit $\zeta=\pm1$ corresponds to a trivial fermionic state. The model interpolates between the two limiting situations.

The Hamiltonian in the low-energy model, Eq.~\eqref{eq:double_PMM}, has 4 energy-degenerate ground states, given by either an even or an odd fermion parity, that we will denote as $|p_L,p_R\rangle$, with $p_j=0,1$ for even and odd parity. The total parity of the system is a good quantum number, so states with different total parity do not couple. Below, we restrict the discussion to the $|0,0\rangle$, $|1,1\rangle$ states, although similar results can be obtained for the odd fermion parity states. In the even subspace, the Hamiltonian in Eq.~\eqref{eq:double_PMM} can be rewritten as
\begin{equation}
    \left.H_{\rm Low-E}^{A,B}\right|_{\rm even}=-\frac{\xi_{+}}{2}\sigma_z+\frac{\lambda_{AB}}{2}\left(1-\zeta_A\zeta_B\right)\sigma_y\,,
    \label{eq:double_PMM_sigmas}
\end{equation}
where $\xi_{+}=\xi_{A}+\xi_{B}$ and we have assumed that $\zeta_j$ are real. Here,  $\sigma$ are the Pauli matrices in the 2-dimensional space given by the $|0,0\rangle$, $|1,1\rangle$ states, described by $\sigma_z=-i\gamma_A\tilde{\gamma}_A=-i\gamma_A\tilde{\gamma}_A$, $\sigma_y=i\gamma_A\gamma_B=-i\tilde{\gamma}_A\tilde{\gamma}_B$, and $\sigma_x=-i\gamma_A\tilde{\gamma}_B=-i\tilde{\gamma}_A\gamma_B$. Therefore, a finite coupling between the PMMs in the same system splits its energy degeneracy and can be used to initialize the system in a state with well-defined fermion parity on the left and the right, {\it i.e.} either the $|0,0\rangle$ or $|1,1\rangle$ state. The tunnel coupling between the sub-systems can be used to induce coherent transition between these 2 quantum states and demonstrate the coherent control on PMM systems.
\end{example}

\begin{example}{Majorana transmon qubit with minimal Kitaev chains}
The system in Fig.~\ref{fig:6.3.2.1}(a) has been analyzed in a transmon setup in Ref.~\cite{Pino_arXiv2023}, see fig.~\ref{fig:PMM_transmon}, that is an extension of the Majorana transmon qubit introduced in Sec.~\ref{Sec:topoQubits}. The minimal Hamiltonian describing the model is given by
\begin{equation}
    H=4E_C(n-n_g)^2+E_J\cos(\phi)+V^{J}_{\rm PMM}\,,
\end{equation}
where the first term accounts for the charging energy, with $n=-i\frac{\partial}{\partial \phi},$ and the second for the tunneling of Cooper pairs via the trivial Josephson junction. The third term describes the Josephson coupling between the two PMM systems that allows for tunneling processes of a single electron across the junction, proportional to $\cos(\phi/2)$, that change the fermion parity of the left and right systems (the total fermion parity remains unchanged). Expressions for $V^{J}_{\rm PMM}$ are given in Ref.~\cite{Pino_arXiv2023}.

\begin{figure}[t]
\includegraphics[width=1\linewidth]{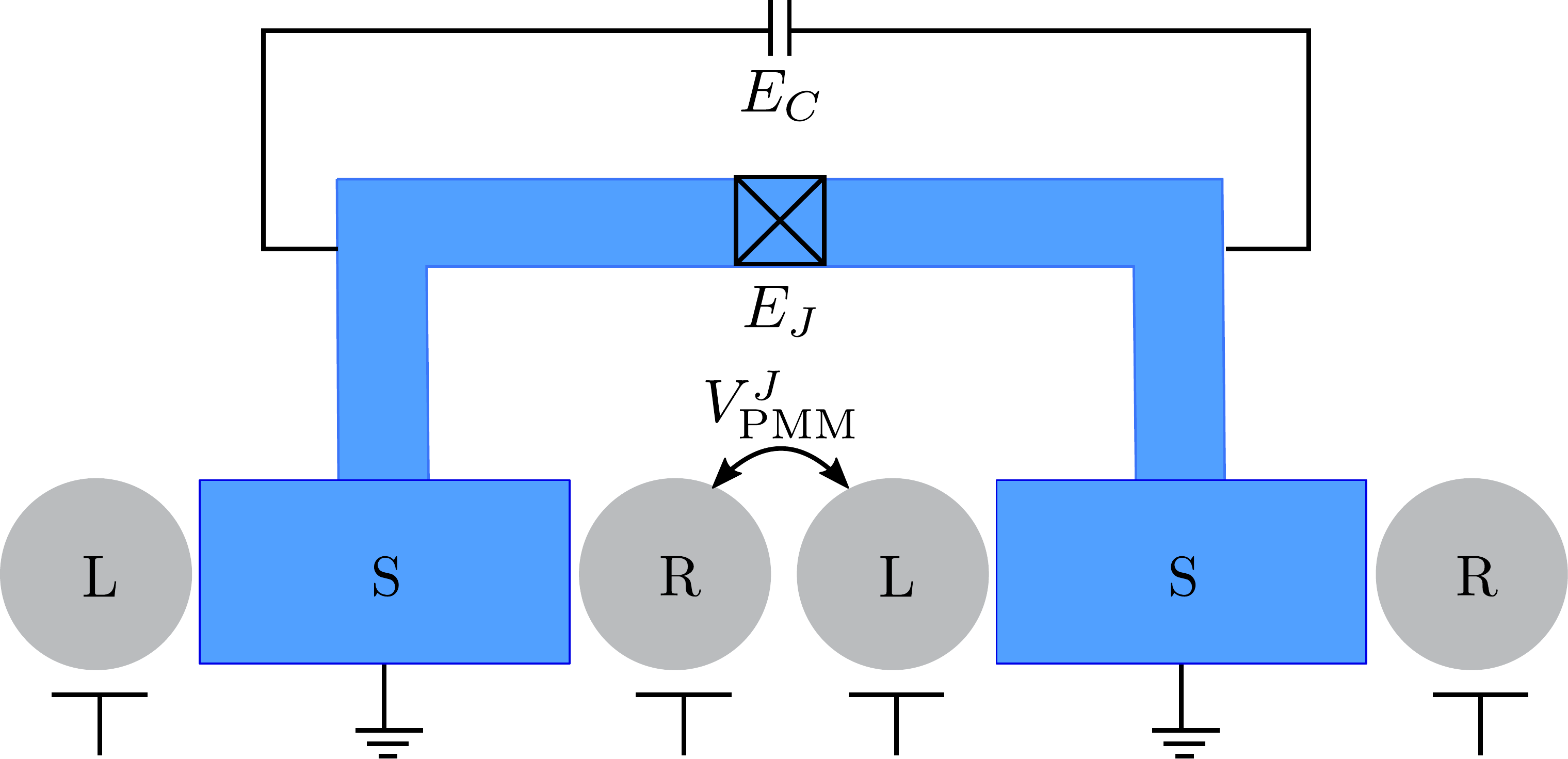}
\caption{Sketch of a PMM transmon qubit studied in Ref.~\cite{Pino_arXiv2023}, where two tunnel-coupled PMM systems are embedded into a SQUID with a trivial Josephson junction.}
\label{fig:PMM_transmon}
\end{figure}

In this setup, the coupling to a microwave resonator that allows measuring the excitation spectrum of the system. The $4\pi$ Josephson effect appears in the system as some transitions that are suppressed close to phase difference $\phi=\pi$ that only appear in sweet spots with well-localized PMMs. In addition, the microwave response can be used to determine the MP, see Eq.~\eqref{eq:MP_definition} and the discussion around for a definition, being a way to quantitatively assess the localization of PMMs.
\end{example}

\subsubsection{Testing Majorana fusion rules with PMMs}
\label{sec:}
Majorana bound states are non-abelian quasiparticle that have non-trivial exchange statistics. Measuring non-abelian properties of MBSs is an open challenge in the field that would constitute the demonstration of a new non-local quasiparticle with a topological origin. Alongside with non-trivial exchange properties, MBSs have non-trivial fusion rules, see Sec.~\ref{Sec:fusion} for a description. The outcome of fusing two MBSs, {\it i.e.} coupling two MBSs, is either an electron or no electron (a hole). The probability of having either of the two possible outcomes depend solely on the state of the system. Usually, fusion experiments aim at initializing and measuring MBSs in orthogonal basis, so the the measured outcome is either an electron or hole with equal probability.

Majorana fusion rules can be also tested in artificial Kitaev chains made out of QDs coupled via superconducting segment. In the minimal version, 4 QDs, similar to the device sketched in Fig.~\ref{fig:6.3.2.1}(a) or (b). Shifting the energy of the QDs allow to initialize the system, while the tunnel between the left and right PMMs allows to couple the inner MBSs. Non-local charge measurement in a pair of QDs allows to infer the outcome of the fusion protocol~\cite{Liu2022_PMMfusion,tsintzis2023roadmap}. We note that some fusion protocols are not sensitive to important details, like the MP. Therefore, fusion experiments are believed to not be a conclusive demonstration of Majorana non-Abelian properties. However, in the case of PMMs, fusion experiments are crucial milestones for more sophisticated experiments probing non-abelian exchange properties, as the ones we discuss in the following section.

\subsection{Non-abelian experiments with PMMs}
\label{sec:PMM_nonAbelian}
In the previous Sec.~\ref{sec:5}, we have analyzed different ways to identify Majorana sweet spots in minimal Kitaev chains, mostly based on transport measurements using metallic electrodes. However, there is a limit on the information that local probes can provide about non-local Majoranas. For this reason, there is always a risk that some trivial states can mimic the properties of PMMs. Additionally, metallic probes are a source of quasiparticles, undesired for quantum applications.

Non-Abelian properties are unique for MBSs, independently on whether they are protected or fine tuned in minimal Kitaev chains. A definitive demonstration of these properties remain as a main challenge in the field to confirm the existence of non-local Majorana states, that will have deep implications in the for quantum information storage and processing. Non-Abelian properties imply that physically exchanging two Majorana states rotates the state of the system in a non-trivial way within the degenerate ground state manyfold, see Sec.~\ref{sec:3} for a discussion. However, spatially exchanging Majorana states is challenging and several protocols have been proposed to demonstrate Majorana's non-abelian properties in nanowire systems, see Sec.~\ref{Sec:MajoranaBraid}. These protocols are based on operations whose result is the same as a physical braid for PMMs, while they give a different result for other subgap states. In this section, we describe three different protocols to demonstrate non-Abelian properties of PMMs, based on the bulk proposals presented above.

\begin{example}{Charge-transfer based braiding with PMMs}

\begin{figure}[t]
\includegraphics[width=1\linewidth]{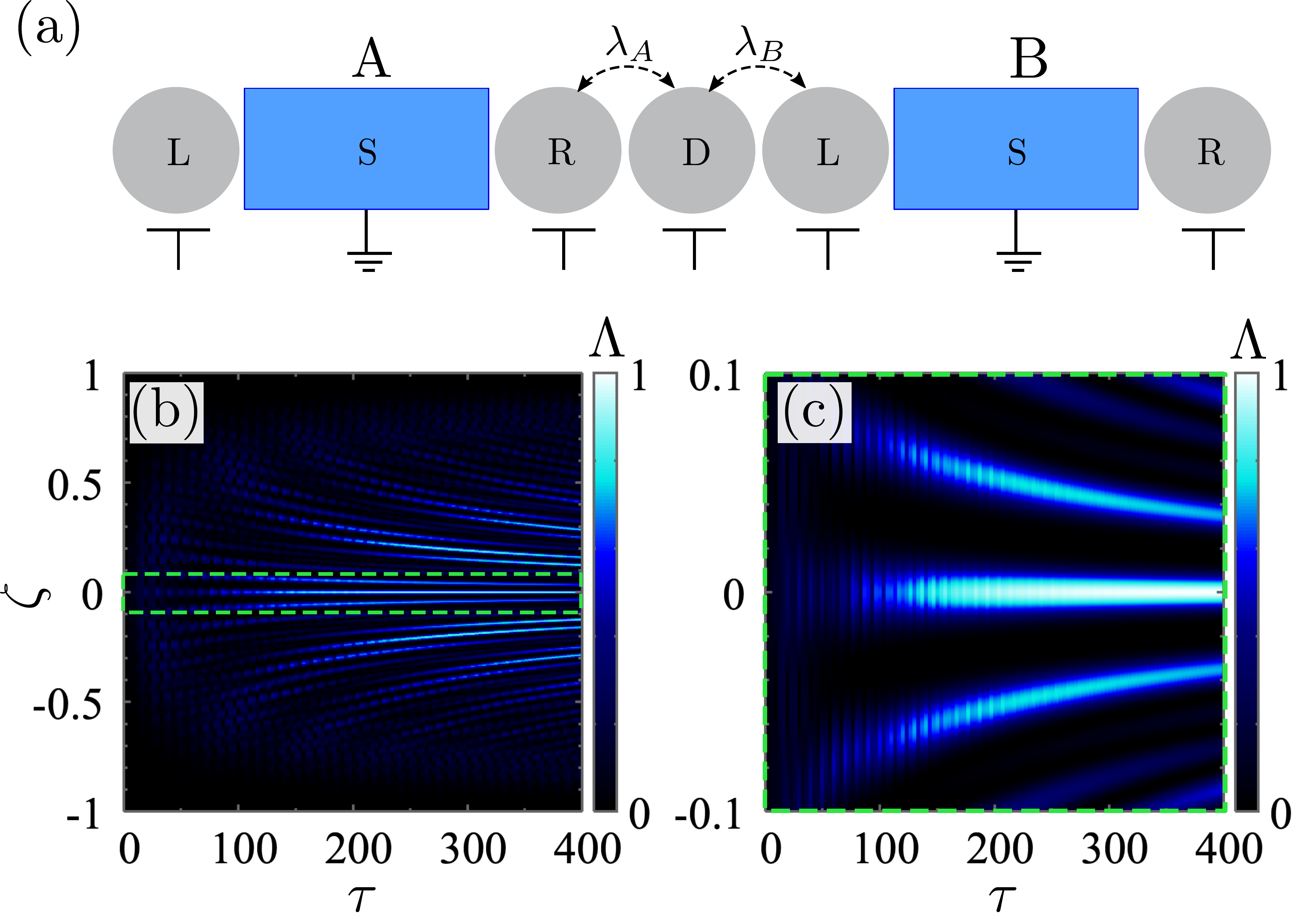}
\caption{(a) Device to demonstrate PMM non-Abelian properties based on charge-transfer to a central QD, $D$. Panels (b) and (c) show the visibility: probability of ending up in the target states after the braiding and the reference protocol (see text), as a function of the Majorana overlap in the same PMM system, $\zeta$, and operation time, $\tau$. Reprinted and adapted from Ref.~\cite{tsintzis2023roadmap} under CC-BY-4.0 license, \textcopyright~2022, The Author(s).}
\label{fig:PMM_chargeBraid}
\end{figure}

Protocols based on transferring charges between a quantum dot and Majorana states provide a conceptually simple way of testing non-Abelian properties of MBSs~\cite{Flensberg_PRL2011,Souto_PRB2020,Krojer_PRB2022}. The basic device for this proposal using minimal Kitaev chains is sketched in Fig.~\ref{fig:PMM_chargeBraid}(a), where two PMM systems couple to a single QD. For this protocol, we consider that the system is subject to a strong magnetic field, in such a way that the QDs' levels are spin-polarized. When the energy of the dentral QD ($D$) is ramped from negative to positive energies, a single electrons is transferred between the QD to the PMM systems. In this picture, we have fixed the zero energy reference to the chemical potential of the superconductors. If the QD couples to only one of the PMMs, tuned to a sweet spot, a charge transfer operation will result on a parity change, mathematically represented as $C_j=\gamma_j$ in the Majorana basis. In contrast, if the QD couples to both PMM systems, $A$ and $B$, the electron can tunnel with equal probability to any of the two, resulting on an operation $F_{AB}=(\gamma_A+\gamma_B)\sqrt{2}$. Now, the operation $B_{AB}=F_{AB}C_A$ leads the same result as a braiding the inner PMMs of the two subsystems, $\gamma_A$ and $\gamma_B$. Therefore, the sequences $F_{AB}C_A$ and $C_AF_{AB}$ lead to different states, due to the non-Abelian character of PMMs. In contrast, the same protocol with trivial fermionic states will result in the same state for both sequences, independently from the order of the charge-transfer operations. In order to measure the state using the parity to charge conversion described in Sec.~\ref{sec:PMM_readout}, it is convenient to add a third charge transfer operation, so the two protocols read as $F_{AB}C_AF_{AB}$ and $F_{AB}F_{AB}C_A$. So, if the system is initialized in $|0,0\rangle$, the outcome of protocols 1 and 2 is $|0,1\rangle$ and  $|1,0\rangle$. 

Figure~\ref{fig:PMM_chargeBraid}(b) shows results of the two protocols as a function of the PMM localization, using the model introduced in Eq.~\eqref{eq:double_PMM}~\cite{tsintzis2023roadmap}. The figure shows the \emph{visibility}, product of the probabilities of measuring the expected outcome after the two protocols~\cite{Krojer_PRB2022}, $\Lambda=P_{|1,0\rangle}^{(1)}P_{|0,1\rangle}^{(2)}$. As shown, the protocol only works independently from the protocol time, $\tau$, for relatively small deviations from the ideal Majorana case with two-well separated PMMs, $\zeta=0$. For some operation times, it is possible that the protocol gives false positives at finite $\zeta$, originated from accumulated dynamical phases. These false positives are easy to rule out as they depend on $\tau$. Therefore, charge-transfer based braiding can, in theory, distinguish PMMs from other trivial crossing. However, the short range of $\zeta$ values where it works can make it challenging to tune and keep the system to a very high MP sweet spot.
\end{example}

\begin{example}{Measurement-based braiding with PMMs}
Measurements provide a way to effectively braid MBSs in topological superconducting wires, as previously discussed in Sec.~\ref{Sec:MajoranaBraid} for topological MBSs, see Fig.~\ref{fig:measurementBraid} for a sketch of the protocol. This proposal has been adapted to PMM systems in Ref.~\cite{tsintzis2023roadmap}, where its outcome has been studied as a function of different parameters. 
The measurement-based braiding could in principle be used to demonstrate PMM braiding. However, the interpretation might become difficult because the outcome depends, not only on the PMM localization parameter, $\zeta$, but also on the device parameters.

\begin{figure}[t]
\includegraphics[width=1\linewidth]{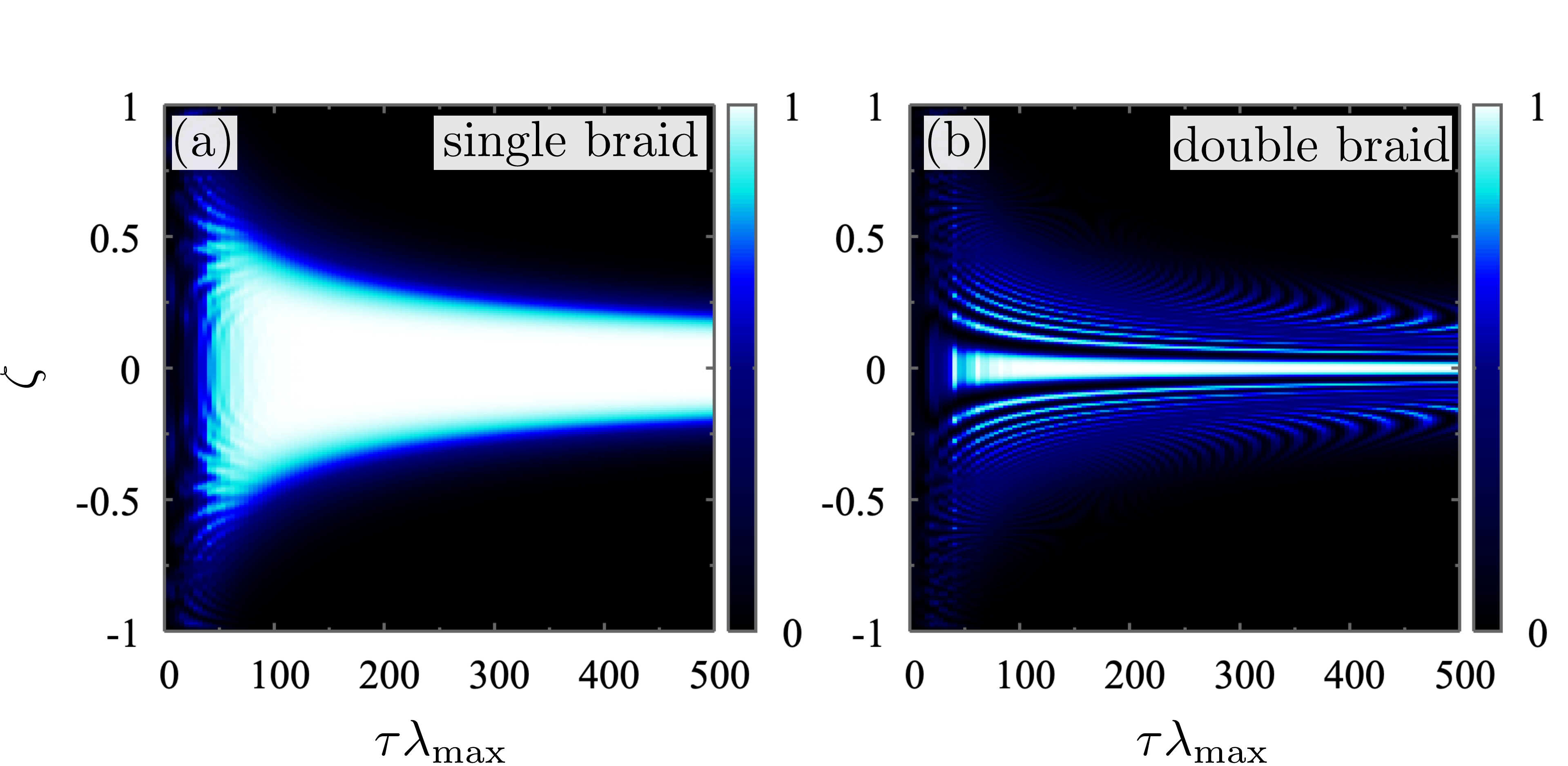}
\caption{Hybridization-induced braiding for a PMM system. Probability of measuring the target parity in both PMM systems after a single (a) and a double (b) braid protocol. Reprinted and Adapted
from Ref.~\cite{tsintzis2023roadmap} under CC-BY-4.0 license, \textcopyright~2022, The Author(s).}
\label{fig:PMM_starBraid}
\end{figure}
\end{example}

\begin{example}{Hybridization-based braiding with PMMs}
We conclude the section discussing the hybridization-induced braiding for PMMs, already introduced in Sec.~\ref{Sec:MajoranaBraid} for topological MBSs, see also Fig.\ref{fig:starBraid}. The introduced protocol, based on switching on and off the coupling between neighboring MBSs, provides the same result as spatially braiding two MBSs. In the ideal case, the system that is initialized $|0,0\rangle$ basis ends up in the state $(|0,0\rangle i\pm|1,1\rangle)/\sqrt{2}$, where the sign depends on the braid direction. For fast enough protocols, the main effect of a non-perfect Majorana localization is a dynamical phase that accumulates in the system. A projective measurement on the fermion parity state of the PMMs is not sensitive to the relative phase between the  $|0,0\rangle$ and $|1,1\rangle$ states, explaining the relatively large plateau in Fig.~\ref{fig:PMM_starBraid}(a). For sufficiently large $\zeta$ or slow operations, also the relatives weights between $|0,0\rangle$ and $|1,1\rangle$ states become different. The accumulated phase can be revealed after a second braid, as a deviation from the expected $|1,1\rangle$ result, Fig.~\ref{fig:PMM_starBraid}(b).
\end{example}

Therefore, PMMs in minimal Kitaev chains are promising systems to demonstrate the non-Abelian properties of Majorana states, a main challenge in the field. This will pave the way toward topological quantum computation, allowing also to explore the combination of non-Abelian quasiparticles with other systems. However, they fine-tuned nature of PMMs lead to a sensitivity to some sources of noise. This is an obstacle toward the scaling up of the system toward realistic applications. Longer Kitaev can help at solving this issue.

\section*{Acknowledgements}
We would like to thank all our co-workers in the topic of superconducting devices. 
We acknowledge financial support from the Horizon Europe Framework Program of the European Commission through the European Innovation Council Pathfinder grant no. 101115315 (QuKiT), the Spanish Ministry of Science through Grants PID2022-140552NA-I00, PID2021- 125343NB-I00 and TED2021-130292B-C43 funded by MCIN/AEI/10.13039/501100011033, "ERDF A way of making Europe", the Spanish CM “Talento Program” (project No. 2022-T1/IND-24070), and European Union NextGenerationEU/PRTR. Support by the CSIC Interdisciplinary Thematic Platform (PTI+) on Quantum Technologies (PTI-QTEP+) is also acknowledged.

\bibliographystyle{unsrt}
\bibliography{bibliography}

\end{document}